\documentclass[journal=jpclcd,manuscript=article,layout=twocolumn]{achemso}

\usepackage{graphicx} 
\usepackage{siunitx}
\usepackage{booktabs, lipsum}
\usepackage{amsmath}
\usepackage{cuted}
\mciteErrorOnUnknownfalse

\newcommand{\onlinecite}[1]{\hspace{-1 ex} \citenum{#1}} 

\title{Environmentally driven symmetry-breaking quenches dual fluorescence in proflavine}

\author{Kye E. Hunter}
\affiliation{Department of Chemistry, Oregon State University, Corvallis, Oregon 97331, United States}
\author{Yuezhi Mao}
\affiliation{Department of Chemistry, San Diego State University, San Diego, California 92182, United States}
\author{Alex W. Chin}
\affiliation{Sorbonne Universit\'{e}, CNRS, Institut des NanoSciences de Paris, 4 place Jussieu, 75005 Paris, France}
\author{Tim J. Zuehlsdorff}
\affiliation{Department of Chemistry, Oregon State University, Corvallis, Oregon 97331, United States}
\email{tim.zuehlsdorff@oregonstate.edu}

\date{\today}

\begin{document}

\maketitle

\begin{tocentry}
\includegraphics[width=\textwidth]{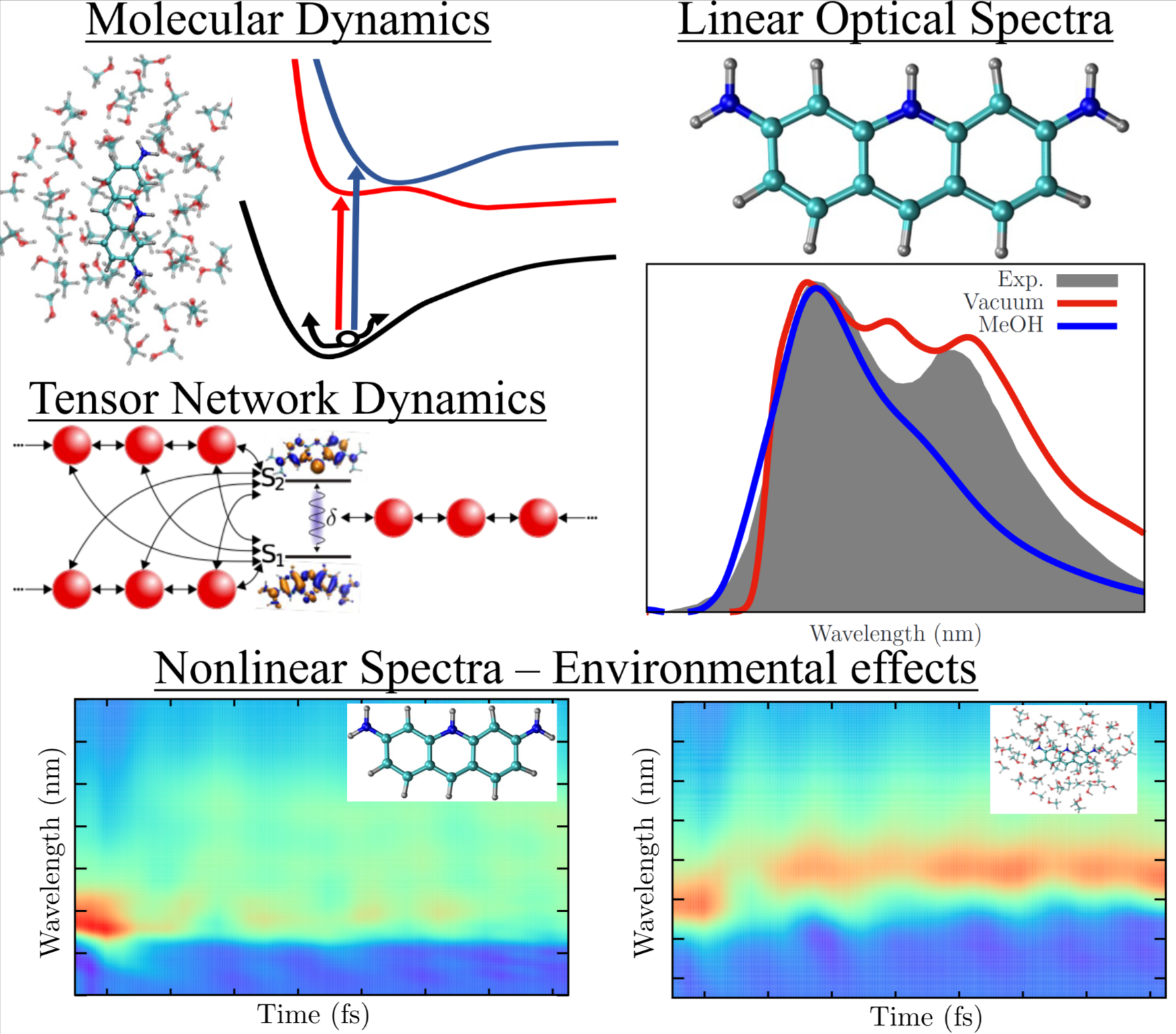}
\end{tocentry}

\begin{abstract}
Nonadiabatic couplings between several electronic excited states are ubiquitous in many organic chromophores and can significantly influence optical properties. A recent experimental study demonstrated that the proflavine molecule exhibits surprising dual fluorescence in the gas phase, that is suppressed in polar solvent environments. Here, we uncover the origin of this phenomenon by parameterizing a linear-vibronic coupling (LVC) Hamiltonian from spectral densities of system-bath coupling constructed along molecular dynamics trajectories, fully accounting for interactions with the condensed-phase environment. The finite-temperature absorption, steady-state emission, and time-resolved emission spectra are then computed using powerful, numerically exact tensor network approaches. We find that the dual fluorescence in vacuum is driven by a single well-defined coupling mode, but is quenched in solution due to dynamic solvent-driven symmetry-breaking that mixes the two low-lying electronic states. We expect the computational framework developed here to be widely applicable to the study of non-Condon effects in complex condensed-phase environments. 
\end{abstract}

Nonadiabatic effects due to the interactions of multiple electronic excited states are ubiquitous in many molecular systems, ranging from biological light-harvesting complexes to quantum dots and organic optoelectronic materials\cite{Tamura2013,Northey2018,alvertis2019switching,Arsenault2020,Levine2023}. They arise when nuclear motion mixes adiabatic electronic states, the true eigenstates of the electronic Hamiltonian, representing a breakdown of the Born-Oppenheimer approximation\cite{Curchod2018,Birchner2018}. Signatures of such nonadiabatic effects\cite{Domcke1981,Domcke2012} can be commonly observed in linear and nonlinear optical spectra of molecules and molecular assemblies, with prominent examples including dark (dipole-forbidden) transitions contributing to spectral lineshapes through intensity borrowing effects\cite{Orlandi1973,Domcke2012,Aranda2021,Dunnett2021}, anomalous emission properties such as observed for molecules undergoing twisted intramolecular charge-transfer (TICT)\cite{Grabowski2003,Park2013,Curchod2017}, or molecules violating Kasha's rule\cite{Demchenko2017,Proflavine_exp,alvertis2019switching,braun}. First-principles modeling of these effects is highly challenging, as explicit couplings between the electronic and nuclear degrees of freedom have to be accounted for\cite{Curchod2018,antikasha}. Additionally, most systems of interest are embedded in condensed-phase environments such as solvents or proteins, and understanding how environmental degrees of freedom influence energy relaxation processes in multiple coupled electronic states poses a major theoretical challenge\cite{Burghardt2004,Santoro2021,Cerezo2023b}. In this work, we introduce a first-principles approach, based on combining molecular dynamics (MD) sampling of vibronic couplings\cite{Zuehlsdorff2019b} with tensor-network methods\cite{Chin2010,kloss2018time,Schroder2019,Dunnett2021}, that is capable of capturing environmental interactions on several coupled excited states in absorption, fluorescence, and time-resolved spectroscopy experiments. We showcase the strengths of the approach by uncovering the origin of recently reported dual fluorescence in the proflavine molecule\cite{Proflavine_exp}.

\begin{figure*}
    \centering
    \includegraphics[width=0.9\textwidth]{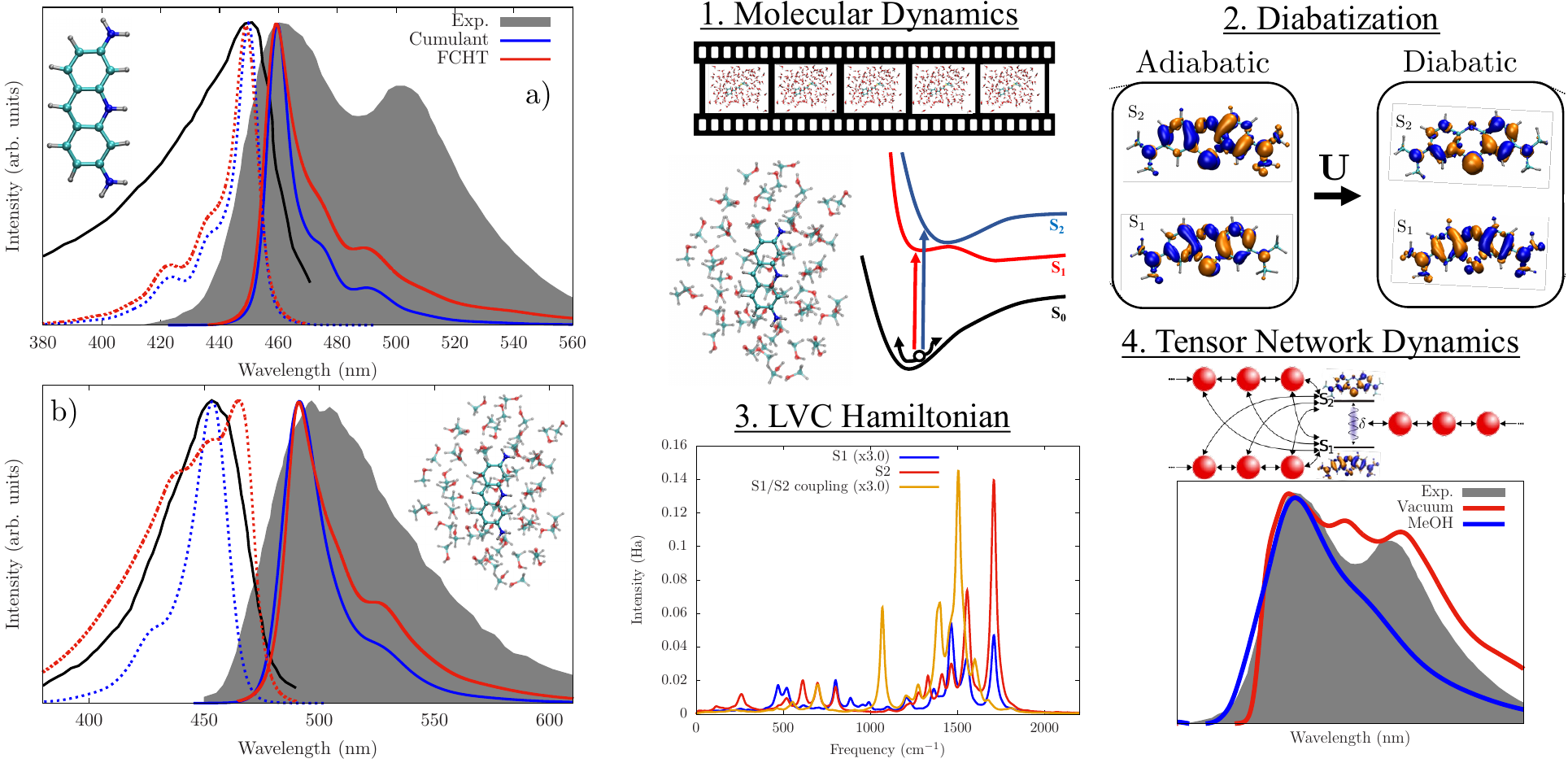}
    \caption{Left: Experimental absorption and fluorescence lineshapes of proflavine\cite{Proflavine_exp} in vacuum (a) and methanol (b) in comparison with computed spectra in the cumulant (blue) and FCHT (red) approach. Right: Schematic outlining the computational framework employed in this work. S$_1$ and S$_2$ energy gaps are sampled along ground state MD trajectories (1), which, after a diabatization (2), yields a parameterization of the LVC Hamiltonian (3). Non-perturbative quantum dynamics of the system is performed using tensor network approaches (4) to obtain absorption and fluorescence lineshapes in the presence of strong non-adiabatic coupling between the electronic states. }
     \label{main_fig:1}
\end{figure*}

Proflavine is a small cationic organic molecule with C$_\textrm{2v}$ symmetry (see Fig.~\ref{main_fig:1}), whose photophysical properties have been well characterized by experimental\cite{Donald2011,Kumar2012,Arden-Jacob2013,Proflavine_exp} and theoretical studies\cite{Kostjukova2022,Savenko2023} in various condensed-phase environments. In polar solvents, the absorption and fluorescence lineshapes are featureless and close to mirror images of each other, but show an unusually large Stokes shift for a small, rigid molecule that should undergo only minimal nuclear reorganization upon excitation (0.24~eV and 0.36~eV in methanol and water respectively\cite{Proflavine_exp}). As demonstrated by recent experimental work\cite{Proflavine_exp}, in the gas phase the mirror symmetry of absorption and fluorescence lineshapes of the molecule is broken, with the emission spectrum exhibiting strong excitation wavelength-dependent dual fluorescence (see Fig.~\ref{main_fig:1}). The origin of this dual fluorescence has been ascribed to a second excited state interacting with the bright, locally excited state responsible for the absorption spectrum. Reported\cite{Proflavine_exp} time-dependent density-functional theory (TDDFT)\cite{Runge1984,Casida1995} calculations have revealed two close-lying excited states, one dark and one bright, in the Condon region, lending support to the interpretation of the experimental data in terms of interacting excited states. However, an explanation for the precise mechanism of dual fluorescence, its strong suppression in a variety of condensed-phase environments, as well as other anomalous spectroscopic features such as the large Stokes shift in solution, has so far been lacking. To arrive at such an explanation, first-principles modeling of the finite-temperature quantum dynamics of the molecule in vacuum and its condensed-phase environment is needed. Additionally, and apart from a just better understanding of proflavine, such an approach could also provide general, microscopic insights into molecular dual fluorescence, which has emerging applications in areas such as medical imaging of drug uptake \cite{xue2015probe,cellimaging}, single component, white-light OLEDS \cite{yuan2019fluorescence}, and sensing \cite{qian2017high}.  

The absorption and emission lineshape of a complex condensed phase system can expressed in terms of Fourier transforms of quantum correlation functions of the transition dipole operator: 
\pagebreak
\begin{strip}
\begin{eqnarray} \nonumber
\sigma_\textrm{abs}(\omega)&=&\alpha_\textrm{abs}(\omega) \mathcal{FT}\left[\left \langle  \hat{\mu}^-(t) \hat{\mu}^+(0)\right\rangle_{\rho_0}\right] \\
\sigma_\textrm{emi}(\omega)&=&\alpha_\textrm{emi}(\omega) \mathcal{FT}\left[\left \langle \hat{\mu}^+(t) \hat{\mu}^-(0) \right \rangle_{\rho_\textrm{ex}}\right] 
\label{eqn:abs_emi}
\end{eqnarray}
\end{strip}
where $\hat{\mu}^-=\sum_{i=1}^{{N}_\textrm{ex}} \mu_{0 i}|S_{0}\rangle \langle S_i |$ and $\hat{\mu}^+=\left(\hat{\mu}^-\right)^\dagger$ are the transition dipole operators causing transitions between the electronic ground state $|S_0\rangle $ and the excited states $\{ |S_i \rangle \}$, and $\mu_{0i}$ denotes the electronic transition dipole moment between the ground state and state $i$. The frequency-dependent prefactors $\alpha_\textrm{emi}$ and $\alpha_\textrm{abs}$ depend on the experimental conditions\cite{Baiardi2013}  and will be set to 1 for the remainder of this work for convenience. The density matrices $\rho_0$ and $\rho_\textrm{ex}$ denote equilibrium (relaxed) density matrices of the nuclear degrees of freedom for the electronic ground and excited states, respectively. 

If the excited states are decoupled, Eqn.~\ref{eqn:abs_emi} can be approximately evaluated in a number of ways, with standard methods including the Franck-Condon Herzberg-Teller (FCHT) approach\cite{Baiardi2013,deSouza2018} and the cumulant method\cite{Mukamel-book,Zuehlsdorff2019b}. In FCHT, chromophore nuclear degrees of freedom are approximated as harmonic, and any condensed-phase environment is generally modeled through polarizable continuum models (PCMs)\cite{Cammi_2005}, although schemes exist to approximately include the effect of direct solute-solvent interactions\cite{Cerezo2015,Zuehlsdorff2018,Zuehlsdorff2018b,cerezo2019}. In the cumulant scheme, coupling to nuclear motion is expressed in terms of \emph{spectral densities} $\mathcal{J}(\omega)$ of system-bath coupling, which can be parameterized directly from correlation functions of vertical excitation energies sampled along molecular dynamics (MD) trajectories\cite{Valleau2012, Shim2012,Lee2016,Lee2016b,Loco2018b,Zuehlsdorff2019b,Cignoni2022}. Importantly, an explicit MD sampling of the system in solution preserves all coupling of environmental degrees of freedom to electronic excitation, and the influence of solvent fluctuations on the optical spectrum is contained in $\mathcal{J}(\omega)$. However, if several excited states are strongly coupled in the Condon region, both the FCHT and the cumulant approach are insufficient in describing linear optical spectra. Fig.~\ref{main_fig:1}~a) and b) show computed FCHT and cumulant spectra for proflavine in vacuum and methanol (see SI Secs.~I.D., I.E. and I.F. for computational details). The predicted spectra are in poor agreement with experiment, failing to reproduce the pronounced broadening of the emission lineshape in methanol and the dual fluorescence in vacuum. These results indicate that the optical properties of proflavine can indeed be ascribed to non-Condon effects due to interacting electronic exited states. 

To account for non-Condon effects, we make use of the linear-vibronic coupling (LVC) Hamiltonian\cite{Koppel1984,Domcke2012} for a three level system, consisting of an electronic ground- and two electronic excited states: 

\begin{strip}
\begin{eqnarray} \nonumber
\hat{H}_\textrm{LVC}&=&\hat{H}_\textrm{BOM}+\hat{H}_\textrm{c} 
\\
&=& \begin{pmatrix}
    {H}_0 & \mu_{01} & \mu_{02} \\
    \mu_{10} & H_1 & 0 \\
    \mu_{20} & 0 & H_2 
    \end{pmatrix}+\sum_j^N
    \begin{pmatrix}
    0 & 0 & 0 \\
    0 & 0 & \Lambda_j \hat{q}_j  \\
    0 & \Lambda_j \hat{q}_j  & 0  \\
    \end{pmatrix}.
    \label{eqn:LVChamiltonian}
\end{eqnarray}
\end{strip}
Here, $\hat{H}_\textrm{BOM}$ is the Spin-Boson or Brownian Oscillator Model Hamiltonian\cite{Zuehlsdorff2019b} describing the linear coupling of each electronic excited state to nuclear degrees of freedom $\{\hat{q}_j\}$, and $\{\Lambda_j\}$ denotes the set of linear couplings between the two electronic excited states (see SI Sec.~I). The LVC Hamiltonian has been widely used to describe non-adiabatic effects in optical spectra and excited state relaxations of small- to medium-sized molecules\cite{Domcke1981,Worth1996,Capano2014,Papai2016, Neville2018,Aranda2021,Zobel2021,Green2021,Segalina2022,Segatta2023}. While the off-diagonal couplings $\{\Lambda_j\}$ pose a significant challenge in computing the non-equilibrium quantum dynamics in this system, for a relatively small number of (chromophore) vibrational modes, response functions such as Eqn.~\ref{eqn:abs_emi} can be calculated  using the numerically exact multi-configuration time-dependent Hartree (MCTDH) method\cite{Beck2000, Meyer2009}. However, to fully capture the interaction between the chromophore and its condensed-phase environment, it is necessary to consider the coupling of the electronic system to an infinite bath of collective chromophore-environment modes. 

We have recently shown\cite{Dunnett2021} that the LVC Hamiltonian of the form of Eqn.~\ref{eqn:LVChamiltonian} can be parameterized from a set of four continuous spectral densities that can be evaluated directly from MD simulations of the system on the ground-state PES:  
 \begin{eqnarray}\label{eq:spectral-density-from-classical-md} \nonumber
    \mathcal{J}_{0\alpha}(\omega) &\approx& \theta(\omega) \frac{\beta \omega }{2} \int \textrm{d}t\ e^{\textrm{i} \omega t} \ C^{\mathrm{cl}}_{0\alpha}(t) \\ \nonumber
    \mathcal{J}_\textrm{cross}(\omega) &\approx& \theta(\omega) \frac{\beta \omega }{2} \int \textrm{d}t\ e^{\textrm{i} \omega t} \ C^{\mathrm{cl}}_\textrm{cross}(t) \\
    \mathcal{J}_{12}(\omega) &\approx& \theta(\omega) \frac{\beta \omega }{2} \int \textrm{d}t\ e^{\textrm{i} \omega t} \ C^{\mathrm{cl}}_{12}(t).
\end{eqnarray}

Here, $\alpha=1,2$ labels diabatic electronic exited states 1 and 2, $C^\textrm{cl}_{0\alpha}=\langle E_{0\alpha}(t) E_{0\alpha}(0) \rangle_\textrm{cl}$, $C^\textrm{cl}_\textrm{cross}=\langle E_{01}(t) E_{02}(0) \rangle_\textrm{cl}$, $C^\textrm{cl}_{12}=\langle \delta_{12}(t) \delta_{12}(0) \rangle_\textrm{cl}$ are classical time-correlation functions, $\theta(\omega)$ denotes the Heaviside step function, and $\beta \omega/2$ is the harmonic quantum correction factor\cite{Egorov1999,Craig2004,Ramirez2004}. The spectral densities $ \mathcal{J}_{0\alpha}(\omega)$ express the coupling of each electronic excited state to nuclear degrees of freedom (due to so-called tuning modes), the cross-correlation spectral density $\mathcal{J}_\textrm{cross}(\omega)$ encodes to what degree the fluctuations in S$_1$ and S$_2$ are \emph{correlated} or \emph{anti-correlated}, and $\mathcal{J}_{12}(\omega)$ contains the off-diagonal coupling between the excited states  (due to coupling modes). A detailed description of the parameterization of the LVC Hamiltonian in terms of spectral densities can be found in SI Sec.~I. We note that if $\mathcal{J}_{12}(\omega)=0$, evaluating the linear optical spectrum for the LVC Hamiltonian parameterized in this way reduces to the well-known cumulant approach.  

The quantities $\{E_{01}(t),E_{02}(t), \delta_{12}(t)\}$ are a set of \emph{diabatic} excitation energies and the diabatic coupling that can be constructed from \emph{adiabatic} excitation energies computed with time-dependent density functional theory (TDDFT)\cite{Runge1984,Casida1995} along each snapshot of the MD trajectory following a diabatization procedure\cite{Subotnik2015}. In this work, a simple (quasi)-diabatization scheme based on the transition dipole moments of the two adiabatic electronic excited states\cite{Medders2017} is used (see SI Sec.~I~C). Importantly, the MD sampling of the fully solvated system encodes all coupling to the nuclear degrees of freedom of the solute and the solvent in the continuous spectral densities, in contrast to approaches based on parameterizing the LVC Hamiltonian from static calculations of the chromophore in a frozen (decoupled) solvent environment sampled from MD\cite{Green2021,Segalina2022,Cerezo2023}. 

To compute the finite-temperature quantum dynamics for the system subject to the spectral densities in Eqn.~\ref{eq:spectral-density-from-classical-md}, we make use of the thermalized time-evolving density operator with orthogonal polynomials (T-TEDOPA) formalism\cite{tamascelli2019efficient}. First, the continuous spectral densities describing the system-bath interaction are mapped to a $1D$ or quasi-$1D$ (tree) chain Hamiltonian\cite{Chin2010,Schroder2019,Dunnett2021}. This allows for an efficient representation of the wave function in terms of a matrix-product state (MPS)\cite{Haegeman2011,Haegeman2016,Schroder2019, kloss2018time}, and the system dynamics can then be evaluated using the powerful tensor-network approaches such as the one-site time-dependent variational principle (1TDVP) which is particularly suited for treating the long-range couplings that appear when inter-environment correlations are present (see SI Sec.~VI) \cite{Haegeman2011,Haegeman2016,Dunnett2021,PhysRevA.104.052204}.  Second, we make use of the effective spectral density approach of Tamascelli \emph{et al.}\cite{tamascelli2019efficient}, where finite-temperature quantum dynamics can be obtained from an initial pure state evolving under an effective, thermalized spectral density \cite{10.3389/fchem.2020.600731, PhysRevB.108.195138}.  A schematic of the computational framework used in this work can be found in Fig.~\ref{main_fig:1} (see also SI Sec.~II). 

Recently, some of the authors have shown that the outlined approach can be used to compute finite-temperature linear absorption spectra of a chromophore in its condensed-phase environment, under the influence of a conical intersection between a bright and a dark state near the Condon region.\cite{Dunnett2021}. The dipole-dipole response function necessary for computing $\sigma_\textrm{abs}(\omega)$ can be obtained by propagating a superposition of the electronic ground state and the two electronic excited states weighted by their respective electronic transition dipole moments under the LVC Hamiltonian, and measuring the non-Hermitian operator $\hat{\mu}^-$ with respect to the time-evolved state (see SI Sec.~II~D). To uncover the origin of dual fluorescence in the proflavine molecule, in this work we extend the formalism to the evaluation of emission spectra. It can be shown (see SI Sec.~II~E), that the steady-state emission spectrum $\sigma_\textrm{emi}(\omega)$ can be constructed by considering the superposition of a relaxed excited state wavefunction $\Psi_\textrm{ex}$ and the same wavefunction projected to the electronic ground state through the action of $\hat{\mu}^-$. Measuring the expectation value of the non-hermitian operator $\hat{\mu}^+$ with respect to the superposition state evolving under the LVC Hamiltonian then yields the appropriate dipole-dipole correlation function to compute $\sigma_\textrm{emi}(\omega)$. The initial relaxed excited state wavefunction $\Psi_\textrm{ex}$ can be obtained from the final state of the simulation constructing the absorption lineshape. Additionally, since the initial state for the emission spectrum $\Psi_\textrm{ex}$ explicitly depends on the relaxation time $t_\textrm{delay}$ in the electronic excited states, varying  $t_\textrm{delay}$ gives access to time-resolved emission spectra, allowing us to probe the ultrafast evolution of the stimulated emission band in a transient absorption experiment. 

\begin{figure*}
    \centering
    \includegraphics[width=0.7\textwidth]{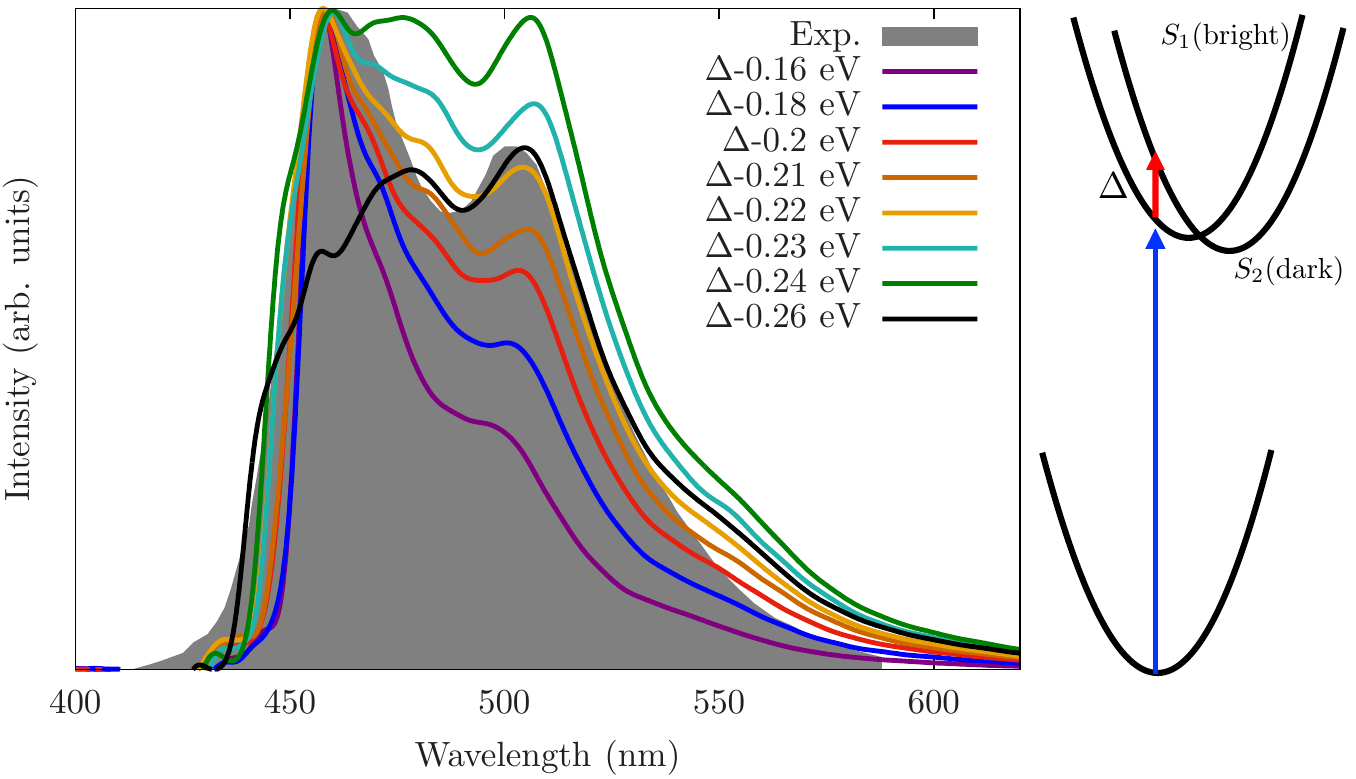}
    \caption{Computed fluorescence spectrum for proflavine in vacuum, where $\Delta_{12}$, the S$_1$-S$_2$ gap in the Condon region, is varied systematically. All spectra are shifted and scaled to align with the experimental lineshape\cite{Proflavine_exp}. Fluorescence spectra are computed after a 100~fs relaxation on the excited state PES.}
     \label{main_fig:2}
\end{figure*}

We are now in a position to study the origin of pronounced dual fluorescence of the proflavine molecule in vacuum and the mechanism of its suppression in solvent environments. Mixed quantum mechanical/molecular mechanical (QM/MM)\cite{Warshel1976} simulations of proflavine in vacuum, acetonitrile and methanol were carried out and spectral densities were constructed from 20~ps of trajectory per system, sampled every 2~fs for a total of 10,000 snapshots. All vertical excitation energies were computed within full TDDFT (i.e. not relying on the Tamm-Dancoff approximation\cite{Hirata1999}) and the CAM-B3LYP\cite{Yanai2004} functional was used throughout. A full description of the computational details can be found in SI Sec. ID and E, and all parameters entering the resulting LVC Hamiltonian are listed in SI Sec. IV. 

Along the sampled MD trajectory in vacuum, we find that the bright S$_1$ and the dark S$_2$ states are indeed close in energy when computed with the CAM-B3LYP functional, with an average separation of 0.234 eV in the Condon region. The energy gap is found to be highly dependent on the choice of DFT functional, with B3LYP predicting a significantly smaller splitting at the ground state optimized geometry\cite{Proflavine_exp}. We have performed benchmark calculations of a single optimized ground state geometry using the equation-of-motion coupled-cluster singles and doubles (EOM-CCSD) method\cite{Stanton1993equation, Comeau1993equation, Krylov2008equation}, resulting in an S$_1$-S$_2$ gap of 0.394~eV. Since the average S$_1$-S$_2$ gap controls the location of the conical intersection between the two states in the Condon region, it is expected to have a strong influence on the system dynamics upon excitation. We explore this sensitivity in Fig.~\ref{main_fig:2}, by systematically changing the average gap  $\Delta$ from the computed TDDFT value, leaving all other system parameters in the LVC Hamiltonian unchanged. 

As shown in Fig.~\ref{main_fig:2}, the emission lineshape of proflavine is highly sensitive to the average S$_1$-S$_2$ gap in the Condon region. Reducing the gap, corresponding to moving the conical intersection between the two states closer to the Condon region, systematically increases the shoulder of the emission lineshape. With $\Delta-0.22$~eV (i.e., the average TDDFT gap further narrowed by 0.22 eV), we obtain a spectrum with two distinct peaks in excellent agreement with the experimental lineshape exhibiting dual fluorescence. Under this parameterization, about 80\% of the population transfers from the bright S$_1$ to the dark S$_2$ state within 50~fs upon excitation (see SI Sec.~VI). A comparison of the lineshape obtained to the pure cumulant spectrum displayed in Fig.~\ref{main_fig:1} demonstrates that the optical properties of proflavine in vacuum can indeed be ascribed to strong non-Condon effects caused by two interacting excited states. Interestingly, using the $\Delta$ parameter predicted from EOM-CCSD calculations yields very little population transfer to S$_2$ and only minor non-Condon effects in the resulting spectrum (see SI Sec.~IV and VII), indicating that the method significantly overestimates the S$_1$-S$_2$ splitting in the Condon region. We note that although EOM-CCSD typically yields an error within 0.15 eV for molecular excitation energies, \cite{Loos2018mountaineering,Loos2020mountaineering} these errors, when propagated to the S$_1$-S$_2$ gap, can have a significant impact on the simulated dynamics due to the above-discussed sensitivity. The large size of the proflavine molecule makes comparison with other higher-order methods to determine the appropriate splitting from first principles computationally unfeasible. While EOM-CCSD and CAM-B3LYP show some discrepancy in the computed S$_1$-S$_2$ splitting, both methods yield similar diabatic couplings for selected MD snapshots (see SI Sec.~V), suggesting that they predict similar curvatures of the diabatic PESs. 

Since different TDDFT functionals and higher-order methods like EOM-CCSD predict a range of S$_1$-S$_2$ energy gaps in the Condon region, for the purpose of this work we base calculations on the TDDFT gap (computed using CAM-B3LYP), reduced by a constant shift of 0.22~eV, following the good agreement with experiment obtained in Fig.~\ref{main_fig:2}. For the solvated systems in methanol and acetonitrile, the average TDDFT gap is corrected for by the same amount. Additionally, since the position of conical intersections can depend sensitively on environmental polarization effects\cite{Glover2024}, we perform calculations on a number of solvated snapshots where the solvent environment is treated fully quantum mechanically at the TDDFT level, rather than as fixed classical point charges (see SI Sec.~I.H). For methanol and acetonitrile, solvent polarization effects are found to increase the average S$_1$-S$_2$ gap by 62 and 63~meV, respectively. All results for solvated systems presented in the main text are obtained by correcting the computed average TDDFT S$_1$-S$_2$ gap both for the underestimated splitting in vacuum and solvent polarization effects. We stress that the corrected S$_1$-S$_2$ gap in the Condon region is the only adjustable parameter used in this work, and all other quantities entering the LVC Hamiltonian are directly obtained from the MD sampling of the system in vacuum or in the condensed phase.

\begin{figure*}
    \centering
    \includegraphics[width=1.0\textwidth]{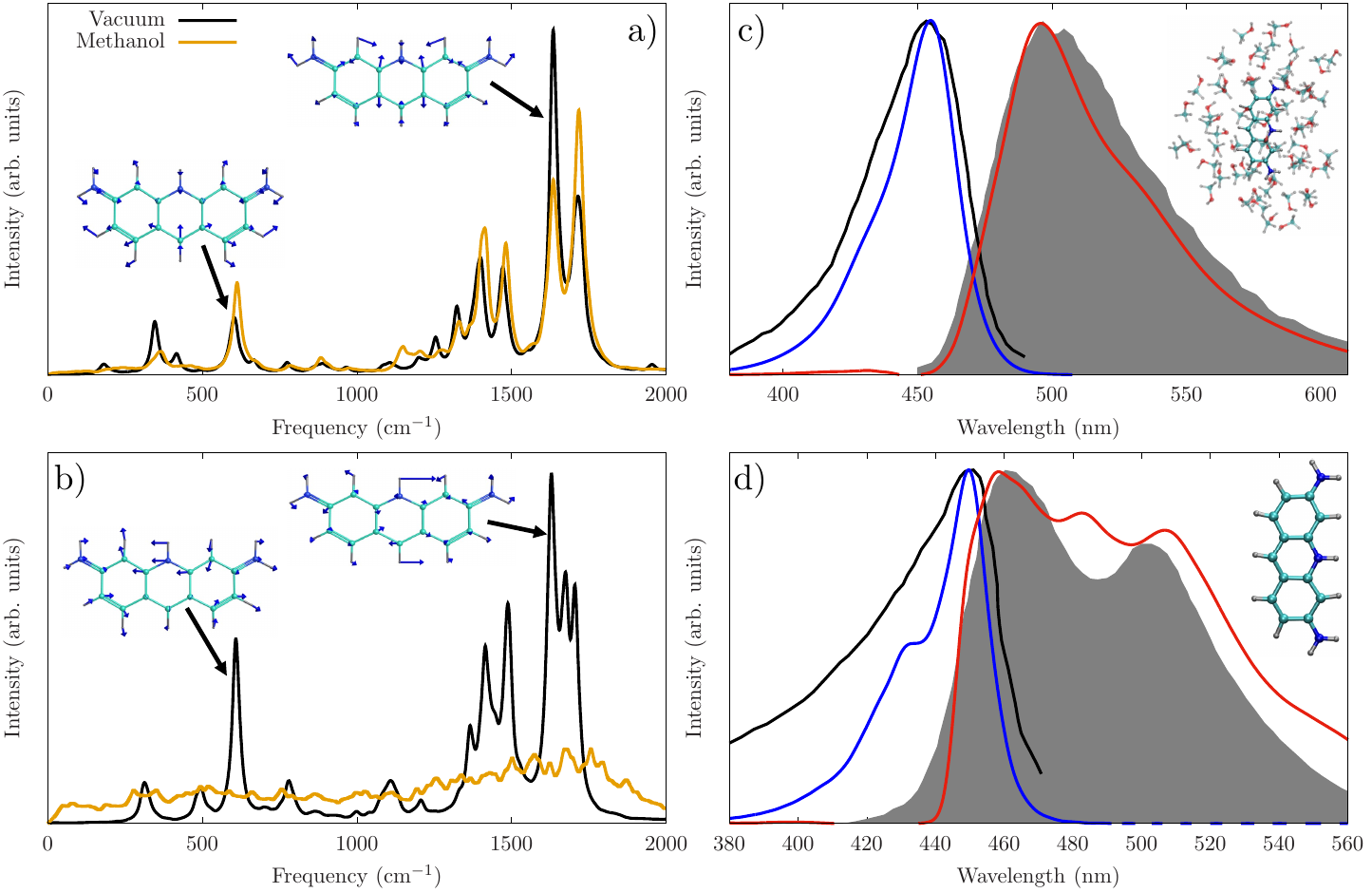}
    \caption{a) Spectral density (SD) for the dark S$_2$ state as computed in vacuum and MeOH, with dominant peaks assigned to normal modes b) SD for the S$_1$-S$_2$ coupling as computed in vacuum and MeOH, with dominant peaks assigned to normal modes c) Computed absorption (blue) and steady-state emission (red) spectrum for proflavine in MeOH in comparison with experimental data. d) Same as c) but for proflavine in vacuum. Emission spectra are computed as an average over 15 spectra with excited state relaxation times $t_\textrm{delay}$ ranging from 100-170~fs. All computed spectra are shifted by a constant amount such that the absorption spectrum aligns with the experimental spectrum\cite{Proflavine_exp}. An additional shift of -0.071~eV is applied to the emission spectrum of proflavine in MeOH to match the experimental spectrum position. }
     \label{main_fig:3}
\end{figure*}

Fig.~\ref{main_fig:3} shows  absorption and steady-state fluorescence spectra for proflavine in methanol (c) and vacuum (d) computed using the T-TEDOPA methodology outlined in this work, as well as the corresponding S$_2$ and coupling spectral densities $\mathcal{J}_{02}$ (a) and $\mathcal{J}_{12}$ (b). Additional results for proflavine in acetonitrile, which closely resemble the spectra in methanol, can be found in SI Sec.~VI.  Computed linear spectra for both methanol and vacuum are found to be in excellent agreement with experiment\cite{Proflavine_exp}, with the vacuum results showing dual fluorescence that strongly breaks the mirror symmetry of absorption and fluorescence lineshapes, and the results in methanol yielding broad, featureless spectra. No artificial broadening has been added to the computed spectra, such that the broad lineshapes are direct consequences of the low-frequency solvent interactions encoded in the spectral densities, as well as the finite-temperature effects accounted for in the T-TEDOPA formalism. The main discrepancy between computed and experimental lineshapes is a slight underestimation of the Stokes shift in methanol by about 71~meV. We ascribe this underestimation to two main sources: 1) an insufficient relaxation time of only up to 165~fs on the excited state PES before computing the emission lineshape results in an incomplete energy relaxation via the low-frequency solvent modes contained in the spectral density and b) a lack of dynamic solvent polarization effects that are known to enhance low frequency regions of the spectral density in condensed-phase systems\cite{Zuehlsdorff2020}.

Interestingly, while the dual fluorescence is suppressed in methanol, the spectral lineshape still exhibits strong non-Condon contributions (see SI Sec.~VII) and a similar amount of S$_1$ population undergoes ultrafast relaxation to the S$_2$ state upon excitation as in vacuum. The quenching of dual fluorescence thus cannot be ascribed to a reduced coupling between the two excited states. To gain insight into the mechanism at play, we instead turn to the spectral densities (see Fig.~\ref{main_fig:3}~a) and b)). We find that the spectral densities $\mathcal{J}_{01}$ and $\mathcal{J}_{02}$ containing the tuning modes causing fluctuations in the S$_1$ and S$_2$ energies only undergo minor changes due to the influence of the condensed-phase solvent environment. However, $\mathcal{J}_{12}$ coupling the two excited states is extremely sensitive to solvent interactions. The sharp peaks in vacuum that can be directly assigned to specific molecular coupling modes disappear in solution and the entire spectral density becomes broad and featureless. This change in the coupling spectral density can be shown to arise from the direct electrostatic interactions between the chromophore excited states and the solvent environment, rather than solvent-induced changes in the chromophore motion (see SI Sec.~VIII). 

We ascribe the dramatic changes to the coupling spectral density in solution to solvent-driven symmetry breaking. A normal mode analysis of proflavine in vacuum (SI Sec.~I.F) shows that the tuning modes contained in $\mathcal{J}_{01}$ and $\mathcal{J}_{02}$ can be assigned to symmetric stretching modes, whereas the coupling modes responsible for mixing the two excited states are due to asymmetric stretching modes breaking the C$_\textrm{2v}$ symmetry of the molecule (see Fig.~\ref{main_fig:3} b)). Due to the structure of the proflavine molecule, any uncorrelated solvent fluctuations around the two --NH$_2$ sites cause C$_\textrm{2v}$ symmetry breaking, resulting in the broad featureless coupling spectral density mixing the two excited states. On the other hand, solvent fluctuations have little influence on the $\mathcal{J}_{01}$ and $\mathcal{J}_{02}$ spectral density, as the solvent fluctuations in the first solvation shell generally do not preserve C$_\textrm{2v}$ symmetry. To demonstrate that the broad featureless coupling spectral density is indeed responsible for the quenching of dual fluorescence, we compute absorption and fluorescence spectra based on the trajectory of proflavine in methanol, but with the condensed-phase environment stripped away when computing vertical excitation energies and spectral densities (see SI Sec.~VIII). The coupling spectral density computed without the effect of the electrostatic solvent environment is found to closely resemble the spectral density of proflavine in vacuum, and the resulting emission spectrum displays a strongly enhanced dual-fluorescence shoulder. 

\begin{figure*}
    \centering
    \includegraphics[width=1.0\textwidth]{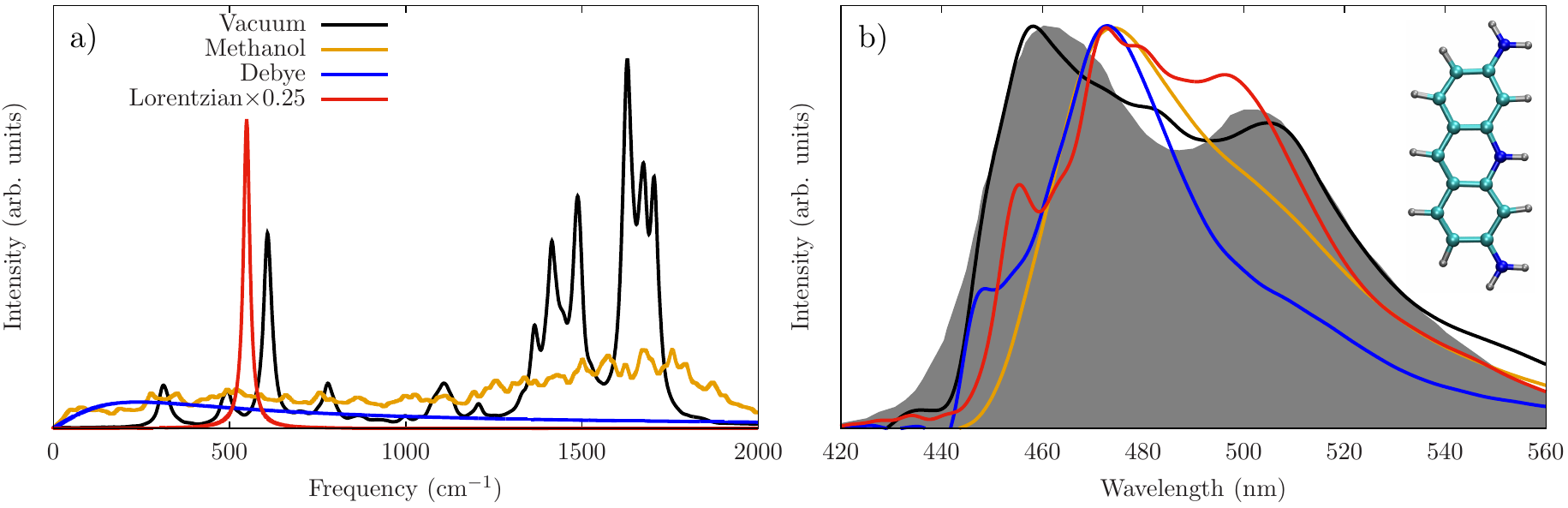}
    \caption{a) Coupling spectral density (SD) for proflavine in vacuum and MeOH, in comparison with two model SDs: A Debye type SD and a Lorentzian peak centered at $\omega=0.0025~\textrm{Ha}=549~\textrm{cm}^{-1}$. The Lorentzian and Debye SDs contain the same total reorganization energy as the vacuum SD. The Lorentzian peak is scaled by a factor of 0.25 to ease comparison. b) Resulting fluorescence spectra for proflavine in vacuum, MeOH, and proflavine in vacuum with the coupling SD replaced by the two model SDs, computed for a time delay of 100~fs, in comparison with the experimental lineshape\cite{Proflavine_exp}.}
     \label{main_fig:4}
\end{figure*}

To analyze the role of the coupling spectral density on the non-equilibrium quantum dynamics and the resulting steady-state absorption and fluorescence lineshapes, we perform several calculations on proflavine in vacuum using model coupling spectral densities (see Fig.~\ref{main_fig:4} and SI Sec.~IX). All model spectral densities are chosen to have a constant total reorganization energy to keep the total coupling between the excited states unaltered, and all other Hamiltonian parameters apart from $\mathcal{J}_{12}$ are unchanged. Specifically, we consider model spectral densities of the Debye type to model solvent coupling over a broad range of frequencies and Lorentzian spectral densities corresponding to a single molecular coupling mode. We find that the broad Debye spectral density, while causing significant population transfer to S$_2$, results in a spectrum in close agreement with the experimental lineshape of proflavine in methanol, lacking any pronounced dual fluorescence. Interestingly, the spectrum computed under the Debye spectral density retains a small peak at 450~nm that we ascribe to quenched fluorescence from the bright S$_1$ state. This quenching, resulting in a red-shifted main emission peak, is likely responsible for the anomalous large Stokes shift of proflavine in solution. For a Lorentzian spectral density centered at 549~cm$^{-1}$, close to the dominant coupling mode in the mid-frequency range observed for proflavine in vacuum, the emission spectrum gains an intense shoulder in the region of the second fluorescence peak in the experimental lineshape. Additionally, the onset peak at 450~nm is strongly enhanced. Further model calculations (SI Sec.~IX) show the spectrum to be highly sensitive to the position of the Lorentzian peak, with a significantly higher or lower frequency resulting in a quenching of the dual fluorescence. 

\begin{figure*}
    \centering    \includegraphics[width=1.0\textwidth]{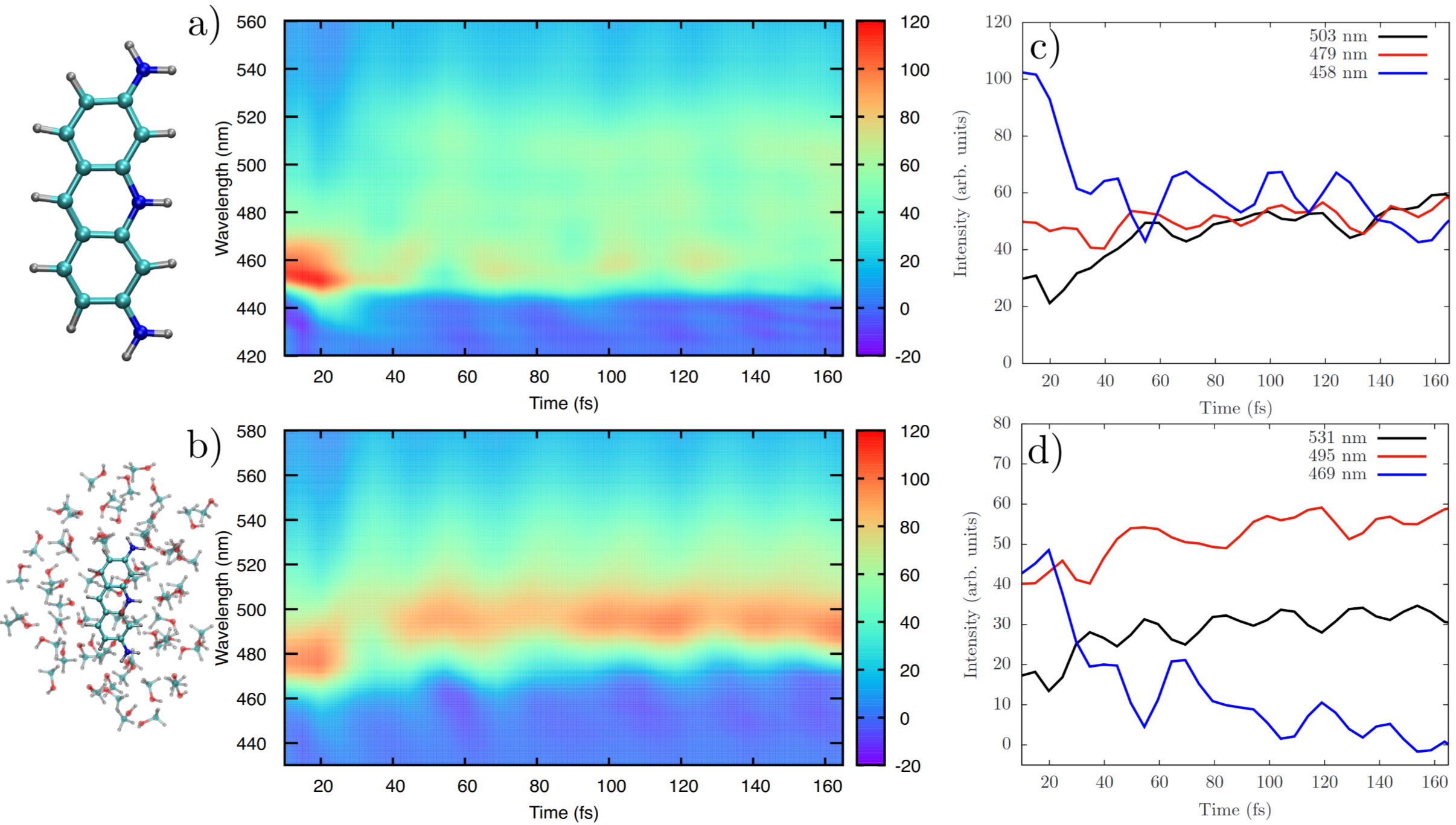}
    \caption{Time-resolved stimulated emission spectrum of proflavine as simulated in a) vacuum and b) Methanol. The plots in c) and d) represent slices through the nonlinear spectra a) and b) at specific wavelengths.}
     \label{main_fig:5}
\end{figure*}

We can conclude that the dual fluorescence properties of proflavine in vacuum are largely driven by an intense  mid-frequency coupling mode at approximately 610~cm$^{-1}$. In solution, this coupling mode is suppressed due to a wide range of solvent fluctuations breaking the C$_\textrm{2v}$ symmetry of the molecule and resulting in the mixing of the two electronic excited states over a broad frequency range. While the dual fluorescence is suppressed in solution, both the vacuum and the condensed-phase lineshapes are strongly influenced by non-Condon effects and the emitting state in both cases is of a highly mixed character, in line with previous experimental interpretations\cite{Kumar2012,Proflavine_exp}. Observed spectral features, such as the large Stokes shift in solution and the significant spectral broadening in both vacuum and methanol, can be directly ascribed to the mixing of the two low-lying excited states. 

The computational approach outlined in this work also allows for the evaluation of time-resolved fluorescence spectra, and a comparison of proflavine in vacuum and methanol can be found in Fig.~\ref{main_fig:5}. We note that the strong dual fluorescence features in vacuum are established after only 40~fs, indicating ultrafast relaxation of the excited state wavepacket to a mixed S$_1$-S$_2$ state. After the initial relaxation the resulting spectral lineshape is largely unchanged, but the relative intensities of the two fluorescent peaks (at approximately 458 nm and 503 nm respectively, see Fig.~\ref{main_fig:5}~c) undergo anti-correlated oscillations that persist through the entirety of the simulation out to 165~fs. For proflavine in methanol, after an initial fast relaxation over approximately 40~fs, a single intense emission peak (at approximately 495 nm, see Fig.~\ref{main_fig:5}~d) is established. The signal undergoes a continuous, slow dynamic Stokes shift, that is not completed at the end of the 165~fs of simulated quantum dynamics. Like in vacuum, the red shoulder of the spectrum at 531~nm exhibits oscillating intensity fluctuations through the entirety of the simulation. The oscillatory features in the emission spectrum are found to be strongly quenched if the S$_1$-S$_2$ energy gap in the Condon region is increased by just 40~meV (see SI Sec.~X) and thus represent spectroscopic signatures of the coherent vibronic coupling between the two electronic excited states upon relaxation. 

In summary, we have presented a general approach to model the finite-temperature absorption, steady-state fluorescence, and time-resolved fluorescence of systems with multiple coupled excited states in arbitrary condensed-phase environments. Coupling of the states to nuclear degrees of freedom is represented through spectral densities that can be computed directly from MD simulations of the system in its ground state equilibrium and the non-perturbative quantum dynamics of the system are computed using a numerically exact tensor network approach. We have demonstrated the strengths of the approach by unveiling the origin of recently observed strong dual fluorescence in the proflavine chromophore in vacuum, and its quenching in polar solvent environments. Dual fluorescence in vacuum was found to be mainly driven by a mid-frequency asymmetric stretching mode coupling the two low-lying excited states. In solution, a wide range of solvent fluctuations were found to mix the two lowest excited states, resulting in a featureless coupling spectral density, a quenched dual fluorescence and an increased Stokes shift. Our calculations demonstrate that both in vacuum and solution, the absorption, steady-state and time-resolved fluorescence spectra are strongly influenced by non-Condon effects, and in both cases the relaxed emitting state is of a highly mixed character. The proflavine molecule serves as an example of how non-Condon effects in optical spectra can be highly sensitive to the condensed-phase environment, especially if solvent fluctuations cause  a symmetry-breaking that mixes two excited states. We expect similar environmental coupling effects to play an important role in a wide range of systems, from solvated dyes to pigment-protein complexes, and recent work has explicitly pointed out the role of symmetry/symmetry breaking in the photophysics of light-harvesting dendrimer systems \cite{galiana23}. The capability of the outlined approach to treat arbitrarily structured spectral densities makes it well-suited to study these types of systems from first principles, as well as other systems whose non-equilibrium properties arise from delicate interplays of different dynamical quantum effects driven by open system physics \cite{antikasha}. Further methodological developments in the direction of time-resolved multidimensional spectroscopies, as begun here, also raise the possibility of exploring how these processes might be controlled or directed by external control of vibrations or electronic coherences via emerging experimental approaches such as mixed IR-UV/vis excitation or polaritonic effects \cite{gallop2024ultrafast, garcia2021manipulating,sokolovskii2023multi}.

\section{Acknowledgements}
T.J.Z acknowledges startup funding from Oregon State University. Y.M. acknowledges the support from San Diego State University startup fund and Seed Grant Program. The EOM-CCSD calculations used resources of the National Energy Research Scientific Computing Center (NERSC), a DOE Office of Science User Facility supported by the Office of Science of the U.S. Department of Energy
under Contract No.~DE-AC02-05CH11231 using NERSC award
DDR-ERCAP0024796. AWC wishes to acknowledge support from ANR Project ACCEPT (Grant No. ANR-19-CE24-0028)

\section{Data Availability}
The underlying data of this publication is made available under the following persistent DOI: 10.5281/zenodo.10699238. 
The python code used for generating the spectral densities from molecular dynamics data, as well as performing the chain mapping of the LVC Hamiltonian can be found the same DOI. 
All tensor network simulations were performed using a modified version of the MPSDynamics.jl Julia package\cite{dunnett_angus_2021_5106435} capable of running on GPUs that is available under the following DOI: 10.5281/zenodo.10712009.

\section{Supplementary Material}
See the supplementary material for a detailed description of the LVC Hamiltonian, its parameterization using TDDFT, and a derivation of the formalism to construct finite-temperature absorption, steady-state and time-resolved fluorescence spectra. Some supplementary results, such as benchmark calculations carried out with EOM-CCSD, the simulated lineshape of proflavine in acetonitrile, the emission spectrum in vacuum under several model coupling spectral densities, and the stimulated emission band of proflavine in vacuum with an increased S$_1$-S$_2$ gap are also presented. 

\bibliography{main}

\clearpage
\onecolumn
\setcounter{equation}{0}
\setcounter{figure}{0}
\setcounter{table}{0}
\setcounter{page}{1}
\setcounter{section}{0}
\setcounter{subsection}{0}
\makeatletter
\renewcommand{\theequation}{S\arabic{equation}}
\renewcommand{\thefigure}{S\arabic{figure}}
\renewcommand{\bibnumfmt}[1]{$^{\mathrm{S#1}}$}
\renewcommand{\citenumfont}[1]{S#1}
\renewcommand{\thetable}{S\arabic{table}}
\renewcommand{\section}[1]{
  \stepcounter{section}
  \setcounter{subsection}{0}
  \vspace{0.5\baselineskip}\par\noindent\textbf{
    \Roman{section}.\hspace*{0.2in}\uppercase{#1}
  }\vspace{0.5\baselineskip}\par
}
\renewcommand{\subsection}[1]{
  \stepcounter{subsection}
  \par\vspace{0.7\baselineskip}\noindent\textbf{
  \Alph{subsection}.\hspace*{0.2in}#1
  }\par\vspace{0.7\baselineskip}
}

{
\setlength{\parskip}{0pt}
\setlength{\baselineskip}{1.5em}

{\sffamily\noindent\textbf{Supplementary material for ``Environmentally driven symmetry-breaking quenches dual fluroescence in Proflavine''}}

\begin{quote}
{\sffamily{Kye E. Hunter}$^1$, {Yuezhi Mao}$^2$, {Alex W. Chin}$^3$, {Tim J. Zuehlsdorff}$^1$}

$^{1)}$ \emph{Department of Chemistry, Oregon State University, Corvallis, Oregon 97331, United States}

$^{2)}$ \emph{Department of Chemistry, San Diego State University, San Diego, California 92182, United States}

$^{3)}$ \emph{Sorbonne Universit\'{e}, CNRS, Institut des NanoSciences de Paris, 4 place Jussieu, 75005 Paris, France}


(Dated: \today)
\end{quote}

}

\pagebreak

\appendix






\section{Parameterizing the LVC Hamiltonian from MD}
In this work, the absorption, time-resolved fluorescence and steady-state fluorescence spectra of proflavine are computed by modeling the molecule as a three-level system, with a ground state and two (diabatic) excited states that are explicitly coupled by nuclear vibrations. The system Hamiltonian is taken to be the well-known linear vibronic coupling (LVC) Hamiltonian\cite{Koppel1984}. However, to account for the condensed phase environment from first principles, the LVC Hamiltonian is not parameterized, as is commonly done, from static calculations of a few selected reference structures\cite{Segalina2022}, but instead parameters are constructed directly from \emph{spectral densities} computed from molecular dynamics (MD) simulations. Some of the authors have recently outlined an approach to achieve such a parameterization, an have successfully applied it to the modeling of non-adiabatic effects in the absorption lineshape of methylene blue in water\cite{Dunnett2021}. Here, we summarize the main features of the computational approach of Ref.~\onlinecite{Dunnett2021} as it applies to the current work. 

\subsection{The linear vibronic coupling (LVC) Hamiltonian for a three-state model}
We describe proflavine as a three-level system of an electronic ground state and two (diabatic) electronic excited states coupled to nuclear motion described by $N$ harmonic oscillator modes. The LVC Hamiltonian can be defined in the following way:

\begin{equation}
    \hat{H}_\textrm{LVC}=\hat{H}_\textrm{BOM}+\sum_j^N
    \begin{pmatrix}
    0 & 0 & 0 \\
    0 & 0 & \Lambda_j \hat{q}_j  \\
    0 & \Lambda_j \hat{q}_j  & 0  \\
    \end{pmatrix}.
    \label{sieqn:LVChamiltonian}
\end{equation}
Here, $\hat{H}_{\textrm{BOM}}$ is the Brownian Oscillator Model (BOM) Hamiltonian taking the following form: 

\begin{equation}
 \hat{H}_\textrm{BOM}=   \begin{pmatrix}
    {H}_0 & \mu_{01} & \mu_{02} \\
    \mu_{10} & H_1 & 0 \\
    \mu_{20} & 0 & H_2 
    \end{pmatrix},
\end{equation}
where $\mu_{0n}=\mu^\dagger_{n0}$ denotes the transition dipole operator between the electronic ground and $n^{th}$ excited state, and $H_0$, $H_1$ and $H_2$ are the nuclear Hamiltonians of the electronic ground and the two excited states respectively. No dipole-induced transitions between the two excited states are considered in this work. Instead, transitions between the two excited states S$_1$ and S$_2$ can be induced by the off-diagonal couplings to nuclear motion included in Eqn.~\ref{sieqn:LVChamiltonian}. In the LVC Hamiltonian, this non-adiabatic coupling is taken to be linear with nuclear coordinates $\left \{\hat{q}_i\right \}$, and is described by the coupling parameters $\left\{\Lambda_i\right \}$. For proflavine, we only consider off-diagonal couplings $\left\{\Lambda_i\right \}$ between S$_1$ and S$_2$, not the electronic ground state and the two excited states. As such, non-radiative relaxation from the electronic excited states to the ground state is not included in the Hamiltonian of Eqn.~\ref{sieqn:LVChamiltonian}. Since the main focus of this work is modeling of absorption and time-resolved fluorescence spectroscopy in proflavine driven by short timescale dynamics, non-radiative relaxations to the electronic ground state can be safely discarded from the system dynamics.

The nuclear Hamiltonians $H_0$, $H_1$ and $H_2$ describing the diagonal coupling of the diabatic electronic states to nuclear motion are taken to be harmonic, with the excited state potential energy surfaces (PESs) linearly displaced with respect to the ground state surface: 

\begin{eqnarray}
H_0(\boldsymbol{\hat{p}},\boldsymbol{\hat{q}})&=&\sum_j^N  \left(\frac{\hat{p}_j^2}{2}+\frac{1}{2}\omega^2_j \hat{q}^2_j \right) \\
H_1(\boldsymbol{\hat{p}},\boldsymbol{\hat{q}})&=&\sum_j^N  \left(\frac{\hat{p}_j^2}{2}+\frac{1}{2}\omega^2_j \left(\hat{q}_j-K_j^{\{1\}}\right)^2 \right)+\omega^\textrm{ad}_{01} \\
H_2(\boldsymbol{\hat{p}},\boldsymbol{\hat{q}})&=&\sum_j^N  \left(\frac{\hat{p}_j^2}{2}+\frac{1}{2}\omega^2_j \left(\hat{q}_j-K_j^{\{2\}}\right)^2 \right) +\omega^\textrm{ad}_{02}.
\end{eqnarray}

Here, $\omega_j$ is the vibrational frequency of normal mode $j$, $\textbf{K}^{\{1\}}$ and $\textbf{K}^{\{2\}}$ denote the linear displacement vectors of the S$_1$ and S$_2$ PES minima with respect to the ground state minimum, and $\omega^\textrm{ad}_{01}$ and $\omega^\textrm{ad}_{02}$ are the adiabatic energy gaps of the two electronic excited states relative to the ground state respectively. 

\subsection{LVC parameters from spectral densities}

The parameters defining the LVC Hamiltonian, the set of vibrational frequencies $\{\omega_i\}$, the displacement vectors between ground- and excited state PES minima $\textbf{K}^{\{1\}}$ and $\textbf{K}^{\{2\}}$, as well as the couplings between diabatic states $\{\Lambda_i\}$, can be encoded in form of \emph{spectral densities} of system-bath coupling, such that:

\begin{eqnarray}
\mathcal{J}_{01}(\omega) &=&\frac{\pi}{2}\sum^N_j \omega^3_j \left(K^{\{1\}}_j\right)^2\delta (\omega -\omega_j),  \\ 
\mathcal{J}_{02}(\omega) &=&\frac{\pi}{2}\sum^N_j \omega^3_j \left(K^{\{2\}}_j\right)^2\delta (\omega -\omega_j), \\
    \mathcal{J}_{12}(\omega)&=&\frac{\pi}{2} \sum_j^N \frac{\Lambda^2_j}{\omega} \delta(\omega-\omega_j).
\end{eqnarray}

Here, $\mathcal{J}_{01}(\omega)$ and   $\mathcal{J}_{02}(\omega)$ contain vibronic couplings to the S$_1$ and S$_2$ diabatic states due to \emph{tuning modes}, whereas $\mathcal{J}_{12}(\omega)$ is the coupling spectral density encoding the vibrational modes that couple the diabatic excited states together. While the above spectral densities are written in terms of a sum over a common set of $N$ modes, it is commonly observed in studies of molecular systems that the set of tuning modes and coupling modes belong to different symmetry groups\cite{Koppel1984}. 

We note that both tuning and coupling spectral densities are positive semi-definite functions. This means that while the magnitudes of the vibronic couplings in the LVC Hamiltonian are encoded in the three spectral densities, their relative signs are not. In particular, tuning modes can be \emph{correlated} or \emph{anti-correlated} between the S$_1$ and the S$_2$ state, which can have a significant influence on the excited state relaxation dynamics\cite{Dunnett2021}. To represent correlations between the tuning modes, we introduce the cross spectral density:
\begin{equation}
\mathcal{J}_\textrm{cross}(\omega)=\frac{\pi}{2}\sum^N_j \omega^3_j K^{\{1\}}_j K^{\{2\}}_j\delta (\omega -\omega_j),
\end{equation}
as well as the normalized cross spectral density $\tilde{\mathcal{J}}_\textrm{cross}(\omega)$:
\begin{equation}
\tilde{\mathcal{J}}_\textrm{cross}(\omega)=\frac{\mathcal{J}_\textrm{cross}(\omega)}{\sqrt{\mathcal{J}_{01}(\omega)\mathcal{J}_{02}(\omega)}}.
\end{equation}

We note that $\tilde{\mathcal{J}}_\textrm{cross}(\omega)$ can take values of -1 or 1 for $\omega=\{\omega_j\}$, with 1 indicating that $K^{\{1\}}_j$  and $K^{\{2\}}_j$ have the same sign (are \emph{correlated}) and -1 indicating an opposite sign (\emph{anti-correlated} fluctuations). 

In principle, a similar function can be constructed to encode the relative signs of the diabatic couplings $\{\Lambda_i\}$ with respect to $\left\{K^{\{1\}}_j\right\}$ or $\left\{K^{\{2\}}_j\right\}$. However, following the principle that coupling and tuning modes often represent distinct sets of vibrational modes as they have different symmetries, nuclear vibrations causing fluctuations in the S$_1$ and S$_2$ states are taken to be \emph{uncorrelated} with fluctuations driving transitions between the two states for the purpose of this work. Under this assumption, the four spectral densities $\mathcal{J}_{01}(\omega)$, $\mathcal{J}_{02}(\omega)$, $\mathcal{J}_{12}(\omega)$, and $\tilde{\mathcal{J}}_\textrm{cross}(\omega)$ fully specify the LVC Hamiltonian. 

\subsection{Spectral densities from MD:  Quantum correction factors (QCFs) and diabatization}
\label{si_sec:diabatization}

For isolated systems in vacuum, spectral densities can be directly constructed by summing contributions from the $3N_\textrm{Atom}-6$ vibrational normal modes within the harmonic approximation. While this approach neglects anharmonic effects, it has been demonstrated to produce reliable LVC Hamiltonians for small rigid molecules\cite{Aranda2021, Aleotti2021, Segatta2023}. In condensed phase systems however, a large number of vibrational modes can couple to the energy gap fluctuations and the S$_1$-S$_2$ couplings considered in this work. When parameterizing the LVC Hamiltonian from normal mode calculations, solvent degrees of freedom are generally discarded and the solvent environment is represented collectively through polarizable continuum models (PCMs)\cite{Cammi_2005}. Some recent approaches to modeling non-adiabatic effects in spectral lineshapes directly include interactions between the system and an atomistic solvent environment\cite{Aranda2021,Segatta2023}, but generally rely on some form of timescale separation between the fast chromophore and slow environmental degrees of freedom, for example by parameterizing the LVC Hamiltonian for chromophore degrees of freedom inside a frozen solvent pocket. 

Following our previous work\cite{Dunnett2021}, we instead extract LVC Hamiltonian parameters directly from molecular dynamics (MD) calculations of the chromophore in its condensed phase environment, accounting for both anharmonic effects and environmental coupling through the MD sampling. To do so, we recognize that the spectral densities parameterizing the LVC Hamiltonian can be written in terms of equilibrium quantum correlation functions, such that: 
\begin{equation}\label{eq:spectral-density}
    \mathcal{J}_{01}(\omega) = \textrm{i}\theta(\omega) \int \textrm{d}t\ e^{\textrm{i} \omega t} \ \mathrm{Im}\ C_{01}(t),
\end{equation}
where $\theta(\omega)$ is the Heaviside step function and the quantum autocorrelation function $C_{01}(t)$ is a function of the energy gap fluctuation operator $\delta U_{01}$,
\begin{equation}\label{eq:energy-gap-fluctuation-autocorrelation-function}
    C_{01}(t) = \langle \delta U_{01}(\hat{\mathbf{q}}, t) \delta U_{01}(\hat{\mathbf{q}}, 0) \rangle_{\rho_{0}},
\end{equation}
with $\delta U_{01}=\left(H_1-H_0\right)-\omega_{01}^\textrm{av}=U_{01}-\omega_{01}^\textrm{av}$ defined as the deviation from the mean difference $\omega_{01}^\textrm{av}$ between the ground and excited state nuclear Hamiltonians in thermal equilibrium. The brackets $\langle \rangle$ indicate thermal averaging with respect to the equilibrium ground state density matrix $\rho_0$.

Eqn.~\ref{eq:spectral-density} defines the S$_1$ spectral density in terms of a quantum correlation function, which is generally inaccessible in complex condensed phase systems. However, the quantum correlation function can be approximately reconstructed from its classical counterpart using \emph{quantum correction factors}\cite{Bader1994,Egorov1999, Kim2002b, Craig2004, Ramirez2004}. In this work, we make use of the harmonic quantum correction factor, that can be obtained by associating the classical correlation function with the Kubo-transformed version of the quantum correlation function obeying the same set of symmetries. The spectral density can then be written as: 
 \begin{equation}\label{eq:spectral-density-from-classical-md}
    \mathcal{J}_{01}(\omega) \approx \theta(\omega) \frac{\beta \omega }{2} \int \textrm{d}t\ e^{\textrm{i} \omega t} \ C^{\mathrm{cl}}_{01}(t),
\end{equation}
where $\beta=1/k_\textrm{B}T$ and $\theta(\omega)$ is again the Heaviside step function. The classical correlation function $C^{\mathrm{cl}}_{01}(t)$ can be straightforwardly constructed from fluctuations in vertical excitation energies computed along ground-state molecular dynamics simulations of the system in equilibrium. In this work, we use mixed quantum mechanical-classical mechanical (QM/MM) simulations\cite{Warshel1976} to sample the ground state dynamics, with proflavine treated at the DFT level and the surrounding environment through a classical force field. Vertical excitation energies are then computed using time-dependent DFT (TDDFT). We note that constructing $\mathcal{J}(\omega)$ from MD via Eqn.~\ref{eq:spectral-density-from-classical-md} yields a continuous function in the frequency domain, such that the system-bath interactions correspond to a coupling to an infinite number of displaced harmonic oscillator modes. All solvent interactions sampled from MD are encoded in the low frequency part of $\mathcal{J}_{01}(\omega)$. 

While Eqn.~\ref{eq:spectral-density-from-classical-md} provides a pathway for computing the $\mathcal{J}_{01}$ and $\mathcal{J}_{02}$ spectral densities for two uncoupled excited states, the LVC Hamiltonian describes a situation where S$_1$ and S$_2$ are two explicitly coupled \textit{diabatic} states. Calculating vertical excitation energies along MD trajectories, for example using TDDFT, however yields $\emph{adiabatic}$ states, i.e. states that are eigenstates of the electronic Hamiltonian, but change \emph{character} and thus their electronic transition dipole moment along nuclear coordinates. 

To obtain diabatic energies of S$_1$ and S$_2$ required to construct spectral densities $\mathcal{J}_{01}$ and $\mathcal{J}_{02}$, as well as their explicit couplings, it is necessary to approximately diabatize the transitions along the MD trajectory. A wide range of diabatization schemes exist, based on the dipole, quadrupole, or the localization of the excited state\cite{Subotnik2008}. For the purpose of this work, where a bright excited state is coupled to a dark one, we use a scheme introduced by Medders \emph{et al.}\cite{Medders2017} based on maximizing the difference between the transition dipoles of the two diabatic states, which we have successfully applied previously to the case of methylene blue. For each step along the MD trajectory, we construct the matrix  
 \begin{equation}
     \textbf{D}(t)=
     \begin{pmatrix}
     \boldsymbol{\mu}_{01} (t)\cdot \boldsymbol{\mu}_{01}(t) & \boldsymbol{\mu}_{01} (t)\cdot \boldsymbol{\mu}_{02}(t) \\
     \boldsymbol{\mu}_{01}(t)\cdot \boldsymbol{\mu}_{02}(t) & \boldsymbol{\mu}_{02}(t)\cdot \boldsymbol{\mu}_{02}(t)
     \end{pmatrix} 
 \end{equation}
where $\boldsymbol{\mu}_{01}$ and $\boldsymbol{\mu}_{02}$ denote the transition dipole moments between ground and first and ground and second excited state for the \emph{adiabatic} electronic states respectively. Diagonalizing $\textbf{D}(t)$ yields a $2\times2$ matrix of eigenvectors that can be used to rotate the S$_1$ and S$_2$ energies into the diabatic representation, where the off-diagonal element contains the diabatic coupling. Thus for every time-step we obtain $\left\{E_{01}(t), E_{02}(t), \delta_{12}(t)\right\}$ which, after constructing \textrm{diabatic} S$_1$ and S$_2$ correlation functions and the autocorrelation function of S$_1$-S$_2$ coupling, allows us to compute $\mathcal{J}_{01}(\omega)$, $\mathcal{J}_{02}(\omega)$, and $\mathcal{J}_{12}(\omega)$. Additionally, the S$_1$-S$_2$ cross correlation function can be computed from $E_{01}(t)$ and $E_{02}(t)$, giving us access to $\tilde{\mathcal{J}}_\textrm{cross}(\omega)$ and completing the parameterization of the LVC Hamiltonian from MD. 

We note that the sign of the adiabatic transition dipole moments and thus the sign of the diabatic coupling $\delta_{12}$ is arbtrary for a given TDDFT snapshot, as the transition dipole moment is not a physical observable. Discontinuities due to sign flips between snapshots in $\delta_{12}(t)$ causes artifacts in the computed coupling spectral density. However, in practice this problem can be straightforwardly avoided by aligning the transition dipoles in each snapshot along a common reference, yielding a $\delta_{12}(t)$ that varies smoothly in time.

\subsection{Computational details: Vacuum}
For proflavine in vacuum, a 22~ps MD trajectory was generated using the TeraChem code\cite{Ufimtsev2009}. For the dynamics, proflavine was treated at the DFT level, using a 6-31+G* Gaussian basis set\cite{Dunning1990} and the CAM-B3LYP functional\cite{Yanai2004}. The temperature was kept at 300~K using a Langevin thermostat with a collision frequency of 1~ps$^{-1}$. A time-step of 0.5~fs was used throughout. The first 2~ps of trajectory were discarded to allow for system equilibration. 

From the 20~ps of equilibrated trajectory, snapshots were extracted every 2~ps and the two lowest excited states were computed using time-dependent DFT in the linear-response formalism as implemented in the TeraChem code\cite{Isborn2011}. The resulting 10,000 snapshots were, following the diabatization procedure outlined in Sec.~\ref{si_sec:diabatization}, used to compute spectral densities that parameterize the LVC Hamiltonian. 

\subsection{Computational details: Solvated systems}
For the solvated systems, mixed quantum mechanical/molecular mechanical (QM/MM)\cite{Warshel1976} calculations were carried out using the interface\cite{Isborn2012} between the Amber\cite{amber} and TeraChem. proflavine was treated at the QM level using DFT, with the 6-31+G* Gaussian basis set and the CAM-B3LYP functional\cite{Yanai2004}. For proflavine in methanol (MeOH), a standard AMBER force field was used to describe the solvent environment, whereas for proflavine in acetonitrile, a solvent force field was generated using AmberTools. 

For acetonitrile, a cubic solvent box containing 3977 molecules was created. Following a 20~ps temperature equilibration in the NVT ensemble under periodic boundary conditions with a time-step of 2.0~fs, a 500~ps pressure equilibration in the NPT ensemble was carried out. Proflavine in its ground-state optimized geometry was solvated in a 30~\AA\, sphere generated from the equilibrated solvent box. For MeOH, a 30~\AA\, solvent sphere was instead created using a pre-equilibrated solvent box available in Amber. A Langevin thermostat with a collision frequency of 1~ps$^{-1}$ was used throughout to keep the system at a constant temperature of 300~K. 

A pure MM equilibration of proflavine in the 30~\AA\, solvent sphere was carried out at 300~K in open boundary conditions for 50~ps, where the proflavine force field was constructed using AmberTools. The timestep was selected to be 0.5~fs to match the timestep used in QM/MM calculations. After completing the 50~ps run, proflavine was switched to the QM Hamiltonian described by DFT and mixed QM/MM dynamics were carried out for 22~ps using a 0.5~fs timestep. The first 2~ps of trajectory were discarded to allow for an additional equilibration of the system upon switching proflavine from the MM to the QM Hamiltonian. 

From the resulting 20~ps of usable trajectory, snapshots were extracted every 2~fs, and vertical excitation energies were computed using full TDDFT (i.e. without relying on the Tamm-Dancoff approximation\cite{Hirata1999}), treating the chromophore at the QM level and the solvent environment as classical point charges taken directly from the Amber force fields. A 6-31+G* basis set and the CAM-B3LYP functional was used throughout to match the level of theory used to describe the QM Hamiltonian during the system dynamics. This was done to avoid well known-issues in computing spectral densities arising from mismatches between the Hamiltonians used for propagating the system and for evaluating the vertical excitation energies\cite{Lee2016,Lee2016b}. The resulting vertical excitation energies for 10,000 snapshots were, following the diabatization procedure outlined in Sec.~\ref{si_sec:diabatization}, used to compute spectral densities that parameterize the LVC Hamiltonian in the condensed phase. 

The total cost of parameterizing the LVC Hamiltonian for a single system using QM/MM dynamics and the calculation of vertical excitation energies was approximately 1000 GPU hours on NVIDIA RTX 2080Ti GPUs. 

\subsection{Normal mode and Franck-Condon Herzberg-Teller calculations}
To analyze the vibrational frequencies corresponding to specific features in the S$_1$, S$_2$ and coupling spectral densities respectively, vibrational normal modes of proflavine in vacuum were computed using Gaussian\cite{gdv}. The ground-state optimized structure and the ground-state frequencies were computed at the 6-31+G*/CAM-B3LYP level of theory, matching the level of theory used for the dynamics and the computation of vertical excitation energies. 

Additionally, in Fig.~1 in the main text, the experimental spectra of proflavine in vacuum and MeOH are compared to cumulant spectra\cite{Mukamel-book,Zuehlsdorff2019b} computed from the \emph{adiabatic} (i.e. uncoupled) S$_1$ and S$_2$ states, as well as spectra computed in the Franck-Condon Herzberg Teller (FCHT) scheme\cite{Baiardi2013,deSouza2018}, where the ground and excited state PESs are approximated as harmonic. For the FCHT calculations, only the bright S$_1$ state was considered. Optimized geometries and normal modes were obtained at the 6-31+G*/CAM-B3LYP level of theory in vacuum and in MeOH represented by a polarizable continuum model (PCM). FCHT spectra for absorption and fluorescence were computed at 300~K using the inbuilt functionality in Gaussian\cite{gdv} that constructs the finite-temperature linear response function in the time-domain\cite{Baiardi2013}. 

\subsection{Comparing to higher order methods: EOM-CCSD}
As observed in our previous work\cite{Dunnett2021}, we expect the dynamics and resulting spectra of the LVC Hamiltonian to be highly sensitive to the exact model system parameters, especially the average gap between the S$_1$ and the S$_2$ state in the Condon region, as this controls the position of the conical intersection between the states. Since standard TDDFT functionals can produce significant errors in the relative energies and orderings of excited states, we carried out a number of benchmark calculations on the optimized ground state geometry using the equation-of-motion coupled-cluster singles-doubles (EOM-CCSD)~\cite{Stanton1993equation, Comeau1993equation, Krylov2008equation} level of theory. The EOM-CCSD calculations were performed using the Q-Chem 6.0 package \cite{Epifanovsky2021} using the 6-31+G(d) basis set,\cite{Frisch1984self} and the results were computed both (i) in the gas phase and (ii) in acetonitrile described by the conductor-like polarizable continuum model (C-PCM) \cite{Truong1995new,Barone1998quantum,Lange2010smooth} with $\varepsilon=37.5$. Note that since non-equilibrium PCM for EOM-CCSD calculations is not available in Q-Chem, here the presence of solvent only affected the ground-state molecular orbitals.

\subsection{Correcting the S$_1$-S$_2$ gap in solution: Polarization effects }
\label{si_sec:corrected_gap}
For the calculation of vertical excitation energies and resulting LVC parameterizations in solution, the solvent environment was treated as classical point charges, thus neglecting dynamic polarization effects. To assess the influence these polarization effects have for proflavine in methanol and acetonitrile, we selected 50 snapshots evenly spaced every 400~fs from both of the solvated trajectories. For the selected snapshots, vertical excitation energies were computed using TDDFT, but the QM region is expanded to contain all solvent molecules with an atom within 6~\AA\, of any atom of the proflavine molecule. The remaining solvent molecules were treated as classical point charges. 

For all 100 snapshots, vertical excitation energies were diabatized using the standard procedure outlined in SI Sec.~\ref{si_sec:diabatization}. The energy difference between diabatic energies for a pure MM treatment of the solvent environment and the expanded 6~\AA~QM region was evaluated and averaged across the 50 snapshots from each trajectory. This allowed us to evaluate a correction to $\Delta_{12}$, the average S$_1$-S$_2$ energy difference in ground-state thermal equilibrium, due to solvent polarization effects. For both acetonitrile and methanol, it was found that solvent polarization effects at the QM level \emph{increase} the average S$_1$-S$_2$ gap, by 63~meV in acetonitrile and 62~meV in methanol respectively. 

Thus, to approximately account for the effect of dynamic polarization of the solvent environment on the system dynamics, we increase the $\Delta_{12}$ parameter entering the LVC Hamiltonian parameterization by 63~meV (acetonitrile) and 62~meV (methanol) compared to the force-field based TDDFT calculations respectively. We note that this approximate approach ignores the influence of environmental polarization effects on the spectral densities themselves, which can be significant\cite{Zuehlsdorff2020}. However, since the results obtained for proflavine so sensitively depend on the S$_1$-S$_2$ gap in the Condon region, we expect that the most important influence of polarization effects on the resulting spectra is captured within the approximate scheme, at much reduced cost compared to evaluating the vertical excitation energy for the expanded 6~\AA~QM region for all snapshots (total cost for evaluating 50 snapshots with a 6~\AA~QM region for MeOH is around 350~GPU hours, comparable to the total cost of generating SDs for proflavine in MeOH when the environment was treated at the MM level).

\section{Solving the LVC Hamiltonian parameterized from MD}
Here we summarize how to compute absorption spectra of the LVC Hamiltonian parameterized from MD using the T-TEDOPA\cite{tamascelli2019efficient} formalism (see Ref.~\onlinecite{Dunnett2021}). Additionally, we outline how the approach is extended in the current work to also give access to steady-state and time-resolved emission spectra.

\subsection{The T-TEDOPA formalism as applied to the LVC Hamiltonian}

Following the approach outlined in Ref.~\onlinecite{Dunnett2021} represent the LVC Hamiltonian in a way amenable to efficient simulations, we first decompose it into three parts:
\begin{equation}
\hat{H}_\textrm{LVC}=\hat{H}_\textrm{S}+\hat{H}_\textrm{I}+\hat{H}_\textrm{B},
\end{equation}
where $\hat{H}_\textrm{S}$ is the system Hamiltonian of the electronic states given by
\begin{equation}
    \hat{H}_\textrm{S}=\sum_{\alpha=1}^2 \left(\lambda_{0\alpha}^\textrm{R}+\omega^\textrm{ad}_{0\alpha} \right)| S_\alpha \rangle\langle S_\alpha|+\delta\left(|S_1\rangle\langle S_2| +  \textrm{h.c.}\right )
\end{equation}
where h.c. denotes the Hermitian conjugate, $\lambda_{0\alpha}^\textrm{R}$ is the reorganization energy of state $\alpha$, and $\omega^\textrm{ad}_{0\alpha}$ is again the adiabatic energy gap between the ground state minimum and the excited state minimum for state $\alpha$. Thus $\Delta_{0\alpha}=\omega^\textrm{ad}_{0\alpha}+\lambda_{0\alpha}^\textrm{R}$ is the energy gap between the ground state and the diabatic excited state $\alpha$ at the ground state minimum (in the Condon region). We also include a constant coupling factor $\delta$ between the two diabatic excited states in the Hamiltonian. In practice, this factor is dependent on the choice of diabatic states, and one can always find a representation where $\delta=0$. In our present work, we find that using the diabatization scheme based on the transition dipole moments of the adiabatic states, $\delta$ is close to zero for all systems studied. 

$\hat{H}_\textrm{B}$ denotes the bath Hamiltonian describing free motion of the nuclei and $\hat{H}_\textrm{I}$ is the interaction Hamiltonian that that describes all couplings of the excited states to nuclear degrees of freedom. It is convenient to split this Hamiltonian into two parts:
\begin{equation}
\hat{H}_\textrm{I}=\hat{H}^\textrm{EL}_\textrm{I}+\hat{H}^\delta_\textrm{I}
\end{equation}
with $\hat{H}^\textrm{EL}_\textrm{I}$ denoting the energy level interaction Hamiltonian that accounts for fluctuations in S$_1$ or S$_2$ energy levels encoded in the spectral densities $\mathcal{J}_{01}$ and $\mathcal{J}_{02}$, and $\hat{H}^\delta_\textrm{I}$ is the interaction Hamiltonian responsible for fluctuations in the S$_1$-S$_2$ coupling matrix encoded in the spectral density $\mathcal{J}_{12}$. In this work, we consider the fluctuations in S$_1$, S$_2$, and the coupling to arise from three distinct bosonic baths. While the assumption of separate baths for $\hat{H}^\textrm{EL}_\textrm{I}$ and $\hat{H}^\delta_\textrm{I}$ is justified by the fact that tuning and coupling modes in molecular systems often exhibit different symmetries\cite{Koppel1984}, the same is not true for the tuning modes encoded in  $\mathcal{J}_{01}$ and $\mathcal{J}_{02}$. However, the assumption of three separate bosonic corresponds to the most flexible representation of the system Hamiltonian, and specifically allows us to model arbitrary cross-correlations between fluctuations in S$_1$ and fluctuations in S$_2$ (see SI Sec.~\ref{si_sec:bath_correlations}). 

To simulate the quantum dynamics of the system at finite temperatures would in principle require an averaging over a large number of initial states. However, this state-averaging can be avoided by applying the T-TEDOPA mapping of Tamascelli \emph{et. al.}\cite{tamascelli2019efficient} to each bath, where the bath spectral density becomes explicitly \emph{temperature-dependent}:
\begin{equation}
\mathcal{J}^\beta(\omega)=\frac{\mathcal{J}^\textrm{ext}(\omega)}{2}\left[1+\textrm{coth}\left(\frac{\beta \omega}{2}\right)\right].
\end{equation}
Here, $\mathcal{J}^\textrm{ext}(\omega)=\textrm{sign}(\omega)\mathcal{J}(|\omega|)$ denotes the antisymmetrized spectral density extended over the entire frequency range. The temperature mapping allows us to compute properties of the system at finite temperature by evolving an initial pure state under the influence of \emph{thermal} spectral densities $\mathcal{J}^\beta(\omega)$.  

\subsection{Bath correlations}
\label{si_sec:bath_correlations}
\begin{figure*}
    \centering
    \includegraphics[width=0.7\textwidth]{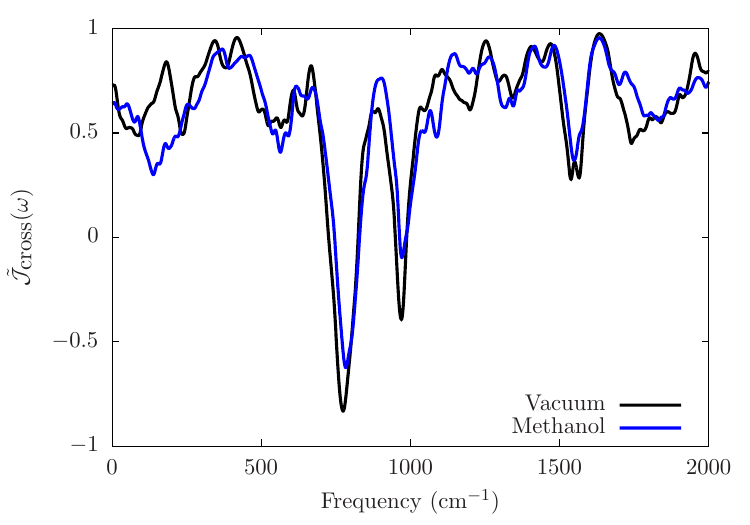}
    \caption{The normalized S$_1$-S$_2$ cross correlation spectral density $\tilde{\mathcal{J}}_\textrm{cross}(\omega)$ for proflavine in vacuum and methanol. }
     \label{fig:normalized_cross_corr_comparison}
\end{figure*}

To describe the relative sign of the S$_1$ and S$_2$ fluctuations in the LVC Hamiltonian in terms of spectral densities, we have introduced the normalized cross correlation spectral density $\tilde{\mathcal{J}}_\textrm{cross}(\omega)$. We note that for an LVC Hamiltonian constructed for a discrete set of modes, such as when parameterized from ground state normal mode calculations, each mode produces either correlated or anti-correlated S$_1$-S$_2$ fluctuations, corresponding to a value of +1 or -1 for $\tilde{\mathcal{J}}_\textrm{cross}(\omega)$. However, when parameterizing the LVC Hamiltonian directly from MD, the fluctuations in the system are mapped onto an infinite number of modes represented by continuous spectral densities. In this case, $\tilde{\mathcal{J}}_\textrm{cross}(\omega)$ is no longer constrained to values of $\pm 1$ for a given frequency $\omega$, but can take any intermediate value. Fig.~\ref{fig:normalized_cross_corr_comparison} shows $\tilde{\mathcal{J}}_\textrm{cross}(\omega)$  for proflavine in vacuum and methanol, demonstrating that S$_1$-S$_2$ fluctuations are largely positively correlated, but mostly remain well below a fully correlated value of +1. Additionally, for a number of frequencies, especially around 750~cm$^{-1}$, fluctuations are actually strongly anti-correlated.

To account for cross correlations, we modify the energy level interaction Hamiltonian $\hat{H}^\textrm{EL}_\textrm{I}$ to take the generalized form

\begin{equation}
\hat{H}^\textrm{EL}_\textrm{I}=\sum_k\sum_{\alpha\beta} g_k^{\alpha \beta} \left|S_\alpha\right\rangle\left\langle S_\alpha\right|\left(a^\dagger_{\beta k}+a_{\beta k}\right)
\end{equation}
where the parameters $g_k^{\alpha \beta}$ are chosen to reproduce the spectral densities $\mathcal{J}_{01}(\omega)$,  $\mathcal{J}_{02}(\omega)$ and $\tilde{\mathcal{J}}_\textrm{cross}(\omega)$. The chosen form of $\hat{H}^\textrm{EL}_\textrm{I}$ can represent arbitrary correlations between S$_1$ and S$_2$ energy gap fluctuations\cite{Dunnett2021}.

\subsection{Chain mapping}
To allow for an efficient simulation of the quantum dynamics of the system, the many-body wavefunction is represented as a matrix-product-state (MPS) or tensor network, that can be effectively propagated using the one-site Time-Dependent Variational Principle (1TDVP)\cite{Haegeman2011,Haegeman2016}. This formalism requires the system Hamiltonian to be in a specific form, namely a chain-like topology with open boundary conditions. Such a chain mapping\cite{Chin2010} of the bath spectral densities can be achieved through a unitary transformation defined by orthogonal polynomials (see the SI of Ref.~\onlinecite{Dunnett2021} for details). The orthogonality of the polynomials, and the fact that they satisfy a three term recurrence relation, means that the chain modes only couple to their nearest neighbors and only the first chain mode couples to the system. 

For the electronic interaction term $\hat{H}_\textrm{I}^\textrm{EL}$, the cross correlations between S$_1$ and S$_2$ additionally introduce long range couplings to the chain coefficients, where each element in the chain mapped $\mathcal{J}_{01}(\omega)$ spectral density couples to S$_2$ and vice versa. The magnitude of the long range couplings is determined by ${\mathcal{J}}_\textrm{cross}(\omega)$. In practice, these long range couplings only introduce a modest increase in computational complexity for the 1TDVP algorithm\cite{Dunnett2021}. 

All chain mappings for the spectral densities considered in this work are carried out using an in-house python code that represents a modified version of the approach used in Ref.~\onlinecite{Dunnett2021} (based on the ORTHOPOL package\cite{gautschi_algorithm_1994}). The python code is freely available on GitHub, and can be found under the following persistent DOI: 10.5281/zenodo.10699238. 

\subsection{Constructing absorption spectra}
The linear absorption spectrum $\sigma(\omega)$ of the system is given by 
\begin{equation}
\sigma(\omega)=\alpha(\omega) \mathcal{FT}\left[C_\textrm{abs}(t)\right]
\end{equation}
where $\mathcal{FT}$ denotes a Fourier transform, $C_\textrm{abs}(t)$ is the dipole-dipole correlation function and $\alpha(\omega)$ is a frequency dependent prefactor necessary when directly comparing computed intensities to experimental data\cite{Baiardi2013}, that will be set to 1 for convenience throughout this work. The dipole-dipole response function $C_\textrm{abs}(t)$ can be expressed as
\begin{equation}
C_\textrm{abs}(t)=\left \langle  \hat{\mu}^-(t) \hat{\mu}^+(0)\right\rangle_{\rho_0} 
\end{equation}
with 
\begin{eqnarray}
    \hat{\mu}^-&=&\mu_{01}\left|S_0\right \rangle\left \langle S_1 \right|+\mu_{02}\left|S_0\right \rangle\left \langle S_2 \right| \\
    \hat{\mu}^+&=&\left(\hat{\mu}^-\right)^\dagger=\mu_{01}\left|S_1\right \rangle\left \langle S_0\right|+\mu_{02}\left|S_2\right \rangle\left \langle S_0 \right|
\end{eqnarray}
and $\hat{\mu}^{\pm}(t)=e^{\textrm{i}\hat{H}_\textrm{LVC}t}\hat{\mu}^\pm e^{-\textrm{i}\hat{H}_\textrm{LVC}t}$. Importantly, there is no term in the total LVC Hamiltonian that couples the electronic excited states and the ground state. 

Following the methodology outlined in Ref.~\onlinecite{Dunnett2021}, we can express the response function as

\begin{eqnarray} \nonumber
C_\textrm{abs}(t)&=&\textrm{Tr}\left[\hat{\mu}^{-}e^{-\textrm{i}\hat{H}_\textrm{LVC}t} \left\{ \hat{\mu}^+ \rho_0\right\} e^{-\textrm{i}\hat{H}_\textrm{LVC}t} \right] \\ \nonumber
&=&\textrm{Tr}_\textrm{S}\left[\hat{\mu}^{-} \textrm{Tr}_\textrm{E} \left \{e^{-\textrm{i}\hat{H}_\textrm{LVC}t}  \hat{\mu}^+ \rho_0\, e^{-\textrm{i}\hat{H}_\textrm{LVC}t} \right\} \right] \\ \nonumber 
&=&\textrm{Tr}_\textrm{S}\left[ \hat{\mu}^{-} \rho_\textrm{R}'(t) \right] \\
&=&\langle \hat{\mu}^{-} \rangle_{\rho_\textrm{R}'(t)} 
\label{eqn:abs_response}
\end{eqnarray}

where $\rho_\textrm{R}' (t)$ is the reduced density matrix that has been evolved from initial state $\hat{\mu}^{+}\rho_0$. However, $\rho_\textrm{R}'$ is not a valid initial condition as it contains only off-diagonal terms. Instead, we define $\rho_\textrm{R}=\left | \Psi_\textrm{R} \right \rangle \left \langle \Psi_\textrm{R} \right |$ with $\Psi_\textrm{R}=c\left( \left |S_0\right\rangle  +\mu_{01}\left |S_1\right\rangle+\mu_{02}\left |S_2\right\rangle\right)$, where $c$ is a normalization constant. Computing the expectation value of $\hat{\mu}^-$ with respect to the time-evolved state $\Psi_\textrm{R}$ yields the same expectation value as in Eqn.~\ref{eqn:abs_response}, as additional cross terms of the form $\left |S_1\right \rangle\left \langle S_2 \right | $ are projected out by the measurement of $\mu^{-}$. Thus to compute a finite temperature absorption spectrum by calculating $\chi_\textrm{abs}(t)$, it is sufficient to propagate a zero-temperature initial wavefunction $\Psi_\textrm{R}$ under the LVC Hamiltonian, with finite temperature effects accounted for through the thermalized spectral densities in the T-TEDOPA formalism.

\subsection{Constructing steady-state and time-resolved fluorescence spectra }

To construct an  emission spectrum, we need to evaluate the dipole-dipole response function
\begin{equation}
C_\textrm{emi}(t)=\left \langle \hat{\mu}^+(t) \hat{\mu}^-(0) \right \rangle_{\rho_\textrm{ex}}
\end{equation}
where $\rho_\textrm{ex}$ denotes final (relaxed) electronic-vibrational density matrix that describes the long-time state of the photo-excited system. Following the same procedure as for $C_\textrm{abs}(t)$, this correlation function can be evaluated as an expectation value of the operator $\hat{\mu}^+$ with respect to the time-evolved pseudo-density matrix $\left(\hat{\mu}^- \rho_\textrm{ex}\right)(t)$. $\left(\hat{\mu}^- \rho_\textrm{ex}\right)$ is again not a valid, i.e. physical density matrix as it contains only off-diagonal elements, i.e. coherences without any corresponding populations. We note that the density matrix $\rho_\textrm{ex}=|\Psi_\textrm{ex}\rangle \langle \Psi_\textrm{ex}|$ corresponding to the relaxed excited electronic-vibrational state is pure. The wavefunction $|\Psi_\textrm{ex}\rangle$ can be obtained as the final MPS wavefunction evolved by the TDVP algorithm from some initial excited state condition. Since $|\Psi_\textrm{ex}\rangle$ is an excited state wavefunction (the LVC Hamiltonian we define contains no off-diagonal couplings between the electronic ground state and the two electronic excited states), it must take the following (unnormalized) general form:
\begin{equation}
|\Psi_\textrm{ex}\rangle = c_1 |S_1\rangle \otimes | E_1\rangle +c_2 |S_2\rangle \otimes | E_2 \rangle 
\end{equation}
where $| E_i \rangle $ is the many-body state of the environment corresponding to diabatic state $| S_i\rangle $ and $c_i$ is a weighting constant. We now consider the action of the dipole operator $\hat{\mu}^-$ on $|\Psi_\textrm{ex}\rangle$:
\begin{equation}
|\phi_\textrm{R}\rangle =\hat{\mu}^- |\Psi_\textrm{ex}\rangle = |S_0\rangle \otimes \left [ c_1 \mu_{01} | E_1\rangle + c_2 \mu_{02} | E_2 \rangle \right ]. 
\end{equation}
The un-normalized state $| \phi_\textrm{R} \rangle$ is a product state on the electronic ground state. Additionally, we note that the pseudo density matrix $\left(\mu^-\rho_\textrm{ex}\right)$ can be experessed as $\left(\hat{\mu}^- \rho_\textrm{ex}\right)=\hat{\mu}^- |\Psi_\textrm{ex}\rangle \langle \Psi_\textrm{ex}|=|\phi_\textrm{R}\rangle \langle \Psi_\textrm{ex}|$. Now considering the superposition $|\chi_\textrm{R}\rangle=|\phi_\textrm{R}\rangle +|\Psi_\textrm{ex}\rangle$, we can define the density matrix $\rho_{\chi_\textrm{R}}=|\chi_\textrm{R}\rangle\langle\chi_\textrm{R}|$:
\begin{equation}
|\chi_\textrm{R}\rangle\langle\chi_\textrm{R}|=|\Psi_\textrm{ex}\rangle\langle\Psi_\textrm{ex}|+|\phi_\textrm{R}\rangle\langle\phi_\textrm{R}|+|\Psi_\textrm{ex}\rangle\langle\phi_\textrm{R}|+|\phi_\textrm{R}\rangle\langle\Psi_\textrm{ex}|
\end{equation}
which contains the required pseudo-density matrix as one of its elements. 

Computing the expectation value of $\hat{\mu}^+$ with respect to the time-evolved state $\chi_\textrm{R}$ we find
\begin{eqnarray} \nonumber
\left \langle \hat{\mu}^+ \right \rangle_{\rho_{\chi_\textrm{R}}(t)}&=&\textrm{Tr}_\textrm{S} \left [\hat {\mu}^+ \textrm{Tr}_\textrm{E} \left \{e^{-\textrm{i}\hat{H}_\textrm{LVC}t}  \rho_{\chi_\textrm{R}} e^{-\textrm{i}\hat{H}_\textrm{LVC}t} \right\} \right] \\ \nonumber 
&=&\textrm{Tr}_\textrm{S} \left [\hat {\mu}^+ \textrm{Tr}_\textrm{E} \left \{e^{-\textrm{i}\hat{H}_\textrm{LVC}t} \left(|\Psi_\textrm{ex}\rangle\langle\Psi_\textrm{ex}|+|\phi_\textrm{R}\rangle\langle\phi_\textrm{R}|+|\Psi_\textrm{ex}\rangle\langle\phi_\textrm{R}|+|\phi_\textrm{R}\rangle\langle\Psi_\textrm{ex}| \right) e^{-\textrm{i}\hat{H}_\textrm{LVC}t} \right\} \right]  \\ \nonumber
&=&\textrm{Tr}_\textrm{S} \left [\hat {\mu}^+ \textrm{Tr}_\textrm{E} \left \{e^{-\textrm{i}\hat{H}_\textrm{LVC}t}  \left(\hat{\mu}^- \rho_\textrm{ex}\right) e^{-\textrm{i}\hat{H}_\textrm{LVC}t} \right\} \right]  \\ 
&=& C_\textrm{emi}(t), 
\end{eqnarray}
where in going from line 2 to line 3 we have exploited the fact that the LVC Hamiltonian does not mix ground and excited states, such that density matrices  $|\phi_\textrm{R}\rangle\langle\phi_\textrm{R}|$ and $|\Psi_\textrm{ex}\rangle\langle\Psi_\textrm{ex}|$ evolve purely on the ground- and excited states respectively and therefore do not contribute the the expectation value. Additionally, acting with $\hat{\mu}^+$ on  $e^{-\textrm{i}\hat{H}_\textrm{LVC}t} |\Psi_\textrm{ex}\rangle $ gives zero, such that the only term contributing to the expectation value is the pseudo-density matrix $\left(\hat{\mu}^- \rho_\textrm{ex}\right)$ as required. 

Thus we can construct a steady-state emission spectrum following an absorption spectrum calculation, by carrying out these steps:

\begin{itemize}
\item 
Construct the state $|\Psi_\textrm{R}\rangle$ and propagate under the LVC Hamiltonian using the 1TDVP algorithm for a time $t_\textrm{delay}$. Evaluating the expectation value of $\hat{\mu}^{-}$ with respect to the evolved state yields $C_\textrm{abs}(t)$ and thus the linear absorption spectrum.
\item 
Take the final state $|\Psi_\textrm{R}(t_\textrm{delay})\rangle$ and form $|\phi_\textrm{R}\rangle =\hat{\mu}^-|\Psi_\textrm{R}(t_\textrm{delay})\rangle  $
\item 
Construct $|\chi_\textrm{R}\rangle $ from $|\phi_\textrm{R}\rangle$ and the component of $|\Psi_\textrm{R}(t_\textrm{delay})\rangle$ involving only the diabatic excited states (again exploiting that there is no coupling between the ground state and the excited states).
\item 
Propagate $|\chi_\textrm{R}\rangle $ under the LVC Hamiltonian and compute the expectation value of $\hat{\mu}^+$, yielding $C_\textrm{emi}(t)$ and the emission spectrum. 
\end{itemize}

For steady-state emission spectra, $t_\textrm{delay}$ has to be chosen sufficiently large to yield a fully relaxed state $|\Psi_\textrm{ex}\rangle $. We find that a $t_\textrm{delay} \approx 100$ fs yields converged absorption spectra and allows the population dynamics of diabatic states S$_1$ and S$_2$ to reach a steady state, but the environment degrees of freedom may still undergo a significant evolution. This means that $|\Psi_\textrm{ex}\rangle $ obtained from different values for $t_\textrm{delay}$ yields slightly different emission lineshapes. To overcome this issue, we compute the steady state emission spectrum as an average of several emission spectra obtained with different delay times (see SI Sec.~\ref{si_sec:comp_details_quantum_dyn}). 

Since $t_\textrm{delay}$ determines the initial state of the simulation of the emission lineshape, we can interpret the dipole-dipole correlation function as being parametrically dependent on this quantity, such that $C_\textrm{emi}\equiv C_\textrm{emi}(t;t_\textrm{delay})$. Varying $t_\textrm{delay}$ thus gives us access to the stimulated emission component of a transient absorption signal. While lacking the ground state bleach and excited state absorption components that accompany real experimental transient absorption spectra, constructing time-resolved fluorescence spectra allows us to monitor dynamic properties of the system such as the Stokes shift and fluorescence intensity quenching due to coupling to the dark S$_2$ state. 

 \section{Computational details of the quantum dynamics}
 \label{si_sec:comp_details_quantum_dyn}
The accuracy of the quantum dynamics as described using the 1TDVP is controlled by three parameters. First, the many-body wavefunction is represented as a matrix-product state (MPS) or tensor-train\cite{Haegeman2011,Haegeman2016,Schroder2019, kloss2018time}, the accuracy of which is determined by a parameter known as the bond-dimension. This internal parameter is kept fixed in the 1TDVP, although approaches exist to allow for simulations with varying bond-dimensions\cite{Dunnett2021b}. Large bond-dimensions allow for a more accurate representation of the many-body wavefunction, but result in higher computational cost and increased memory requirements. In this work, we found that both absorption and fluorescence spectra were well-converged for a bond-dimension of 50. 

Additionally, two controllable approximations are necessary for simulating the chain-mapped LVC Hamiltonian. First, the semi-infinite chain needs to be truncated at a finite length $N$. Second, the local Fock space of each chain mode has to be limited to a finite number of states $d$. In this work, a chain length of $N=150$ and a local Fock space of $d=30$ was used throughout. By monitoring the populations of the individual chain modes, it was confirmed that these settings were sufficient to propagate the MPS for up to 260~fs without appreciable loss in accuracy of the computed absorption and time-resolved fluorescence spectra. 

The system was propagated with a time-step of 10~a.u. The absorption spectrum was computed from 4200~a.u. of propagation time ($\approx 100$~fs). For the time-resolved fluorescence spectra, delay times of up to 6800~a.u. ($\approx 165$~fs) were considered and for each delay time, the fluorescence spectrum was computed for an additional 4200~a.u. of propagation time to match the calculation settings for the absorption spectrum. Steady-state fluorescence spectra presented in Fig.~3 of the main text were computed by averaging all time-resolved fluorescence spectra constructed from delay times of 100~fs to 165~fs. 

All calculations were carried out using a modified version (available under the following persitent DOI: 10.5281/zenodo.10712009) of the algorithm used in Ref.~\onlinecite{Dunnett2021}, with the modification allowing us to use GPUs rather than CPUs. Typical calculation times were around 4~hrs on a single NVIDIA RTX 2080Ti GPU with 11~GB of GPU RAM for 4200 a.u. of propagation time, such that calculations of a single fluorescence spectrum with 4200 a.u. delay time took around 8~hrs on a single GPU. For the time-resolved stimulated emission spectra presented in the main manuscript, the delay time was varied from 0 to 6970~a.u. in 210~a.u.~(5~fs) increments. A total of 35 individual spectra were computed for both proflavine in MeOH and in vacuum, with a combined computational cost of about 500~GPU hours.

\section{System parameters}

\begin{table}
\begin{tabular}{ c|c|c|c } 
  & Vacuum & Acetonitrile & Methanol  \\ \hline
 $ \mu_{01}$ (Ha) & 2.884 & 2.845 & 2.858 \\
 $ \mu_{02}$ (Ha) & 0.291 & 0.282 & 0.268 \\
 $\lambda^\textrm{R}_{S_1}$ (eV) & 0.0624 & 0.0832 & 0.0783\\ 
 $\lambda^\textrm{R}_{S_2}$ (eV) & 0.1934 & 0.2229 & 0.2211 \\ 
 $\lambda^\textrm{R}_{\textrm{c}}$ (eV) & 0.0449 & 0.0785 & 0.0691 \\ 
 $ \Delta_{12}$ (TDDFT, eV) & 0.2339 & 0.2264 & 0.2260 \\
 $ \Delta_{12}$ (EOM-CCSD, eV) & 0.3942 & 0.3689 & 0.3690 \\
 $ \Delta_{12}$ (Shifted, eV) & 0.0139 & 0.0064 & 0.0060 \\
 $ \Delta_{12}$ (Shifted+QM correction, eV) & - & 0.0697 & 0.0682\\
 \hline
\end{tabular}
\caption{ Hamiltonian parameters for all systems studied in this work. $\mu_{01}$ and $\mu_{02}$ denote the average transition dipole moments between the ground state and the two diabatic excited states. $\lambda_{S_1}$, $\lambda_{S_2}$ and  $\lambda_{\textrm{c}}$ are the reorganization energies of the diabatic S$_1$, S$_2$ and diabatic coupling spectral densities respectively. $\Delta_{12}$ denotes the energy gap between the diabatic S$_1$ and S$_2$ states in the Condon region, as computed either from TDDFT for a QM/MM trajectory, EOM-CCSD on a single snapshot in PCM, TDDFT but with the S$_2$ state shifted down in energy by 0.220~eV and the shifted energy gap with an additional correction calculated by treating part of the solvent environment at the QM level. }
\label{si_tab:system_params}
\end{table}

The LVC Hamiltonian parameters (diabatic dipole moments and $\Delta_{12}=\Delta_{02}-\Delta_{01}$) used for all calculations presented in the main text can be found in Table~\ref{si_tab:system_params}. The gap $\Delta_{12}$ entering the T-TEDOPA algorithm is taken to be the mean energy gap between the diabatic S$_1$ and S$_2$ states in ground state thermal equilibrium and thus specifies the separation of the PESs of the two electronic excited states in the Condon region. For $\Delta_{12}$, four different parameterizations are considered. $\Delta_{12}$ (TDDFT) corresponds to the parameterization directly obtained from the diabatic energy fluctuations computed from TDDFT, where the solvent environment is treated as MM point charges. $\Delta_{12}$ (EOM-CCSD) corresponds to the gap computed for the ground state optimized structure of proflavine in vacuum or in PCM\cite{Cammi_2005} using EOM-CCSD. $\Delta_{12}$ (shifted) is given by $\Delta_{12}$ (TDDFT)-0.220~eV, ie. it corresponds to the average gap computed from TDDFT reduced by 0.220~eV to match experimental results. Finally, $\Delta_{12}$ (shifted+QM correction) is the reduced energy gap corrected for solvent polarization effects as outlined in SI Sec.~\ref{si_sec:corrected_gap}. 

Table~\ref{si_tab:system_params} also specifies the total reorganization energies contained in each of the diabatic spectral densities for S$_1$, S$_2$, and the diabatic coupling. The reorganization energies are computed using the following expression:
\begin{equation}
\lambda^\textrm{R}_x =\frac{1}{\pi}\int_0^\infty \frac{\mathcal{J}_x(\omega)}{\omega}\,\textrm{d}\omega.
\end{equation}
We note that the condensed phase environment generally increases the amount of reorganization energy contained in the spectral density, which is especially significant for the coupling reorganization $\lambda^\textrm{R}_c$.

From the reorganization energies of the two electronic states, $\lambda^\textrm{R}_{S_1}$ and $\lambda^\textrm{R}_{S_2}$, as well as the gap between the two states in the Condon region as quantified by $\Delta_{12}$, we can compute the \emph{adiabatic} energy gap between the two excited state minima $\omega_{12}$, which gives insight on the relative energies of the relaxed S$_1$ and S$_2$ states that are relevant for emission. The results can be found in Table~\ref{si_tab:adiabatic_gap}. We note that the TDDFT and EOM-CCSD parameterizations both predict the dark S$_2$ minimum to be higher in energy than the bright S$_1$ minimum. For the shifted gap and the shifted gap with QM correction, the dark S$_2$ state is higher in energy in the Condon region, but becomes the lowest excited state in the system upon relaxation. In general, the shifted S$_2$ minimum relative to the S$_1$ minimum is slightly lower in energy in acetonitrile and methanol, due to the larger reorganization energy contained in S$_2$ in solution. However, when accounting for the QM solvent correction derived by taking into account solvent polarization effects, the magnitude of $\omega_{12}$ decreases slightly. In general, only relatively minor differences are found between the parameterizations in vacuum, acetonitrile and methanol as detailed in Table~\ref{si_tab:system_params} and Table~\ref{si_tab:adiabatic_gap}, suggesting that the reason for the quenching of dual fluorescence in proflavine in polar solvents is not simply due to solvent-induced changes in the relative energies of the PESs. 

\begin{table}
\begin{tabular}{ c|c|c|c } 
  & Vacuum & Acetonitrile & Methanol  \\ \hline
 $ \omega_{12}$ (TDDFT, eV) & 0.1029 & 0.0867 & 0.0832 \\
 $ \omega_{12}$ (EOM-CCSD, eV) & 0.2632 & 0.2292 & 0.2262 \\
 $ \omega_{12}$ (Shifted, eV) & -0.1171 & -0.1333 & -0.1368 \\
 $ \omega_{12}$ (Shifted+QM correction, eV) & - & -0.0700 & -0.0746\\
 \hline
\end{tabular}
\caption{Effective adiabatic energy gap $\omega_{12}$ between the PES minima of the electronic states S$_1$ and S$_2$ for all system parameterizations. Definitions of the four different parameterizations are the same as in Table~\ref{si_tab:system_params}. }
\label{si_tab:adiabatic_gap}
\end{table}

\begin{figure*}
    \centering
    \includegraphics[width=1.0\textwidth]{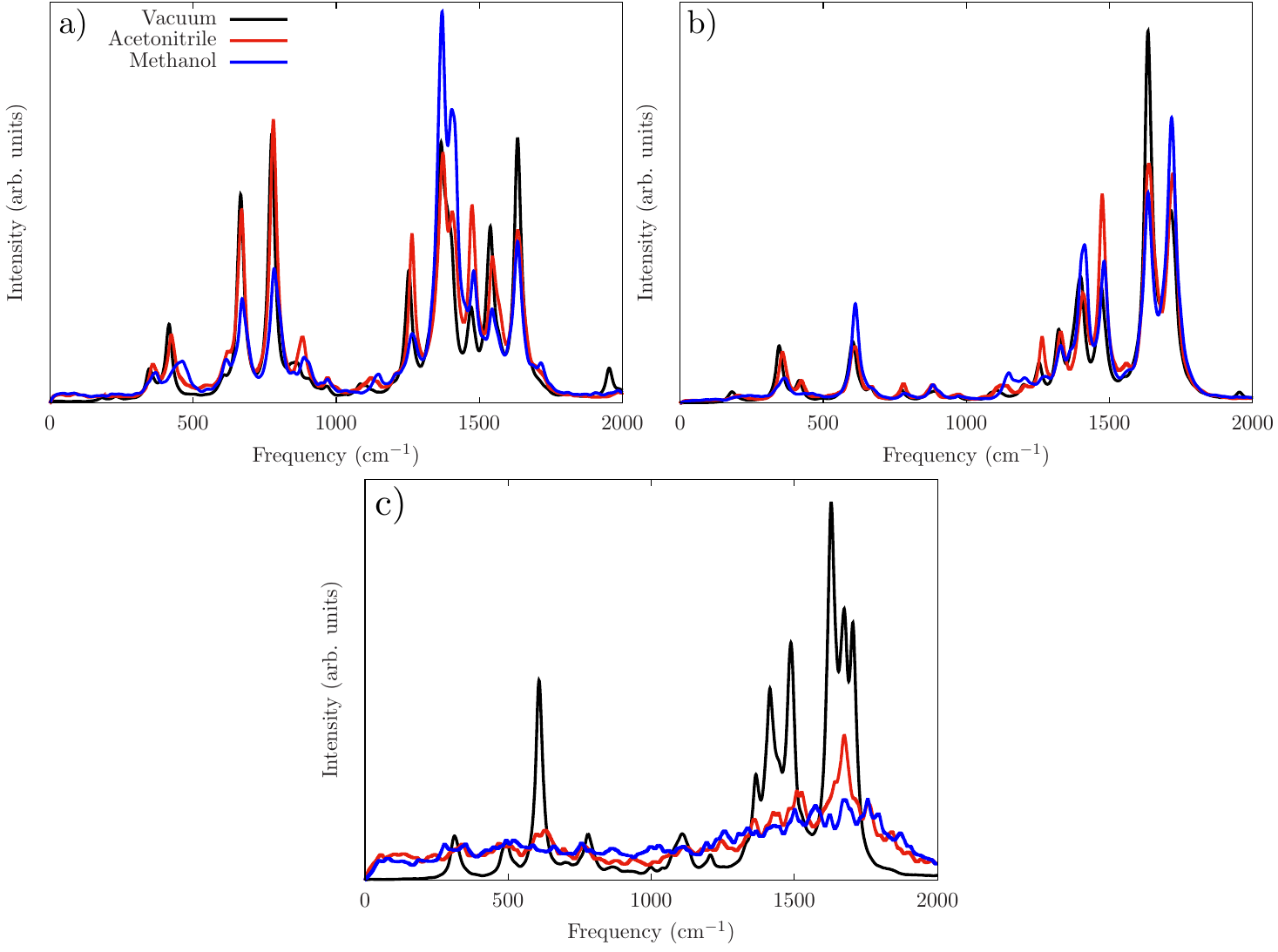}
    \caption{The S$_1$ (a), S$_2$ (b) and coupling (c) spectral densities for proflavine in vacuum, acetonitrile and methanol. }
     \label{fig:all_sds_comparison}
\end{figure*}

Fig.~\ref{fig:all_sds_comparison} shows the spectral densities of the diabatic S$_1$ and S$_2$ states, as well as their coupling. We note the the S$_1$ and S$_2$ spectral densities only undergo minor changes due to the polar condensed phase environment. The condensed phase environment does contribute a low frequency region due to solvent reorganization effects, but the high frequency structure due to chromophore vibrational modes is largely preserved. Conversely, the coupling spectral density becomes broad and featureless in the polar solvent environment for both acetonitrile and methanol. This is especially apparent for the low frequency coupling mode at around 600~cm$^{-1}$ that is present in vacuum, but is completely quenched in solution.

\section{Additional Results: Comparison of EOM-CCSD and TDDFT diabatic energies and couplings}

In the main text, we establish that the quantum dynamics of proflavine are highly sensitive to the S$_1$-S$_2$ gap in the Condon region. Since the exact gap is uncertain due to the observed discrepancies in predictions stemming from different DFT functionals and EOM-CCSD, in this work it is treated as an adjustable parameter. To further justify this treatment, we computed vertical excitation energies and transition dipole moments using CAM-B3LYP and EOM-CCSD for three selected geometries along the trajectory of proflavine in vacuum and performed diabatization. Table \ref{si_tab:EOM_CCSD_TDDFT_diabatic} shows the resulting diabatic S$_1$ and S$_2$ energies, as well as the magnitude of the diabatic coupling $\delta_{12}$. 

\begin{table}
\begin{tabular}{ c|c|c|c|c|c|c| } 
& \multicolumn{3}{c|}{TDDFT} & \multicolumn{3}{c|}{EOM-CCSD}\\
  & E$_1$ (eV) & E$_2$ (eV) & $|\delta_{12}|$ (eV)  & E$_1$ (eV) & E$_2$ (eV) & $|\delta_{12}|$ (eV) \\ \hline
 geom 1  & 3.3321 & 3.6304 & 0.1650 & 3.2013 & 3.6239 & 0.1630 \\
 geom 2  & 3.4878 & 3.7754 & 0.006669 & 3.3936 & 3.8130 & 0.008372 \\
 geom 3  & 3.3439 & 3.4855 & 0.1112 & 3.2075 & 3.4895 & 0.1292 \\
 \hline
\end{tabular}
\caption{Diabatic S$_1$ and S$_2$ energies and diabatic couplings computed for three selected snapshots (geom 1 to 3) of proflavine in vacuum. The value are either computed at the CAM-B3LYP level of theory, or using EOM-CCSD. }
\label{si_tab:EOM_CCSD_TDDFT_diabatic}
\end{table}

We find that for all three snapshots, TDDFT and EOM-CCSD predict rather similar S$_2$ diabatic energies, but EOM-CCSD predicts S$_1$ energies that are around 0.13~eV lower, resulting in an increased S$_1$-S$_2$ gap compared to TDDFT. However, both approaches predict very similar diabatic couplings following the diabatization procedure. This suggests that both EOM-CCSD and TDDFT predict a similar curvature of the diabatic excited states along both tuning and coupling modes. This finding justifies treating the S$_1$-S$_2$ gap in the Condon region as the only adjustable parameter in our model. 

\section{Additional Results: The role of correlated fluctuations}

As described in SI Sec.~\ref{si_sec:bath_correlations}, correlated fluctuations between the S$_1$ and the S$_2$ state are encoded in the cross correlation spectral density $\mathcal{J}_\textrm{cross}(\omega)$ constructed directly from MD. This results in a normalized cross-correlation density $\tilde{\mathcal{J}}_\textrm{cross}(\omega)=\mathcal{J}_\textrm{cross}(\omega)/\sqrt{\mathcal{J}_{S_1}(\omega)\mathcal{J}_{S_2}(\omega)}$ that can take values between -1 (for fully anti-correlated fluctuations) and 1 (for fully correlated fluctuations). We note that for an LVC Hamiltonian with a finite number of modes, where S$_1$ and S$_2$ fluctuations share a common set of tuning modes, $\tilde{\mathcal{J}}_\textrm{cross}(\omega)$ can take values of 1, 0 and -1 for a given mode. Here, a value of 0 corresponds to an uncorrelated scenario in which a mode drives energy gap fluctuations in one of the two states, but not the other. For an infinite set of modes with cross correlations encoded in the continuous spectral density $\mathcal{J}_\textrm{cross}(\omega)$, $\tilde{\mathcal{J}}_\textrm{cross}(\omega)$ can take any value between -1 and 1 for a given frequency $\omega$. The physical interpretation of non-integer values in  $\tilde{\mathcal{J}}_\textrm{cross}(\omega)$ is that a large number of tuning modes exist in a small region $\omega\pm\delta\omega$, some of which are positively and some of which are negatively correlated with respect to S$_1$ and S$_2$. 

\begin{figure*}
    \centering
    \includegraphics[width=1.0\textwidth]{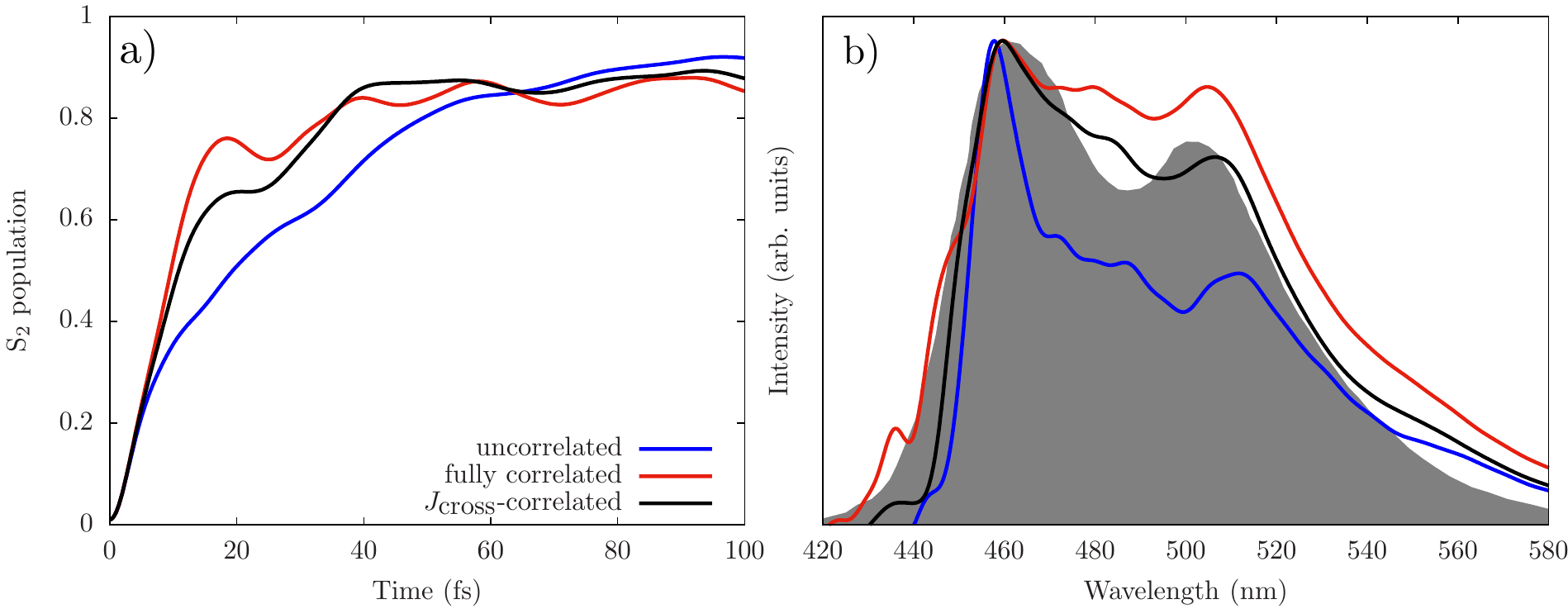}
    \caption{a) S$_2$ population dynamics and b) emission spectrum computed at a 100~fs delay time for proflavine in vacuum, with different degrees of correlation between the S$_1$ and the S$_2$ fluctuations. All spectra are normalized and shifted to align with the experimental spectrum\cite{Proflavine_exp}. }
     \label{fig:bath_correlations_comparison}
\end{figure*}

To analyse in more detail the influence of bath correlations on the fluorescence spectrum in proflavine, we compute time-resolved fluorescence spectra with a delay time of 100~fs for two scenarios: 1) A fully uncorrelated one where $\tilde{\mathcal{J}}(\omega)=0$ and 2) a fully positively correlated one where $\tilde{\mathcal{J}}(\omega)=1$. Results, in comparison to the data obtained when parameterizing $\tilde{\mathcal{J}}(\omega)$ directly from the MD-derived cross-correlation spectral density can be found in Fig.~\ref{fig:bath_correlations_comparison}. 

We find the correlation of tuning modes to play a significant role in both the population dynamics of S$_2$ upon excitation and the fluorescence spectrum. While all scenarios transfer a similar amount of population from the bright S$_1$ to the dark S$_2$ state, the rate of transfer is much slower for uncorrelated S$_1$-S$_2$ fluctuations. In the resulting emission spectrum, the shoulder due to dual fluorescence is significantly reduced in the uncorrelated case. For the fully correlated bath, population transfer is slightly faster than for the cross spectral density parameterized directly from MD and the shoulder due to the dark S$_2$ state is even more pronounced in the resulting fluorescence spectrum. We find that bath correlations, as encoded in our approach outlined in SI Sec.~\ref{si_sec:bath_correlations}, play an important role in determining the spectral lineshape of proflavine in vacuum.  

\section{Additional Results: Comparing proflavine in MeOH and acetonitrile}
In the main text, we focus on proflavine in methanol and in vacuum to analyze the quenching of dual fluorescence of the molecule in polar solvents. Here, to demonstrate the generality of the conclusions drawn, we also present results for proflavine in acetonitrile. 

\begin{figure*}
    \centering
    \includegraphics[width=1.0\textwidth]{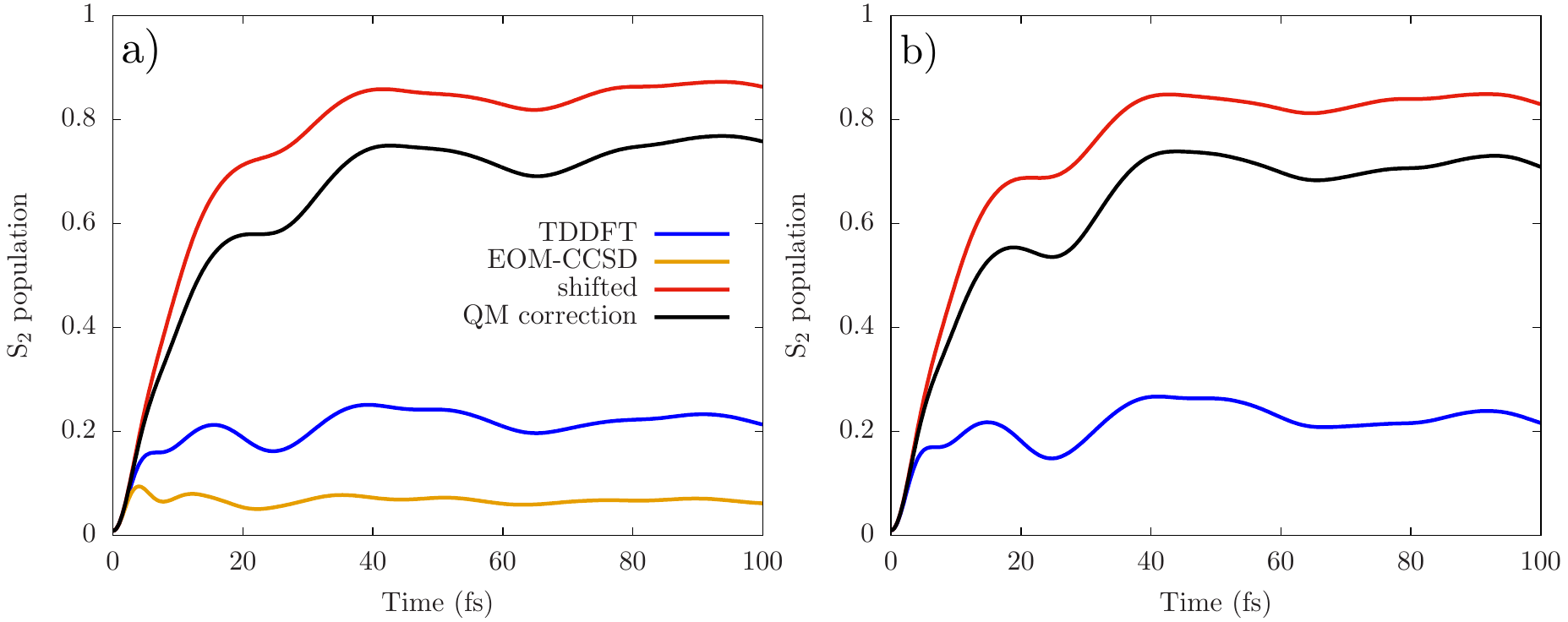}
    \caption{S$_2$ population dynamics for proflavine in a) methanol and b) acetonitrile of the diabatic S$_2$ state upon acting on the ground state wavefunction with the transition dipole operator. Four different parameterizations of the S$_1$-S$_2$ energy gap are considered. }
     \label{fig:pop_dyn_Ace_vs_MeOH}
\end{figure*}

Fig.~\ref{fig:pop_dyn_Ace_vs_MeOH} shows the population dynamics upon excitation into the bright S$_1$ state for both proflavine in methanol and acetonitrile. We note that population dynamics are very similar overall. The large S$_1$-S$_2$ gap in the Condon region for the LVC parameterization directly from TDDFT or EOM-CCSD results in only minimal population transfer to the dark S$_2$ state. On the other hand, reducing the TDDFT gap by the vacuum amount yields a population transfer of more than 80\%\, of the population to S$_2$. The LVC parameterization with a QM-corrected gap, obtained by approximately accounting for QM polarization of the solvent environment, reduces the population transfer slightly but overall shows very similar behavior to the parameterization using the vacuum-shifted gap. 

\begin{figure*}
    \centering
    \includegraphics[width=1.0\textwidth]{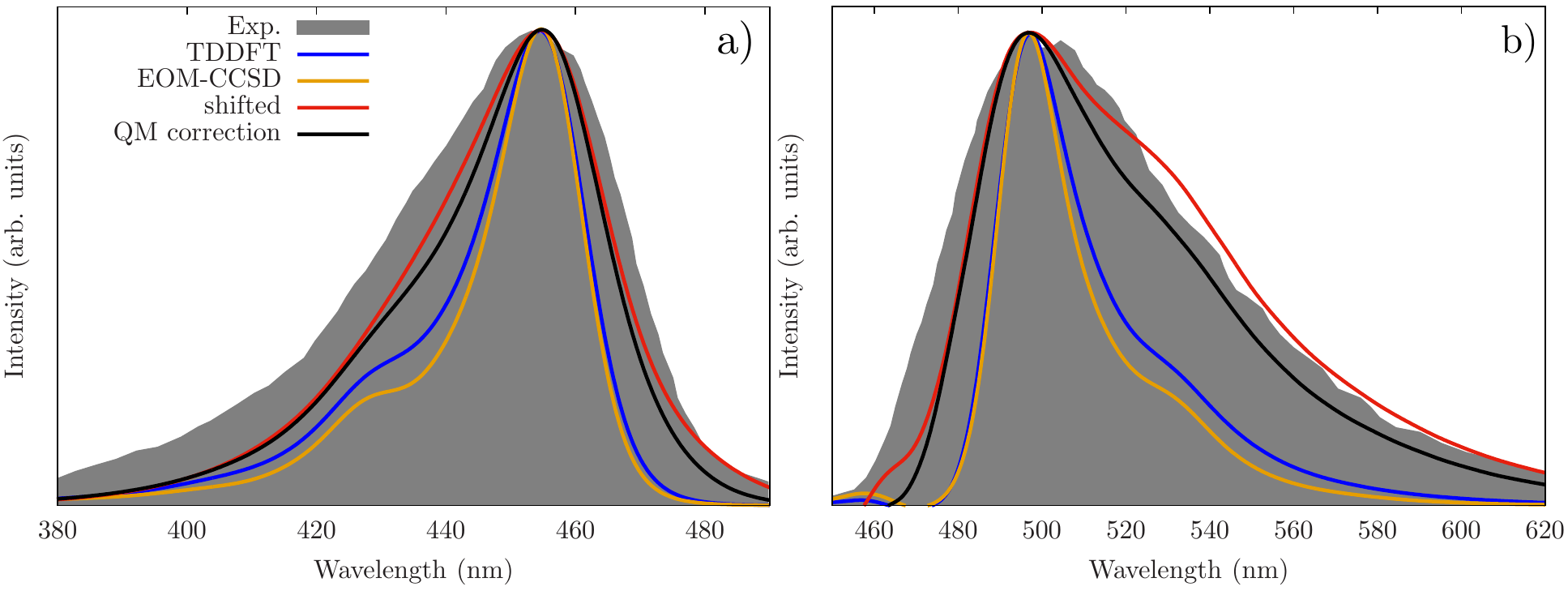}
    \caption{a) Absorption and b) fluorescence spectra for different LVC Hamiltonian parameterizations of proflavine in methanol. Experimental spectra are taken from Ref.~\onlinecite{Proflavine_exp}. Fluorescence spectra are computed after a delay time of 100~fs. All spectra are scaled and shifted to align with the experimental lineshapes. }
     \label{fig:MeOH_abs_fluo}
\end{figure*}

\begin{figure*}
    \centering
    \includegraphics[width=1.0\textwidth]{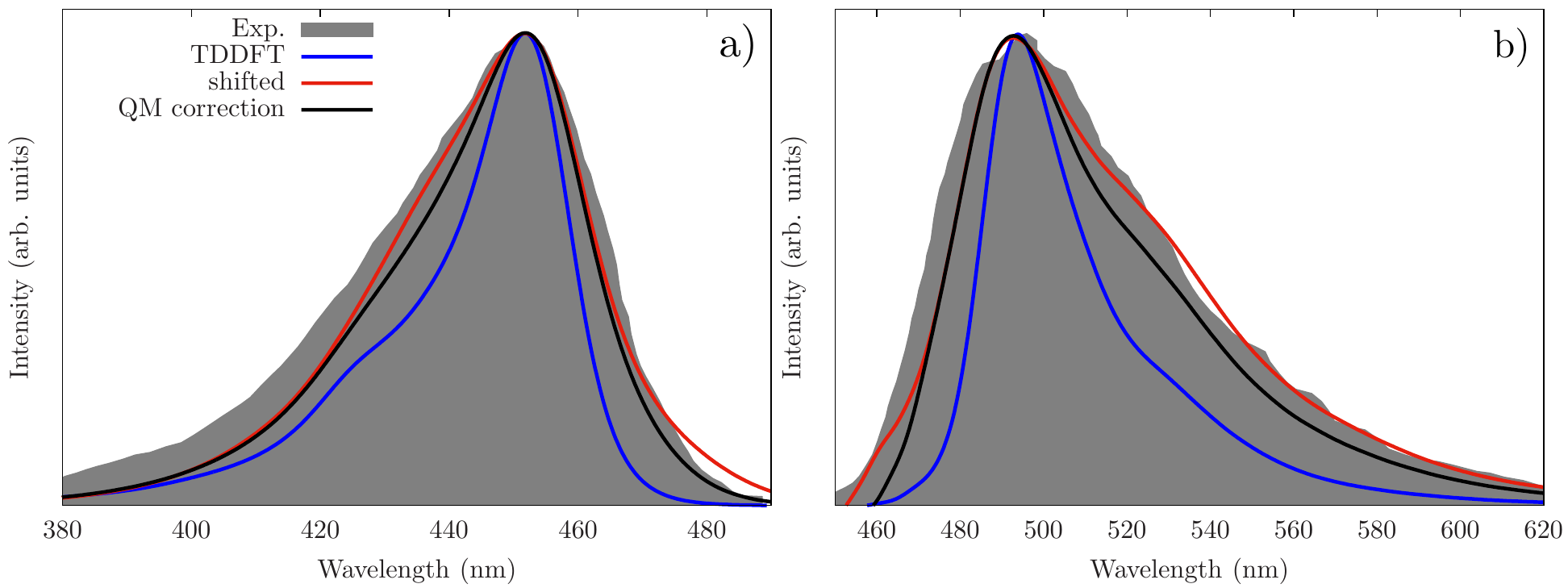}
    \caption{a) Absorption and b) fluorescence spectra for different LVC Hamiltonian parameterizations of proflavine in acetonitrile. Experimental spectra are taken from Ref.~\onlinecite{Proflavine_exp}. Fluorescence spectra are computed after a delay time of 100~fs. All spectra are scaled and shifted to align with the experimental lineshapes. }
     \label{fig:Ace_abs_fluo}
\end{figure*}

Figs.~\ref{fig:MeOH_abs_fluo} and \ref{fig:Ace_abs_fluo} show the absorption and fluorescence spectra of proflavine in methanol and acetonitrile, in comparison with experimental data\cite{Proflavine_exp}. While the TDDFT gap and EOM-CCSD gap parameterizations significantly underestimate the width of the computed spectral lineshapes, the QM corrected LVC parameterization captures both absorption and emission lineshapes in both systems very well. The results indicate that even though proflavine in acetonitrile and methanol does not exhibit the significant dual fluorescence observed in vacuum, the linear spectra are still strongly influenced by non-adiabatic effects that are well-captured by the LVC Hamiltonian.

\section{Additional Results: Analyzing solvent effects in MeOH}

\begin{figure*}
    \centering
    \includegraphics[width=1.0\textwidth]{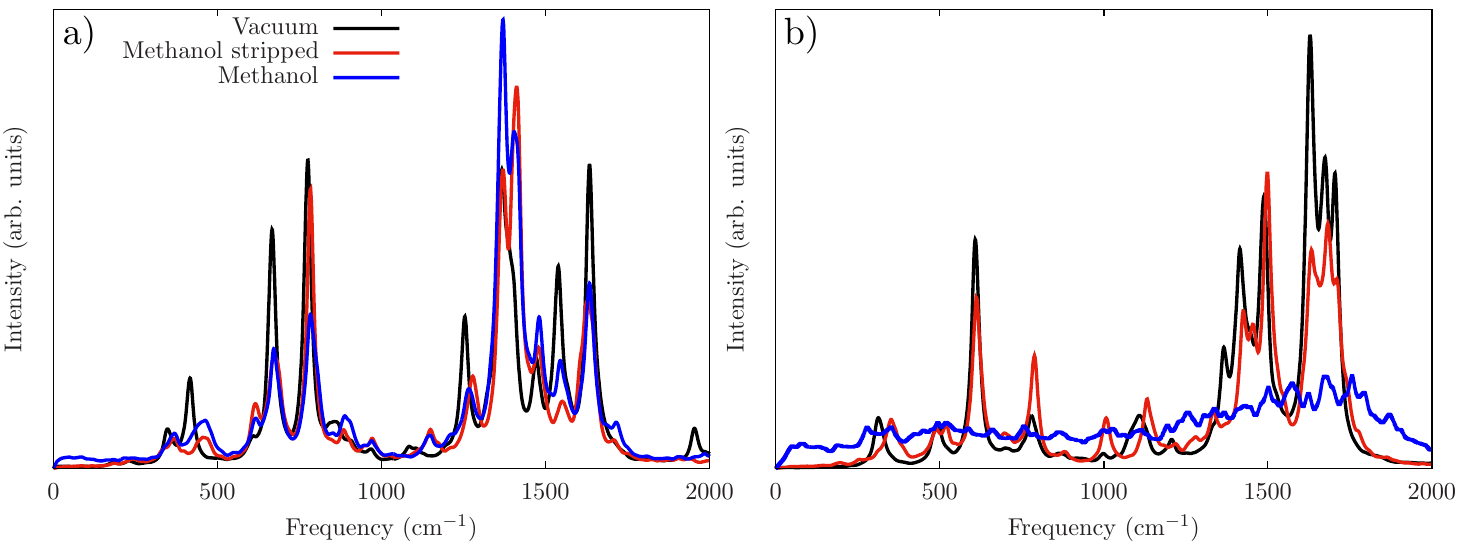}
    \caption{a) S$_1$ and b) coupling spectral density for proflavine in methanol, both where the solvent environment is treated as classical point charges and where all solvent molecules are stripped away when computing the vertical excitation energies, in comparison with spectral densities computed for proflavine in vacuum.  }
     \label{si_fig:sd_MeOH_stripped}
\end{figure*}

To isolate the influence of the polar solvent environment on the simulated emission spectrum of proflavine in MeOH, we perform calculations where, instead of representing the condensed phase environment as classical point charges when computing vertical excitation energies, all solvent molecules are stripped away. Identical snapshots as for the fully solvated calculations are used. Thus, the computed spectral densities retain effects due to solvent-induced changes in the nuclear motion of the chromophore, but effects of the polar solvent environment on vertical excitation energies are removed. The resulting spectral densities for the diabatic S$_1$ state and the S$_1$-S$_2$ coupling, in comparison with the vacuum results, can be found in Fig.~\ref{si_fig:sd_MeOH_stripped}.

We note that, in agreement with previous work\cite{Zuehlsdorff2020} by some of the authors, stripping the solvent environment removes the featureless low frequency contribution in the spectral density due to the removal of explicit solvent coupling. However, the high frequency part of the S$_1$ spectral density largely unchanged, demonstrating that these features are due to high frequency chromophore modes that are only weakly coupled to the condensed phase environment. In stark contrast, the S$_1$-S$_2$ coupling spectral density in MeOH changes from broad and featureless to highly structured upon stripping the solvent environment. The stripped coupling spectral density matches closely the spectral density computed for the vacuum trajectory. These results suggest that the polar solvent environment drives the loss of structure in the coupling spectral density observed for both methanol and acetonitrile. 

\begin{figure*}
    \centering
    \includegraphics[width=1.0\textwidth]{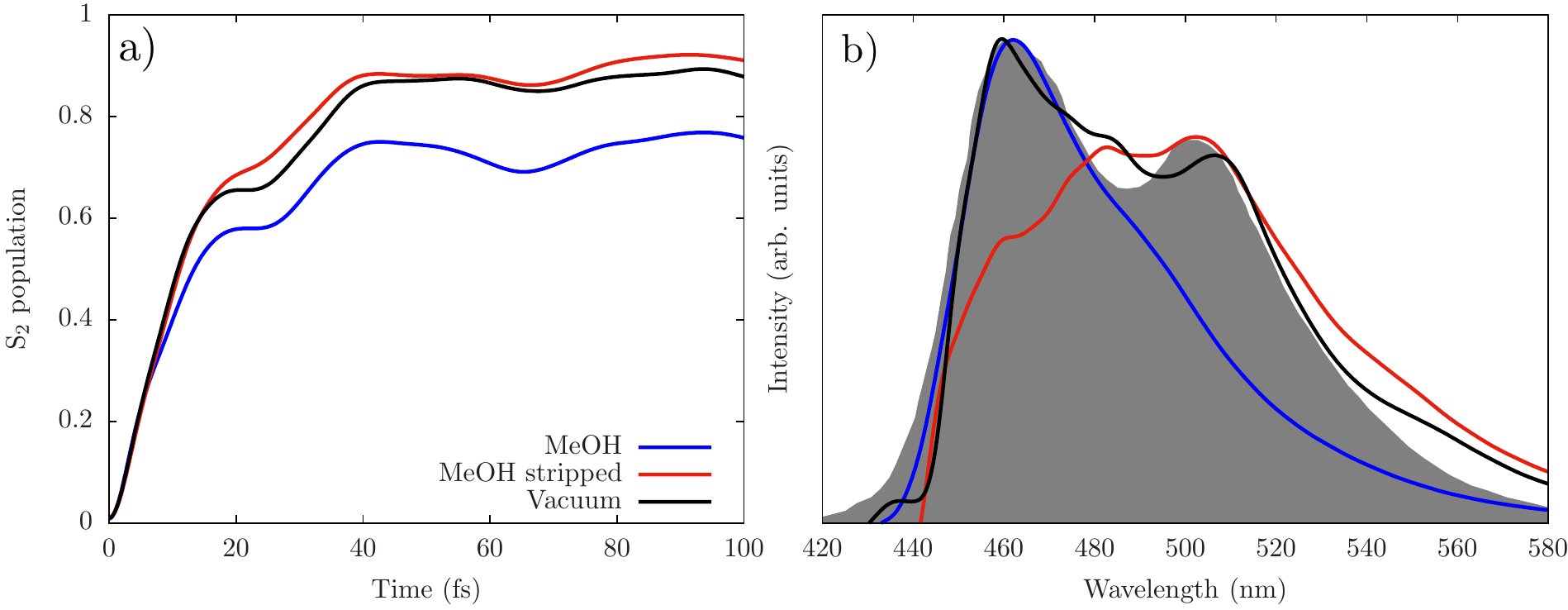}
    \caption{a) S$_2$ population dynamics upon excitation and b) emission spectrum as computed after a 100~fs delay, for proflavine in methanol, both where the solvent environment is treated as classical point charges and where all solvent molecules are stripped away when computing the vertical excitation energies. Vacuum results, as well as the experimental fluorescence spectrum in vacuum\cite{Proflavine_exp}, are displayed for comparison.  All spectra are scaled and shifted to align with the experimental spectrum in vacuum.}
     \label{si_fig:pops_emi_stripped_MeOH_vac}
\end{figure*}

Fig.~\ref{si_fig:pops_emi_stripped_MeOH_vac} shows the fluorescence spectra for proflavine in vacuum, MeOH and the MeOH stripped solvent trajectory, as computed after a 100~fs relaxation in the excited state. Population dynamics during the 100~fs relaxation are also displayed in Fig.~\ref{si_fig:pops_emi_stripped_MeOH_vac}. We note that the vacuum and stripped solvent parameterizations of the LVC Hamiltonian produce almost identical population dynamics, transferring more than 80\%\, of population from the bright S$_1$ state to the dark S$_2$ state within 100~fs. The featureless coupling spectral density in MeOH results in less population transfer to S$_2$ in the same timespan. The influence of solvent polarization effects on the resulting fluorescence lineshape is significant. Whereas the vacuum results reproduce the experimental dual fluorescence lineshape, in MeOH the second fluorescent peak due to the dark S$_2$ state is strongly quenched. Conversely, for the stripped solvent parameterization, the first peak in the fluorescence spectrum due to S$_1$ is quenched, whereas the second peak is enhanced. We conclude that a highly structured coupling spectral density is vital for preserving dual fluorescence in proflavine, and that solvent polarization effects disrupt the mixing of S$_1$ and S$_2$ via well-defined chromophore modes. 

\section{Additional Results: Model spectral densities}

To further investigate how sharp chromophore coupling modes in the spectral density drive dual fluorescence in proflavine, we perform calculations of the system in vacuum, where we replace the coupling spectral density by model spectral densities. Specifically, we consider three different types of models:
\begin{itemize}
\item 
A continuous spectral density of the Debye form, with 
\begin{equation}
\mathcal{J}^{\textrm{Debye}}_{12}(\omega) = 2\lambda^{\textrm{R}} \omega_c \frac{ \omega }{\omega_c^2 + \omega^2},
\end{equation}
where $\omega_c$ is a low frequency cutoff and $\lambda^{\textrm{R}}$ is the reorganization energy. The Debye-type spectral density, due to its continuous low frequency contribution, can be seen as a model for collective environment motion coupling S$_1$ and S$_2$, as seen for proflavine in MeOH. A cutoff frequency of $\omega_c=1.1$~mHa (241~cm$^{-1}$) is chosen for the purpose of the model cacluations.  
\item 
A Deybe spectral density with an additional broad high frequency contribution modeled by a Gaussian peak centered at 7.2~mHa (1580~cm$^{-1}$) with full width-half maximum of 2.5~mHa (549~cm$^{-1}$). Relative weigths of the Debye and the Gaussian contribution are chosen to resemble the broad coupling spectral density of proflavine in MeOH. 
\item 
A narrow Lorentzian peak with a full width-half maximum of 0.1~mHa (22~cm$^{-1}$), centered at some frequency $\omega$. This model represents just a single S$_1$-S$_2$ coupling mode. In total, we simulate spectra for three coupling mode frequencies. A low frequency mode at 0.5~mHa (110~cm$^{-1}$), a mid frequency mode at 2.5~mHa (549~cm$^{-1}$) resembling the position of the sharp coupling mode of proflavine in vacuum (see SI Fig.~\ref{fig:all_sds_comparison}), and a high frequency mode at 7.5~mHa (1646~cm$^{-1}$).
\end{itemize}

\begin{figure*}
    \centering
    \includegraphics[width=1.0\textwidth]{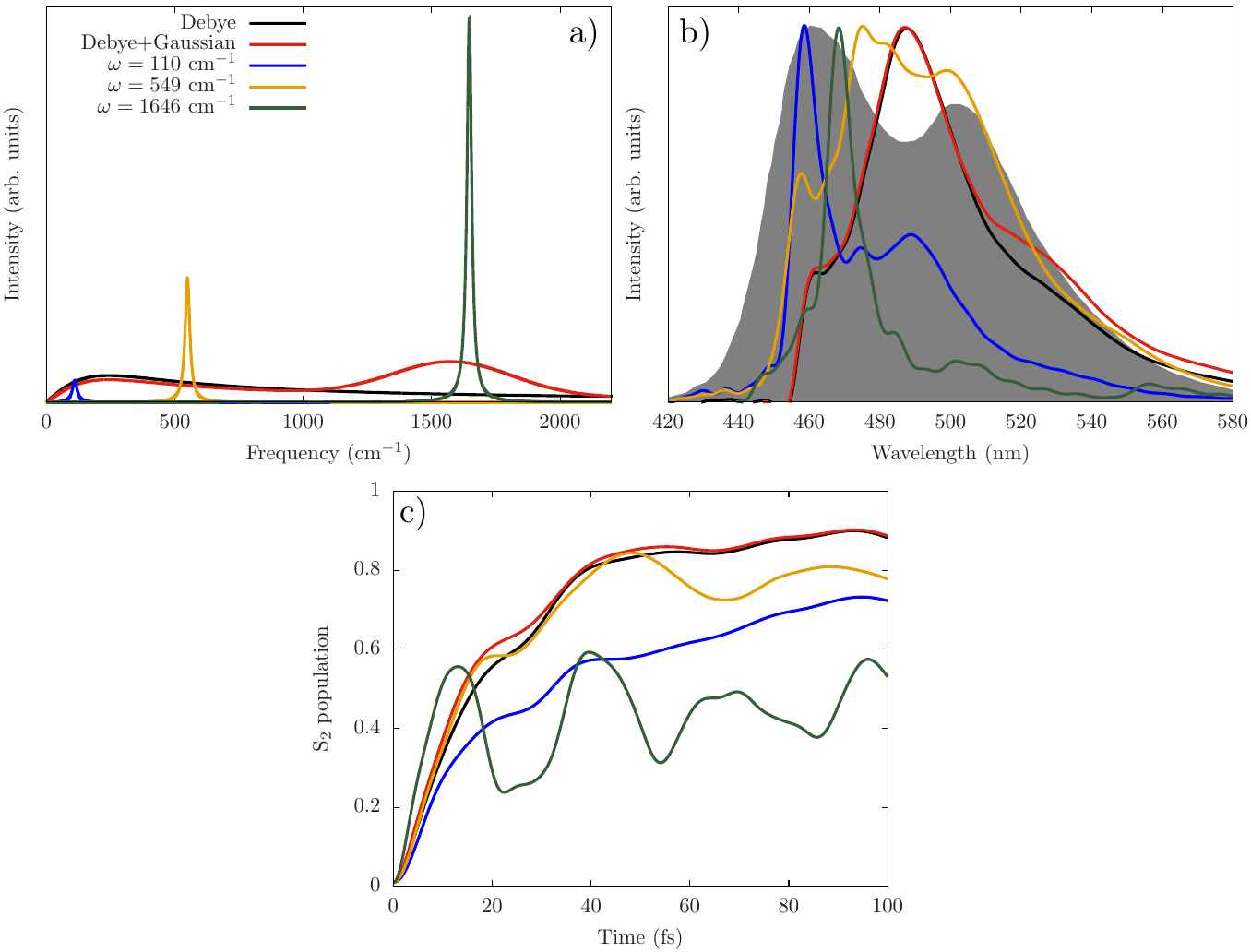}
    \caption{a) Model coupling SDs for proflavine in vacuum, based on the Debye model, the Debye model with an added high frequency Gaussian peak and Lorentzian peaks at different wavelengths. Lorentzian peaks are scaled by a factor of 0.1 to ease comparison; b) Emission spectra computed for a 100~fs delay for all model coupling SDs, in comparison with the experimental lineshape\cite{Proflavine_exp}; c) S$_2$ population for the 100~fs of excited state relaxation.}
     \label{si_fig:model_sds_vscuum}
\end{figure*}

All coupling spectral densities are scaled such that the reorganization energy contained in the coupling modes equals the reorganization energy of $\mathcal{J}_{12}$ calculated for proflavine in vacuum. The spectral densities, fluorescence spectra computed after a 100~fs delay time and S$_2$ population dynamics upon excitation are shown in SI Fig.~\ref{si_fig:model_sds_vscuum}.

We note that different coupling spectral densities result in dramatically different population dynamics and final spectra, with the broad, featureless spectral densities of the Deybe type exhibiting the most population transfer to the dark S$_2$ state upon excitation. The emission spectrum generated is broad and almost featureless, with a sharp onset that can be ascribed to emission from the S$_1$ state, and an intense shoulder at 480~nm likely due to intensity borrowing in the S$_2$ state. Most notably, the spectrum does not show two distinct peaks like in the vacuum case. Adding a high frequency contribution to the coupling SD in form of a broad Gaussian has only minor influence on the spectrum, resulting in a small additional high energy shoulder at 530~nm. 

When considering coupling to a single, well defined mode represented by the Lorentzian spectral density, we find a wide range of behaviors dependent on the frequency of the mode. For a very low frequency mode, population transfer to S$_2$ is slow and gradual, and the resulting spectrum has an intense peak attributable to S$_1$ and a shoulder due to the non-adiabatic coupling to S$_2$. The lineshape is more similar to the experimental spectrum in vacuum than for the broad coupling SDs, but with insufficient intensity in the high energy shoulder and an overall lack of spectral broadening. If the coupling mode is shifted into the high frequency region at 1646~cm$^{-1}$, we instead observe an initially very rapid population transfer to S$_2$, followed by long-lived, large-amplitude oscillations in the population. The total population transfer stays below that of any other model SDs considered here, reaching at most 60\%. The resulting fluorescence spectrum resembles more closely the cumulant spectrum for the uncoupled adiabatic states presented in Fig.~1 of the main text, which is free of any non-adiabatic effects. Only for a mid-frequency mode at 549~cm$^{-1}$, close in energy to an intense mode in the computed coupling SD for proflavine in vacuum, do we observe a fluorescence spectrum that is very broad and has an intense shoulder in the long wavelengths. Additionally, for this single mid-frequency coupling mode, population transfer closely follows the Debye+Gaussian model spectral density, only reaching a slightly lower total population transfer in the long timescale limit. 

We conclude that the presence of an intense mid-frequency coupling mode is crucial for the dual fluorescence behavior in proflavine. While proflavine exhibits similar population transfer to the dark S$_2$ state in methanol and vacuum, as well as the Debye model spectral density and the mid-frequency Lorentzian peak, the resulting fluorescence spectra are markedly different. The strong intensity borrowing effects between S$_1$ and S$_2$ causing the dual fluorescence behavior in proflavine can thus, to a large degree, be ascribed to a well-defined mid-frequency coupling mode that gets strongly quenched for proflavine in polar solvents. Interestingly, our results also demonstrate that if the frequency of the coupling mode is too high, non-adabatic effects are strongly dampened. This is likely related to there being insufficient energy available during the non-equilibrium relaxation of the wavepacket on the excited state PES to sufficiently populate the high-frequency coupling mode.  

\section{Influence of the average S$_1$-S$_2$ gap on stimulated emission}

\begin{figure*}
    \centering
    \includegraphics[width=1.0\textwidth]{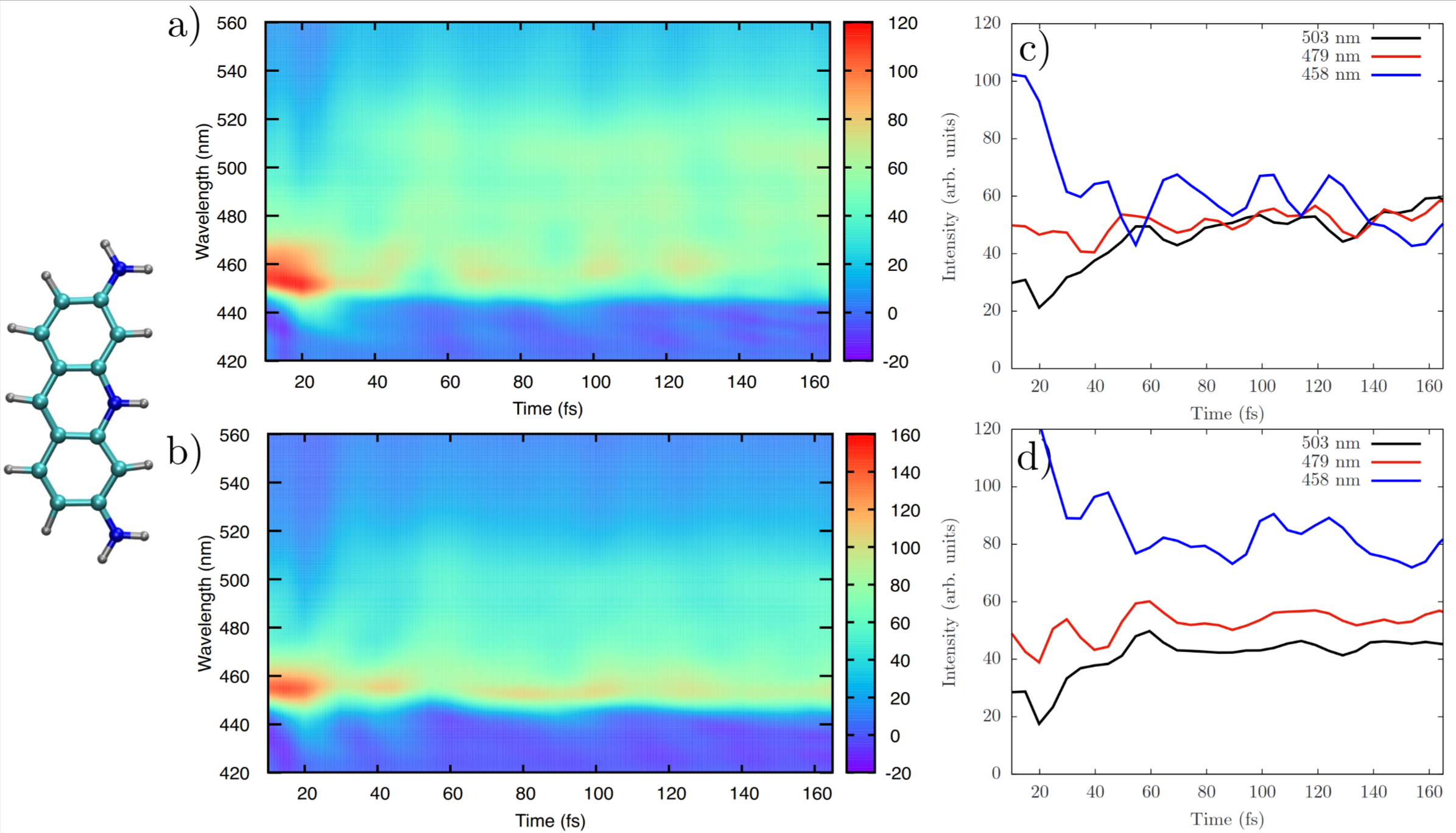}
    \caption{Time-resolved stimulated emission spectrum of proflavine as simulated in a) vacuum with our chosen reduced S$_1$-S$_2$ gap of $\Delta_{12}-0.22$~eV and b) vacuum with an increased S$_1$-S$_2$ gap of $\Delta_{12}-0.18$~eV. The plots in c) and d) represent slices through the nonlinear spectra a) and b) at specific wavelengths.}
     \label{SI_fig:trans_abs_comparsions}
\end{figure*}

To complement the results presented in the main text, we test how the stimulated emission band of proflavine in vacuum changes with respect to changing the energy gap. Specifically, we increase the energy gap from its ideal reduced value of $\Delta_{12}-0.220$ used throughout the main text by 40~meV to $\Delta_{12}-0.180$. The results can be found in Fig.~\ref{SI_fig:trans_abs_comparsions}. 

We note that increasing the gap just by 40~meV noticeably quenches the high wavelength shoulder in the vacuum spectrum. Additionally, while the heights of the main peak and the dual fluorescence shoulder in the standard vacuum parameterization undergo sustained oscillations from 40 to 160~fs (see Fig. \ref{SI_fig:trans_abs_comparsions} c)), these oscillations are strongly quenched once the S$_1$-S$_2$ gap is increased slightly. The results demonstrate again that the dual fluorescence properties of proflavine are highly sensitive to both the form of the coupling spectral density and the precise S$_1$-S$_2$ gap in the Condon region. 

\bibliography{main}

\providecommand{\latin}[1]{#1}
\makeatletter
\providecommand{\doi}
  {\begingroup\let\do\@makeother\dospecials
  \catcode`\{=1 \catcode`\}=2 \doi@aux}
\providecommand{\doi@aux}[1]{\endgroup\texttt{#1}}
\makeatother
\providecommand*\mcitethebibliography{\thebibliography}
\csname @ifundefined\endcsname{endmcitethebibliography}
  {\let\endmcitethebibliography\endthebibliography}{}
\begin{mcitethebibliography}{90}
\providecommand*\natexlab[1]{#1}
\providecommand*\mciteSetBstSublistMode[1]{}
\providecommand*\mciteSetBstMaxWidthForm[2]{}
\providecommand*\mciteBstWouldAddEndPuncttrue
  {\def\EndOfBibitem{\unskip.}}
\providecommand*\mciteBstWouldAddEndPunctfalse
  {\let\EndOfBibitem\relax}
\providecommand*\mciteSetBstMidEndSepPunct[3]{}
\providecommand*\mciteSetBstSublistLabelBeginEnd[3]{}
\providecommand*\EndOfBibitem{}
\mciteSetBstSublistMode{f}
\mciteSetBstMaxWidthForm{subitem}{(\alph{mcitesubitemcount})}
\mciteSetBstSublistLabelBeginEnd
  {\mcitemaxwidthsubitemform\space}
  {\relax}
  {\relax}

\bibitem[Tamura and Burghardt(2013)Tamura, and Burghardt]{Tamura2013}
Tamura,~H.; Burghardt,~I. Ultrafast Charge Separation in Organic Photovoltaics
  Enhanced by Charge Delocalization and Vibronically Hot Exciton Dissociation.
  \emph{J. Am. Chem. Soc.} \textbf{2013}, \emph{135}, 16364--16367, PMID:
  24138412\relax
\mciteBstWouldAddEndPuncttrue
\mciteSetBstMidEndSepPunct{\mcitedefaultmidpunct}
{\mcitedefaultendpunct}{\mcitedefaultseppunct}\relax
\EndOfBibitem
\bibitem[Northey and Penfold(2018)Northey, and Penfold]{Northey2018}
Northey,~T.; Penfold,~T. The intersystem crossing mechanism of an ultrapure
  blue organoboron emitter. \emph{Org. Electron.} \textbf{2018}, \emph{59},
  45--48\relax
\mciteBstWouldAddEndPuncttrue
\mciteSetBstMidEndSepPunct{\mcitedefaultmidpunct}
{\mcitedefaultendpunct}{\mcitedefaultseppunct}\relax
\EndOfBibitem
\bibitem[Alvertis \latin{et~al.}(2019)Alvertis, Lukman, Hele, Fuemmeler, Feng,
  Wu, Greenham, Chin, and Musser]{alvertis2019switching}
Alvertis,~A.~M.; Lukman,~S.; Hele,~T.~J.; Fuemmeler,~E.~G.; Feng,~J.; Wu,~J.;
  Greenham,~N.~C.; Chin,~A.~W.; Musser,~A.~J. Switching between coherent and
  incoherent singlet fission via solvent-induced symmetry breaking. \emph{J.
  Am. Chem. Soc.} \textbf{2019}, \emph{141}, 17558--17570\relax
\mciteBstWouldAddEndPuncttrue
\mciteSetBstMidEndSepPunct{\mcitedefaultmidpunct}
{\mcitedefaultendpunct}{\mcitedefaultseppunct}\relax
\EndOfBibitem
\bibitem[Arsenault \latin{et~al.}(2020)Arsenault, Yoneda, Iwai, Niyogi, and
  Fleming]{Arsenault2020}
Arsenault,~E.~A.; Yoneda,~Y.; Iwai,~M.; Niyogi,~K.~K.; Fleming,~G.~R. Vibronic
  mixing enables ultrafast energy flow in light-harvesting complex II.
  \emph{Nature Commun.} \textbf{2020}, \emph{11}, 1460\relax
\mciteBstWouldAddEndPuncttrue
\mciteSetBstMidEndSepPunct{\mcitedefaultmidpunct}
{\mcitedefaultendpunct}{\mcitedefaultseppunct}\relax
\EndOfBibitem
\bibitem[Hetherington \latin{et~al.}(2023)Hetherington, Mohan T.~M., Tilluck,
  Beck, and Levine]{Levine2023}
Hetherington,~C.~V.; Mohan T.~M.,~N.; Tilluck,~R.~W.; Beck,~W.~F.;
  Levine,~B.~G. Origin of Vibronic Coherences During Carrier Cooling in
  Colloidal Quantum Dots. \emph{J. Phys. Chem. Lett.} \textbf{2023}, \emph{14},
  11651--11658\relax
\mciteBstWouldAddEndPuncttrue
\mciteSetBstMidEndSepPunct{\mcitedefaultmidpunct}
{\mcitedefaultendpunct}{\mcitedefaultseppunct}\relax
\EndOfBibitem
\bibitem[Curchod and Mart\'{i}nez(2018)Curchod, and Mart\'{i}nez]{Curchod2018}
Curchod,~B. F.~E.; Mart\'{i}nez,~T.~J. Ab Initio Nonadiabatic Quantum Molecular
  Dynamics. \emph{Chem. Rev.} \textbf{2018}, \emph{118}, 3305--3336\relax
\mciteBstWouldAddEndPuncttrue
\mciteSetBstMidEndSepPunct{\mcitedefaultmidpunct}
{\mcitedefaultendpunct}{\mcitedefaultseppunct}\relax
\EndOfBibitem
\bibitem[Bircher \latin{et~al.}(2018)Bircher, Liberatore, Browning, Brickel,
  Hofmann, Patoz, Unke, Zimmermann, Chergui, Hamm, Keller, Meuwly, Woerner,
  Vanicek, and Rothlisberger]{Birchner2018}
Bircher,~M.~P.; Liberatore,~E.; Browning,~N.~J.; Brickel,~S.; Hofmann,~C.;
  Patoz,~A.; Unke,~O.~T.; Zimmermann,~T.; Chergui,~M.; Hamm,~P. \latin{et~al.}
  {Nonadiabatic effects in electronic and nuclear dynamics}. \emph{Struct.
  Dyn.} \textbf{2018}, \emph{4}, 061510\relax
\mciteBstWouldAddEndPuncttrue
\mciteSetBstMidEndSepPunct{\mcitedefaultmidpunct}
{\mcitedefaultendpunct}{\mcitedefaultseppunct}\relax
\EndOfBibitem
\bibitem[Domcke \latin{et~al.}(1981)Domcke, K{\"{o}}ppel, and
  Cederbaum]{Domcke1981}
Domcke,~W.; K{\"{o}}ppel,~H.; Cederbaum,~L.~S. {Spectroscopic effects of
  conical intersections of molecular potential energy surfaces}. \emph{Mol.
  Phys.} \textbf{1981}, \emph{43}, 851--875\relax
\mciteBstWouldAddEndPuncttrue
\mciteSetBstMidEndSepPunct{\mcitedefaultmidpunct}
{\mcitedefaultendpunct}{\mcitedefaultseppunct}\relax
\EndOfBibitem
\bibitem[Domcke and Yarkony(2012)Domcke, and Yarkony]{Domcke2012}
Domcke,~W.; Yarkony,~D.~R. {Role of conical intersections in molecular
  spectroscopy and photoinduced chemical dynamics}. \emph{Annu. Rev. Phys.
  Chem.} \textbf{2012}, \emph{63}, 325--352\relax
\mciteBstWouldAddEndPuncttrue
\mciteSetBstMidEndSepPunct{\mcitedefaultmidpunct}
{\mcitedefaultendpunct}{\mcitedefaultseppunct}\relax
\EndOfBibitem
\bibitem[Orlandi and Siebrand(1973)Orlandi, and Siebrand]{Orlandi1973}
Orlandi,~G.; Siebrand,~W. {Theory of vibronic intensity borrowing. Comparison
  of Herzberg-Teller and Born-Oppenheimer coupling}. \emph{J. Chem. Phys.}
  \textbf{1973}, \emph{4513}, 4513--4523\relax
\mciteBstWouldAddEndPuncttrue
\mciteSetBstMidEndSepPunct{\mcitedefaultmidpunct}
{\mcitedefaultendpunct}{\mcitedefaultseppunct}\relax
\EndOfBibitem
\bibitem[Aranda and Santoro(2021)Aranda, and Santoro]{Aranda2021}
Aranda,~D.; Santoro,~F. Vibronic Spectra of pi-Conjugated Systems with a
  Multitude of Coupled States: A Protocol Based on Linear Vibronic Coupling
  Models and Quantum Dynamics Tested on Hexahelicene. \emph{J. Chem. Theory
  Comput.} \textbf{2021}, \emph{17}, 1691--1700\relax
\mciteBstWouldAddEndPuncttrue
\mciteSetBstMidEndSepPunct{\mcitedefaultmidpunct}
{\mcitedefaultendpunct}{\mcitedefaultseppunct}\relax
\EndOfBibitem
\bibitem[Dunnett \latin{et~al.}(2021)Dunnett, Gowland, Isborn, Chin, and
  Zuehlsdorff]{Dunnett2021}
Dunnett,~A.~J.; Gowland,~D.; Isborn,~C.~M.; Chin,~A.~W.; Zuehlsdorff,~T.~J.
  Influence of non-adiabatic effects on linear absorption spectra in the
  condensed phase: Methylene blue. \emph{J. Chem. Phys.} \textbf{2021},
  \emph{155}, 144112\relax
\mciteBstWouldAddEndPuncttrue
\mciteSetBstMidEndSepPunct{\mcitedefaultmidpunct}
{\mcitedefaultendpunct}{\mcitedefaultseppunct}\relax
\EndOfBibitem
\bibitem[Grabowski \latin{et~al.}(2003)Grabowski, Rotkiewicz, and
  Rettig]{Grabowski2003}
Grabowski,~Z.~R.; Rotkiewicz,~K.; Rettig,~W. Structural Changes Accompanying
  Intramolecular Electron Transfer: Focus on Twisted Intramolecular
  Charge-Transfer States and Structures. \emph{Chem. Rev.} \textbf{2003},
  \emph{103}, 3899--4032\relax
\mciteBstWouldAddEndPuncttrue
\mciteSetBstMidEndSepPunct{\mcitedefaultmidpunct}
{\mcitedefaultendpunct}{\mcitedefaultseppunct}\relax
\EndOfBibitem
\bibitem[Park \latin{et~al.}(2013)Park, Kim, and Joo]{Park2013}
Park,~M.; Kim,~C.~H.; Joo,~T. Multifaceted Ultrafast Intramolecular Charge
  Transfer Dynamics of 4-(Dimethylamino)benzonitrile (DMABN). \emph{J. Phys.
  Chem. A} \textbf{2013}, \emph{117}, 370--377\relax
\mciteBstWouldAddEndPuncttrue
\mciteSetBstMidEndSepPunct{\mcitedefaultmidpunct}
{\mcitedefaultendpunct}{\mcitedefaultseppunct}\relax
\EndOfBibitem
\bibitem[Curchod \latin{et~al.}(2017)Curchod, Sisto, and
  Martínez]{Curchod2017}
Curchod,~B. F.~E.; Sisto,~A.; Martínez,~T.~J. Ab Initio Multiple Spawning
  Photochemical Dynamics of DMABN Using GPUs. \emph{J. Phys. Chem. A}
  \textbf{2017}, \emph{121}, 265--276\relax
\mciteBstWouldAddEndPuncttrue
\mciteSetBstMidEndSepPunct{\mcitedefaultmidpunct}
{\mcitedefaultendpunct}{\mcitedefaultseppunct}\relax
\EndOfBibitem
\bibitem[Demchenko \latin{et~al.}(2017)Demchenko, Tomin, and
  Chou]{Demchenko2017}
Demchenko,~A.~P.; Tomin,~V.~I.; Chou,~P.-T. Breaking the Kasha Rule for More
  Efficient Photochemistry. \emph{Chem. Rev.} \textbf{2017}, \emph{117},
  13353--13381\relax
\mciteBstWouldAddEndPuncttrue
\mciteSetBstMidEndSepPunct{\mcitedefaultmidpunct}
{\mcitedefaultendpunct}{\mcitedefaultseppunct}\relax
\EndOfBibitem
\bibitem[Djavani-Tabrizi and Jockusch(2022)Djavani-Tabrizi, and
  Jockusch]{Proflavine_exp}
Djavani-Tabrizi,~I.; Jockusch,~R.~A. Gas-Phase Fluorescence of Proflavine
  Reveals Two Close-Lying, Brightly Emitting States. \emph{J. Phys. Chem.
  Lett.} \textbf{2022}, \emph{13}, 2187--2192\relax
\mciteBstWouldAddEndPuncttrue
\mciteSetBstMidEndSepPunct{\mcitedefaultmidpunct}
{\mcitedefaultendpunct}{\mcitedefaultseppunct}\relax
\EndOfBibitem
\bibitem[Braun \latin{et~al.}(2022)Braun, Borges, Aquino, Lischka, Plasser,
  do~Monte, Ventura, Mukherjee, and Barbatti]{braun}
Braun,~G.; Borges,~J.,~Itamar; Aquino,~A. J.~A.; Lischka,~H.; Plasser,~F.;
  do~Monte,~S.~A.; Ventura,~E.; Mukherjee,~S.; Barbatti,~M. {Non-Kasha
  fluorescence of pyrene emerges from a dynamic equilibrium between excited
  states}. \emph{J. Chem. Phys.} \textbf{2022}, \emph{157}, 154305\relax
\mciteBstWouldAddEndPuncttrue
\mciteSetBstMidEndSepPunct{\mcitedefaultmidpunct}
{\mcitedefaultendpunct}{\mcitedefaultseppunct}\relax
\EndOfBibitem
\bibitem[Veys and Escudero(2022)Veys, and Escudero]{antikasha}
Veys,~K.; Escudero,~D. Anti-Kasha Fluorescence in Molecular Entities: Central
  Role of Electron–Vibrational Coupling. \emph{Acc. Chem. Res.}
  \textbf{2022}, \emph{55}, 2698--2707\relax
\mciteBstWouldAddEndPuncttrue
\mciteSetBstMidEndSepPunct{\mcitedefaultmidpunct}
{\mcitedefaultendpunct}{\mcitedefaultseppunct}\relax
\EndOfBibitem
\bibitem[Burghardt \latin{et~al.}(2004)Burghardt, Cederbaum, and
  Hynes]{Burghardt2004}
Burghardt,~I.; Cederbaum,~L.~S.; Hynes,~J.~T. Environmental effects on a
  conical intersection: A model study. \emph{Faraday Discuss.} \textbf{2004},
  \emph{127}, 395--411\relax
\mciteBstWouldAddEndPuncttrue
\mciteSetBstMidEndSepPunct{\mcitedefaultmidpunct}
{\mcitedefaultendpunct}{\mcitedefaultseppunct}\relax
\EndOfBibitem
\bibitem[Santoro \latin{et~al.}(2021)Santoro, Green, Martinez-Fernandez,
  Cerezo, and Improta]{Santoro2021}
Santoro,~F.; Green,~J.~A.; Martinez-Fernandez,~L.; Cerezo,~J.; Improta,~R.
  Quantum and semiclassical dynamical studies of nonadiabatic processes in
  solution: achievements and perspectives. \emph{Phys. Chem. Chem. Phys.}
  \textbf{2021}, \emph{23}, 8181--8199\relax
\mciteBstWouldAddEndPuncttrue
\mciteSetBstMidEndSepPunct{\mcitedefaultmidpunct}
{\mcitedefaultendpunct}{\mcitedefaultseppunct}\relax
\EndOfBibitem
\bibitem[Cerezo \latin{et~al.}(2023)Cerezo, García-Iriepa, Santoro, Navizet,
  and Prampolini]{Cerezo2023b}
Cerezo,~J.; García-Iriepa,~C.; Santoro,~F.; Navizet,~I.; Prampolini,~G.
  Unraveling the contributions to the spectral shape of flexible dyes in
  solution: insights on the absorption spectrum of an oxyluciferin analogue.
  \emph{Phys. Chem. Chem. Phys.} \textbf{2023}, \emph{25}, 5007--5020\relax
\mciteBstWouldAddEndPuncttrue
\mciteSetBstMidEndSepPunct{\mcitedefaultmidpunct}
{\mcitedefaultendpunct}{\mcitedefaultseppunct}\relax
\EndOfBibitem
\bibitem[Zuehlsdorff \latin{et~al.}(2019)Zuehlsdorff, Montoya-Castillo, Napoli,
  Markland, and Isborn]{Zuehlsdorff2019b}
Zuehlsdorff,~T.~J.; Montoya-Castillo,~A.; Napoli,~J.~A.; Markland,~T.~E.;
  Isborn,~C.~M. {Optical spectra in the condensed phase: Capturing anharmonic
  and vibronic features using dynamic and static approaches}. \emph{J. Chem.
  Phys.} \textbf{2019}, \emph{151}, 074111\relax
\mciteBstWouldAddEndPuncttrue
\mciteSetBstMidEndSepPunct{\mcitedefaultmidpunct}
{\mcitedefaultendpunct}{\mcitedefaultseppunct}\relax
\EndOfBibitem
\bibitem[Chin \latin{et~al.}(2010)Chin, Rivas, Huelga, and Plenio]{Chin2010}
Chin,~A.~W.; Rivas,~{\'{A}}.; Huelga,~S.~F.; Plenio,~M.~B. {Exact mapping
  between system-reservoir quantum models and semi-infinite discrete chains
  using orthogonal polynomials}. \emph{J. Math. Phys.} \textbf{2010},
  \emph{51}\relax
\mciteBstWouldAddEndPuncttrue
\mciteSetBstMidEndSepPunct{\mcitedefaultmidpunct}
{\mcitedefaultendpunct}{\mcitedefaultseppunct}\relax
\EndOfBibitem
\bibitem[Kloss \latin{et~al.}(2018)Kloss, Lev, and Reichman]{kloss2018time}
Kloss,~B.; Lev,~Y.~B.; Reichman,~D. Time-dependent variational principle in
  matrix-product state manifolds: Pitfalls and potential. \emph{Phys. Rev. B}
  \textbf{2018}, \emph{97}, 024307\relax
\mciteBstWouldAddEndPuncttrue
\mciteSetBstMidEndSepPunct{\mcitedefaultmidpunct}
{\mcitedefaultendpunct}{\mcitedefaultseppunct}\relax
\EndOfBibitem
\bibitem[Schr{\"{o}}der \latin{et~al.}(2019)Schr{\"{o}}der, Turban, Musser,
  Hine, and Chin]{Schroder2019}
Schr{\"{o}}der,~F.~A.; Turban,~D.~H.; Musser,~A.~J.; Hine,~N.~D.; Chin,~A.~W.
  {Tensor network simulation of multi-environmental open quantum dynamics via
  machine learning and entanglement renormalisation}. \emph{Nature Commun.}
  \textbf{2019}, \emph{10}, 1--10\relax
\mciteBstWouldAddEndPuncttrue
\mciteSetBstMidEndSepPunct{\mcitedefaultmidpunct}
{\mcitedefaultendpunct}{\mcitedefaultseppunct}\relax
\EndOfBibitem
\bibitem[Donald \latin{et~al.}(2011)Donald, Leib, Demireva, and
  Williams]{Donald2011}
Donald,~W.~A.; Leib,~R.~D.; Demireva,~M.; Williams,~E.~R. Ions in Size-Selected
  Aqueous Nanodrops: Sequential Water Molecule Binding Energies and Effects of
  Water on Ion Fluorescence. \emph{J. Am. Chem. Soc.} \textbf{2011},
  \emph{133}, 18940--18949\relax
\mciteBstWouldAddEndPuncttrue
\mciteSetBstMidEndSepPunct{\mcitedefaultmidpunct}
{\mcitedefaultendpunct}{\mcitedefaultseppunct}\relax
\EndOfBibitem
\bibitem[Kumar \latin{et~al.}(2012)Kumar, Selvaraju, Malar, and
  Natarajan]{Kumar2012}
Kumar,~K.~S.; Selvaraju,~C.; Malar,~E. J.~P.; Natarajan,~P. Existence of a New
  Emitting Singlet State of Proflavine: Femtosecond Dynamics of the Excited
  State Processes and Quantum Chemical Studies in Different Solvents. \emph{J.
  Phys. Chem. A} \textbf{2012}, \emph{116}, 37--45\relax
\mciteBstWouldAddEndPuncttrue
\mciteSetBstMidEndSepPunct{\mcitedefaultmidpunct}
{\mcitedefaultendpunct}{\mcitedefaultseppunct}\relax
\EndOfBibitem
\bibitem[Arden-Jacob \latin{et~al.}(2013)Arden-Jacob, Drexhage, Druzhinin,
  Ekimova, Flender, Lenzer, Oum, and Scholz]{Arden-Jacob2013}
Arden-Jacob,~J.; Drexhage,~K.-H.; Druzhinin,~S.~I.; Ekimova,~M.; Flender,~O.;
  Lenzer,~T.; Oum,~K.; Scholz,~M. Ultrafast photoinduced dynamics of the
  3{,}6-diaminoacridinium derivative ATTO 465 in solution. \emph{Phys. Chem.
  Chem. Phys.} \textbf{2013}, \emph{15}, 1844--1853\relax
\mciteBstWouldAddEndPuncttrue
\mciteSetBstMidEndSepPunct{\mcitedefaultmidpunct}
{\mcitedefaultendpunct}{\mcitedefaultseppunct}\relax
\EndOfBibitem
\bibitem[Kostjukova \latin{et~al.}(2021)Kostjukova, Leontieva, and
  Kostjukov]{Kostjukova2022}
Kostjukova,~L.~O.; Leontieva,~S.~V.; Kostjukov,~V.~V. The vibronic absorption
  spectra and electronic states of proflavine in aqueous solution.
  \emph{Comput. Theor. Chem.} \textbf{2021}, \emph{1197}, 113144\relax
\mciteBstWouldAddEndPuncttrue
\mciteSetBstMidEndSepPunct{\mcitedefaultmidpunct}
{\mcitedefaultendpunct}{\mcitedefaultseppunct}\relax
\EndOfBibitem
\bibitem[Savenko and Kostjukov(2023)Savenko, and Kostjukov]{Savenko2023}
Savenko,~E.~S.; Kostjukov,~V.~V. Theoretical study of the excitation of
  proflavine H-dimers in an aqueous solution: the effect of functionals and
  dispersion corrections. \emph{Phys. Chem. Chem. Phys.} \textbf{2023},
  \emph{25}, 12259--12276\relax
\mciteBstWouldAddEndPuncttrue
\mciteSetBstMidEndSepPunct{\mcitedefaultmidpunct}
{\mcitedefaultendpunct}{\mcitedefaultseppunct}\relax
\EndOfBibitem
\bibitem[Runge and Gross(1984)Runge, and Gross]{Runge1984}
Runge,~E.; Gross,~E. K.~U. Density-Functional Theory for Time-Dependent
  Systems. \emph{Phys. Rev. Lett.} \textbf{1984}, \emph{52}, 997--1000\relax
\mciteBstWouldAddEndPuncttrue
\mciteSetBstMidEndSepPunct{\mcitedefaultmidpunct}
{\mcitedefaultendpunct}{\mcitedefaultseppunct}\relax
\EndOfBibitem
\bibitem[Casida(1995)]{Casida1995}
Casida,~M.~E. \emph{Recent Advances in Density Functional Methods}; 1995; pp
  155--192\relax
\mciteBstWouldAddEndPuncttrue
\mciteSetBstMidEndSepPunct{\mcitedefaultmidpunct}
{\mcitedefaultendpunct}{\mcitedefaultseppunct}\relax
\EndOfBibitem
\bibitem[Xue \latin{et~al.}(2015)Xue, Jin, Zhang, Yang, Huo, Chen, Zou, and
  Liang]{xue2015probe}
Xue,~X.; Jin,~S.; Zhang,~C.; Yang,~K.; Huo,~S.; Chen,~F.; Zou,~G.; Liang,~X.-J.
  Probe-inspired nano-prodrug with dual-color fluorogenic property reveals
  spatiotemporal drug release in living cells. \emph{ACS nano} \textbf{2015},
  \emph{9}, 2729--2739\relax
\mciteBstWouldAddEndPuncttrue
\mciteSetBstMidEndSepPunct{\mcitedefaultmidpunct}
{\mcitedefaultendpunct}{\mcitedefaultseppunct}\relax
\EndOfBibitem
\bibitem[Wu \latin{et~al.}(2022)Wu, Liu, Tian, Groleau, Feng, Yang, Sedgwick,
  Han, Wang, Wang, Huang, Bull, Zhang, Huang, Zang, Li, He, Li, Tang, James,
  and Sessler]{cellimaging}
Wu,~L.; Liu,~J.; Tian,~X.; Groleau,~R.~R.; Feng,~B.; Yang,~Y.; Sedgwick,~A.~C.;
  Han,~H.-H.; Wang,~Y.; Wang,~H.-M. \latin{et~al.}  Dual-Channel Fluorescent
  Probe for the Simultaneous Monitoring of Peroxynitrite and
  Adenosine-5'-triphosphate in Cellular Applications. \emph{J. Am. Chem. Soc.}
  \textbf{2022}, \emph{144}, 174--183, PMID: 34931825\relax
\mciteBstWouldAddEndPuncttrue
\mciteSetBstMidEndSepPunct{\mcitedefaultmidpunct}
{\mcitedefaultendpunct}{\mcitedefaultseppunct}\relax
\EndOfBibitem
\bibitem[Yuan \latin{et~al.}(2019)Yuan, Yuan, Li, Li, Fan, and
  Yang]{yuan2019fluorescence}
Yuan,~T.; Yuan,~F.; Li,~X.; Li,~Y.; Fan,~L.; Yang,~S.
  Fluorescence--phosphorescence dual emissive carbon nitride quantum dots show
  25\% white emission efficiency enabling single-component WLEDs. \emph{Chem.
  Sci.} \textbf{2019}, \emph{10}, 9801--9806\relax
\mciteBstWouldAddEndPuncttrue
\mciteSetBstMidEndSepPunct{\mcitedefaultmidpunct}
{\mcitedefaultendpunct}{\mcitedefaultseppunct}\relax
\EndOfBibitem
\bibitem[Qian \latin{et~al.}(2017)Qian, Dai, Yang, and Yan]{qian2017high}
Qian,~H.-L.; Dai,~C.; Yang,~C.-X.; Yan,~X.-P. High-crystallinity covalent
  organic framework with dual fluorescence emissions and its ratiometric
  sensing application. \emph{ACS Appl. Mater. Inter.} \textbf{2017}, \emph{9},
  24999--25005\relax
\mciteBstWouldAddEndPuncttrue
\mciteSetBstMidEndSepPunct{\mcitedefaultmidpunct}
{\mcitedefaultendpunct}{\mcitedefaultseppunct}\relax
\EndOfBibitem
\bibitem[Baiardi \latin{et~al.}(2013)Baiardi, Bloino, and Barone]{Baiardi2013}
Baiardi,~A.; Bloino,~J.; Barone,~V. General Time Dependent Approach to Vibronic
  Spectroscopy Including Franck--Condon, Herzberg--Teller, and Duschinky
  Effects. \emph{J. Chem. Theory Comput.} \textbf{2013}, \emph{9},
  4097--4115\relax
\mciteBstWouldAddEndPuncttrue
\mciteSetBstMidEndSepPunct{\mcitedefaultmidpunct}
{\mcitedefaultendpunct}{\mcitedefaultseppunct}\relax
\EndOfBibitem
\bibitem[de~Souza \latin{et~al.}(2018)de~Souza, Neese, and
  Izs\'{a}k]{deSouza2018}
de~Souza,~B.; Neese,~F.; Izs\'{a}k,~R. On the theoretical prediction of
  fluorescence rates from first principles using the path integral approach.
  \emph{J. Chem. Phys.} \textbf{2018}, \emph{148}, 034104\relax
\mciteBstWouldAddEndPuncttrue
\mciteSetBstMidEndSepPunct{\mcitedefaultmidpunct}
{\mcitedefaultendpunct}{\mcitedefaultseppunct}\relax
\EndOfBibitem
\bibitem[Mukamel(1995)]{Mukamel-book}
Mukamel,~S. \emph{{Principles of Nonlinear Optical Spectroscopy}}; Oxford
  University Press: New York, 1995\relax
\mciteBstWouldAddEndPuncttrue
\mciteSetBstMidEndSepPunct{\mcitedefaultmidpunct}
{\mcitedefaultendpunct}{\mcitedefaultseppunct}\relax
\EndOfBibitem
\bibitem[Cammi \latin{et~al.}(2005)Cammi, Corni, Mennucci, and
  Tomasi]{Cammi_2005}
Cammi,~R.; Corni,~S.; Mennucci,~B.; Tomasi,~J. Electronic excitation energies
  of molecules in solution: State specific and linear response methods for
  nonequilibrium continuum solvation models. \emph{J. Chem. Phys.}
  \textbf{2005}, \emph{122}, 104513\relax
\mciteBstWouldAddEndPuncttrue
\mciteSetBstMidEndSepPunct{\mcitedefaultmidpunct}
{\mcitedefaultendpunct}{\mcitedefaultseppunct}\relax
\EndOfBibitem
\bibitem[Cerezo \latin{et~al.}(2015)Cerezo, Avila~Ferrer, Prampolini, and
  Santoro]{Cerezo2015}
Cerezo,~J.; Avila~Ferrer,~F.~J.; Prampolini,~G.; Santoro,~F. Modeling Solvent
  Broadening on the Vibronic Spectra of a Series of Coumarin Dyes. From
  Implicit to Explicit Solvent Models. \emph{J. Chem. Theory Comput.}
  \textbf{2015}, \emph{11}, 5810--5825\relax
\mciteBstWouldAddEndPuncttrue
\mciteSetBstMidEndSepPunct{\mcitedefaultmidpunct}
{\mcitedefaultendpunct}{\mcitedefaultseppunct}\relax
\EndOfBibitem
\bibitem[Zuehlsdorff \latin{et~al.}(2018)Zuehlsdorff, Napoli, Milanese,
  Markland, and Isborn]{Zuehlsdorff2018}
Zuehlsdorff,~T.~J.; Napoli,~J.~A.; Milanese,~J.~M.; Markland,~T.~E.;
  Isborn,~C.~M. {Unraveling electronic absorption spectra using nuclear quantum
  effects: Photoactive yellow protein and green fluorescent protein
  chromophores in water}. \emph{J. Chem. Phys.} \textbf{2018}, \emph{149},
  024107\relax
\mciteBstWouldAddEndPuncttrue
\mciteSetBstMidEndSepPunct{\mcitedefaultmidpunct}
{\mcitedefaultendpunct}{\mcitedefaultseppunct}\relax
\EndOfBibitem
\bibitem[Zuehlsdorff and Isborn(2018)Zuehlsdorff, and Isborn]{Zuehlsdorff2018b}
Zuehlsdorff,~T.~J.; Isborn,~C.~M. {Combining the ensemble and Franck-Condon
  approaches for calculating spectral shapes of molecules in solution}.
  \emph{J. Chem. Phys.} \textbf{2018}, \emph{148}, 024110\relax
\mciteBstWouldAddEndPuncttrue
\mciteSetBstMidEndSepPunct{\mcitedefaultmidpunct}
{\mcitedefaultendpunct}{\mcitedefaultseppunct}\relax
\EndOfBibitem
\bibitem[Cerezo \latin{et~al.}(2019)Cerezo, Aranda, Avila~Ferrer, Prampolini,
  and Santoro]{cerezo2019}
Cerezo,~J.; Aranda,~D.; Avila~Ferrer,~F.~J.; Prampolini,~G.; Santoro,~F.
  Adiabatic-molecular dynamics generalized vertical hessian approach: a mixed
  quantum classical method to compute electronic spectra of flexible molecules
  in the condensed phase. \emph{J. Chem. Theory Comput.} \textbf{2019},
  \emph{16}, 1215--1231\relax
\mciteBstWouldAddEndPuncttrue
\mciteSetBstMidEndSepPunct{\mcitedefaultmidpunct}
{\mcitedefaultendpunct}{\mcitedefaultseppunct}\relax
\EndOfBibitem
\bibitem[Valleau \latin{et~al.}(2012)Valleau, Eisfeld, and
  Aspuru-Guzik]{Valleau2012}
Valleau,~S.; Eisfeld,~A.; Aspuru-Guzik,~A. {On the alternatives for bath
  correlators and spectral densities from mixed quantum-classical simulations}.
  \emph{J. Chem. Phys.} \textbf{2012}, \emph{137}, 224103\relax
\mciteBstWouldAddEndPuncttrue
\mciteSetBstMidEndSepPunct{\mcitedefaultmidpunct}
{\mcitedefaultendpunct}{\mcitedefaultseppunct}\relax
\EndOfBibitem
\bibitem[Shim \latin{et~al.}(2012)Shim, Rebentrost, Valleau, and
  Aspuru-Guzik]{Shim2012}
Shim,~S.; Rebentrost,~P.; Valleau,~S.; Aspuru-Guzik,~A. {Atomistic Study of the
  Long-Lived Quantum Coherences in the Fenna-Matthews-Olson Complex}.
  \emph{Biophys. J.} \textbf{2012}, \emph{102}, 649--660\relax
\mciteBstWouldAddEndPuncttrue
\mciteSetBstMidEndSepPunct{\mcitedefaultmidpunct}
{\mcitedefaultendpunct}{\mcitedefaultseppunct}\relax
\EndOfBibitem
\bibitem[Lee and Coker(2016)Lee, and Coker]{Lee2016}
Lee,~M.~K.; Coker,~D.~F. {Modeling Electronic-Nuclear Interactions for
  Excitation Energy Transfer Processes in Light-Harvesting Complexes}. \emph{J.
  Phys. Chem. Lett.} \textbf{2016}, \emph{7}, 3171--3178\relax
\mciteBstWouldAddEndPuncttrue
\mciteSetBstMidEndSepPunct{\mcitedefaultmidpunct}
{\mcitedefaultendpunct}{\mcitedefaultseppunct}\relax
\EndOfBibitem
\bibitem[Lee \latin{et~al.}(2016)Lee, Huo, and Coker]{Lee2016b}
Lee,~M.~K.; Huo,~P.; Coker,~D.~F. {Semiclassical Path Integral Dynamics:
  Photosynthetic Energy Transfer with Realistic Environment Interactions}.
  \emph{Annu. Rev. Phys. Chem.} \textbf{2016}, \emph{67}, 639--668\relax
\mciteBstWouldAddEndPuncttrue
\mciteSetBstMidEndSepPunct{\mcitedefaultmidpunct}
{\mcitedefaultendpunct}{\mcitedefaultseppunct}\relax
\EndOfBibitem
\bibitem[Loco and Cupellini(2018)Loco, and Cupellini]{Loco2018b}
Loco,~D.; Cupellini,~L. {Modeling the absorption lineshape of embedded systems
  from molecular dynamics: A tutorial review}. \emph{Int. J. Quantum Chem.}
  \textbf{2018}, \emph{119}, e25726\relax
\mciteBstWouldAddEndPuncttrue
\mciteSetBstMidEndSepPunct{\mcitedefaultmidpunct}
{\mcitedefaultendpunct}{\mcitedefaultseppunct}\relax
\EndOfBibitem
\bibitem[Cignoni \latin{et~al.}(2022)Cignoni, Slama, Cupellini, and
  Mennucci]{Cignoni2022}
Cignoni,~E.; Slama,~V.; Cupellini,~L.; Mennucci,~B. {The atomistic modeling of
  light-harvesting complexes from the physical models to the computational
  protocol}. \emph{J. Chem. Phys} \textbf{2022}, \emph{156}, 120901\relax
\mciteBstWouldAddEndPuncttrue
\mciteSetBstMidEndSepPunct{\mcitedefaultmidpunct}
{\mcitedefaultendpunct}{\mcitedefaultseppunct}\relax
\EndOfBibitem
\bibitem[K{\"{o}}ppel \latin{et~al.}(1984)K{\"{o}}ppel, Domcke, and
  Cederbaum]{Koppel1984}
K{\"{o}}ppel,~H.; Domcke,~W.; Cederbaum,~L.~S. In \emph{{Multimode Molecular
  Dynamics Beyond the Born-Oppenheimer Approximation}}; Prigogine,~I.,
  Rice,~S.~A., Eds.; Wiley, 1984; Vol. Advances in Chemical Physics, LVII; pp
  59--246\relax
\mciteBstWouldAddEndPuncttrue
\mciteSetBstMidEndSepPunct{\mcitedefaultmidpunct}
{\mcitedefaultendpunct}{\mcitedefaultseppunct}\relax
\EndOfBibitem
\bibitem[Worth \latin{et~al.}(1996)Worth, Meyer, and Cederbaum]{Worth1996}
Worth,~G.~A.; Meyer,~H.~D.; Cederbaum,~L.~S. {The effect of a model environment
  on the S2 absorption spectrum of pyrazine: A wave packet study treating all
  24 vibrational modes}. \emph{J. Chem. Phys.} \textbf{1996}, \emph{105},
  4412--4426\relax
\mciteBstWouldAddEndPuncttrue
\mciteSetBstMidEndSepPunct{\mcitedefaultmidpunct}
{\mcitedefaultendpunct}{\mcitedefaultseppunct}\relax
\EndOfBibitem
\bibitem[Capano \latin{et~al.}(2014)Capano, Chergui, Rothlisberger, Tavernelli,
  and Penfold]{Capano2014}
Capano,~G.; Chergui,~M.; Rothlisberger,~U.; Tavernelli,~I.; Penfold,~T.~J. A
  Quantum Dynamics Study of the Ultrafast Relaxation in a Prototypical
  Cu(I)–Phenanthroline. \emph{J. Phys. Chem. A} \textbf{2014}, \emph{118},
  9861--9869\relax
\mciteBstWouldAddEndPuncttrue
\mciteSetBstMidEndSepPunct{\mcitedefaultmidpunct}
{\mcitedefaultendpunct}{\mcitedefaultseppunct}\relax
\EndOfBibitem
\bibitem[P\'{a}pai \latin{et~al.}(2016)P\'{a}pai, Vank\'{o}, Rozgonyi, and
  Penfold]{Papai2016}
P\'{a}pai,~M.; Vank\'{o},~G.; Rozgonyi,~T.; Penfold,~T.~J. High-Efficiency Iron
  Photosensitizer Explained with Quantum Wavepacket Dynamics. \emph{J. Phys.
  Chem. Lett.} \textbf{2016}, \emph{7}, 2009--2014\relax
\mciteBstWouldAddEndPuncttrue
\mciteSetBstMidEndSepPunct{\mcitedefaultmidpunct}
{\mcitedefaultendpunct}{\mcitedefaultseppunct}\relax
\EndOfBibitem
\bibitem[Neville \latin{et~al.}(2018)Neville, Stolow, and
  Schuurman]{Neville2018}
Neville,~S.~P.; Stolow,~A.; Schuurman,~M.~S. {Vacuum ultraviolet excited state
  dynamics of the smallest ring, cyclopropane. I. A reinterpretation of the
  electronic spectrum and the effect of intensity borrowing}. \emph{J. Chem.
  Phys.} \textbf{2018}, \emph{149}\relax
\mciteBstWouldAddEndPuncttrue
\mciteSetBstMidEndSepPunct{\mcitedefaultmidpunct}
{\mcitedefaultendpunct}{\mcitedefaultseppunct}\relax
\EndOfBibitem
\bibitem[Zobel \latin{et~al.}(2021)Zobel, Heindl, Plasser, Mai, and
  González]{Zobel2021}
Zobel,~J.~P.; Heindl,~M.; Plasser,~F.; Mai,~S.; González,~L. Surface Hopping
  Dynamics on Vibronic Coupling Models. \emph{Acc. Chem. Res.} \textbf{2021},
  \emph{54}, 3760--3771\relax
\mciteBstWouldAddEndPuncttrue
\mciteSetBstMidEndSepPunct{\mcitedefaultmidpunct}
{\mcitedefaultendpunct}{\mcitedefaultseppunct}\relax
\EndOfBibitem
\bibitem[Green \latin{et~al.}(2021)Green, Yaghoubi~Jouybari, Aranda, Improta,
  and Santoro]{Green2021}
Green,~J.~A.; Yaghoubi~Jouybari,~M.; Aranda,~D.; Improta,~R.; Santoro,~F.
  Nonadiabatic Absorption Spectra and Ultrafast Dynamics of DNA and RNA
  Photoexcited Nucleobases. \emph{Molecules} \textbf{2021}, \emph{26},
  1743\relax
\mciteBstWouldAddEndPuncttrue
\mciteSetBstMidEndSepPunct{\mcitedefaultmidpunct}
{\mcitedefaultendpunct}{\mcitedefaultseppunct}\relax
\EndOfBibitem
\bibitem[Segalina \latin{et~al.}(2022)Segalina, Aranda, Green, Cristino,
  Caramori, Prampolini, Pastore, and Santoro]{Segalina2022}
Segalina,~A.; Aranda,~D.; Green,~J.~A.; Cristino,~V.; Caramori,~S.;
  Prampolini,~G.; Pastore,~M.; Santoro,~F. How the Interplay among
  Conformational Disorder, Solvation, Local, and Charge-Transfer Excitations
  Affects the Absorption Spectrum and Photoinduced Dynamics of Perylene Diimide
  Dimers: A Molecular Dynamics/Quantum Vibronic Approach. \emph{J. Chem. Theory
  Comput} \textbf{2022}, \emph{18}, 3718--3736\relax
\mciteBstWouldAddEndPuncttrue
\mciteSetBstMidEndSepPunct{\mcitedefaultmidpunct}
{\mcitedefaultendpunct}{\mcitedefaultseppunct}\relax
\EndOfBibitem
\bibitem[Segatta \latin{et~al.}(2023)Segatta, Ruiz, Aleotti, Yaghoubi, Mukamel,
  Garavelli, Santoro, and Nenov]{Segatta2023}
Segatta,~F.; Ruiz,~D.~A.; Aleotti,~F.; Yaghoubi,~M.; Mukamel,~S.;
  Garavelli,~M.; Santoro,~F.; Nenov,~A. Nonlinear Molecular Electronic
  Spectroscopy via MCTDH Quantum Dynamics: From Exact to Approximate
  Expressions. \emph{J. Chem. Theory Comput.} \textbf{2023}, \emph{19},
  2075--2091\relax
\mciteBstWouldAddEndPuncttrue
\mciteSetBstMidEndSepPunct{\mcitedefaultmidpunct}
{\mcitedefaultendpunct}{\mcitedefaultseppunct}\relax
\EndOfBibitem
\bibitem[Beck \latin{et~al.}(2000)Beck, J\"{a}ckle, Worth, and Meyer]{Beck2000}
Beck,~M.~H.; J\"{a}ckle,~A.; Worth,~G.~A.; Meyer,~H.~D. The Multiconfiguration
  Time-Dependent Hartree (MCTDH) Method: A highly efficient algorithm for
  propagating wavepackets. \emph{Phys. Rep.} \textbf{2000}, \emph{324},
  1--105\relax
\mciteBstWouldAddEndPuncttrue
\mciteSetBstMidEndSepPunct{\mcitedefaultmidpunct}
{\mcitedefaultendpunct}{\mcitedefaultseppunct}\relax
\EndOfBibitem
\bibitem[Gatti and Worth(2009)Gatti, and Worth]{Meyer2009}
Gatti,~F.; Worth,~G.~A. In \emph{Multidimensional Quantum Dynamics: MCTDH
  Theory and Applications}; Meyer,~H.~D., Ed.; Wiley VCH, 2009\relax
\mciteBstWouldAddEndPuncttrue
\mciteSetBstMidEndSepPunct{\mcitedefaultmidpunct}
{\mcitedefaultendpunct}{\mcitedefaultseppunct}\relax
\EndOfBibitem
\bibitem[Egorov \latin{et~al.}(1999)Egorov, Everitt, and Skinner]{Egorov1999}
Egorov,~S.~A.; Everitt,~K.~F.; Skinner,~J.~L. {Quantum Dynamics and Vibrational
  Relaxation}. \emph{J. Phys. Chem. A} \textbf{1999}, \emph{103},
  9494–9499\relax
\mciteBstWouldAddEndPuncttrue
\mciteSetBstMidEndSepPunct{\mcitedefaultmidpunct}
{\mcitedefaultendpunct}{\mcitedefaultseppunct}\relax
\EndOfBibitem
\bibitem[Craig and Manolopoulos(2004)Craig, and Manolopoulos]{Craig2004}
Craig,~I.~R.; Manolopoulos,~D.~E. {Quantum statistics and classical mechanics:
  Real time correlation functions from ring polymer molecular dynamics}.
  \emph{J. Chem. Phys.} \textbf{2004}, \emph{121}, 3368\relax
\mciteBstWouldAddEndPuncttrue
\mciteSetBstMidEndSepPunct{\mcitedefaultmidpunct}
{\mcitedefaultendpunct}{\mcitedefaultseppunct}\relax
\EndOfBibitem
\bibitem[Ramirez \latin{et~al.}(2004)Ramirez, Lopez-Ciudad, P, and
  Marx]{Ramirez2004}
Ramirez,~R.; Lopez-Ciudad,~T.; P,~P.~K.; Marx,~D. {Quantum corrections to
  classical time-correlation functions: Hydrogen bonding and anharmonic floppy
  modes}. \emph{J. Chem. Phys.} \textbf{2004}, \emph{121}, 3973\relax
\mciteBstWouldAddEndPuncttrue
\mciteSetBstMidEndSepPunct{\mcitedefaultmidpunct}
{\mcitedefaultendpunct}{\mcitedefaultseppunct}\relax
\EndOfBibitem
\bibitem[Subotnik \latin{et~al.}(2015)Subotnik, Alguire, Ou, Landry, and
  Fatehi]{Subotnik2015}
Subotnik,~J.~E.; Alguire,~E.~C.; Ou,~Q.; Landry,~B.~R.; Fatehi,~S. The
  requisite electronic structure theory to describe photoexcited nonadiabatic
  dynamics: Nonadiabatic derivative couplings and diabatic electronic
  couplings. \emph{Acc. Chem. Res.} \textbf{2015}, \emph{48}, 1340--1350\relax
\mciteBstWouldAddEndPuncttrue
\mciteSetBstMidEndSepPunct{\mcitedefaultmidpunct}
{\mcitedefaultendpunct}{\mcitedefaultseppunct}\relax
\EndOfBibitem
\bibitem[Medders \latin{et~al.}(2017)Medders, Alguire, Jain, and
  Subotnik]{Medders2017}
Medders,~G.~R.; Alguire,~E.~C.; Jain,~A.; Subotnik,~J.~E. Ultrafast Electronic
  Relaxation through a Conical Intersection: Nonadiabatic Dynamics Disentangled
  through an Oscillator Strength-Based Diabatization Framework. \emph{J. Phys.
  Chem. A} \textbf{2017}, \emph{121}, 1425--1434\relax
\mciteBstWouldAddEndPuncttrue
\mciteSetBstMidEndSepPunct{\mcitedefaultmidpunct}
{\mcitedefaultendpunct}{\mcitedefaultseppunct}\relax
\EndOfBibitem
\bibitem[Cerezo \latin{et~al.}(2023)Cerezo, García-Iriepa, Santoro, Navizet,
  and Prampolini]{Cerezo2023}
Cerezo,~J.; García-Iriepa,~C.; Santoro,~F.; Navizet,~I.; Prampolini,~G.
  Unraveling the contributions to the spectral shape of flexible dyes in
  solution: insights on the absorption spectrum of an oxyluciferin analogue.
  \emph{Phys. Chem. Chem. Phys.} \textbf{2023}, \emph{25}, 5007--5020\relax
\mciteBstWouldAddEndPuncttrue
\mciteSetBstMidEndSepPunct{\mcitedefaultmidpunct}
{\mcitedefaultendpunct}{\mcitedefaultseppunct}\relax
\EndOfBibitem
\bibitem[Tamascelli \latin{et~al.}(2019)Tamascelli, Smirne, Lim, Huelga, and
  Plenio]{tamascelli2019efficient}
Tamascelli,~D.; Smirne,~A.; Lim,~J.; Huelga,~S.~F.; Plenio,~M.~B. Efficient
  simulation of finite-temperature open quantum systems. \emph{Phys. Rev.
  Lett.} \textbf{2019}, \emph{123}, 090402\relax
\mciteBstWouldAddEndPuncttrue
\mciteSetBstMidEndSepPunct{\mcitedefaultmidpunct}
{\mcitedefaultendpunct}{\mcitedefaultseppunct}\relax
\EndOfBibitem
\bibitem[Haegeman \latin{et~al.}(2011)Haegeman, Cirac, Osborne,
  Pi\ifmmode~\check{z}\else \v{z}\fi{}orn, Verschelde, and
  Verstraete]{Haegeman2011}
Haegeman,~J.; Cirac,~J.~I.; Osborne,~T.~J.; Pi\ifmmode~\check{z}\else
  \v{z}\fi{}orn,~I.; Verschelde,~H.; Verstraete,~F. Time-Dependent Variational
  Principle for Quantum Lattices. \emph{Phys. Rev. Lett.} \textbf{2011},
  \emph{107}, 070601\relax
\mciteBstWouldAddEndPuncttrue
\mciteSetBstMidEndSepPunct{\mcitedefaultmidpunct}
{\mcitedefaultendpunct}{\mcitedefaultseppunct}\relax
\EndOfBibitem
\bibitem[Haegeman \latin{et~al.}(2016)Haegeman, Lubich, Oseledets,
  Vandereycken, and Verstraete]{Haegeman2016}
Haegeman,~J.; Lubich,~C.; Oseledets,~I.; Vandereycken,~B.; Verstraete,~F.
  Unifying time evolution and optimization with matrix product states.
  \emph{Phys. Rev. B} \textbf{2016}, \emph{94}, 165116\relax
\mciteBstWouldAddEndPuncttrue
\mciteSetBstMidEndSepPunct{\mcitedefaultmidpunct}
{\mcitedefaultendpunct}{\mcitedefaultseppunct}\relax
\EndOfBibitem
\bibitem[Lacroix \latin{et~al.}(2021)Lacroix, Dunnett, Gribben, Lovett, and
  Chin]{PhysRevA.104.052204}
Lacroix,~T.; Dunnett,~A.; Gribben,~D.; Lovett,~B.~W.; Chin,~A. Unveiling
  non-Markovian spacetime signaling in open quantum systems with long-range
  tensor network dynamics. \emph{Phys. Rev. A} \textbf{2021}, \emph{104},
  052204\relax
\mciteBstWouldAddEndPuncttrue
\mciteSetBstMidEndSepPunct{\mcitedefaultmidpunct}
{\mcitedefaultendpunct}{\mcitedefaultseppunct}\relax
\EndOfBibitem
\bibitem[Dunnett and Chin(2021)Dunnett, and Chin]{10.3389/fchem.2020.600731}
Dunnett,~A.~J.; Chin,~A.~W. Simulating Quantum Vibronic Dynamics at Finite
  Temperatures With Many Body Wave Functions at 0 K. \emph{Frontiers in
  Chemistry} \textbf{2021}, \emph{8}, 1195\relax
\mciteBstWouldAddEndPuncttrue
\mciteSetBstMidEndSepPunct{\mcitedefaultmidpunct}
{\mcitedefaultendpunct}{\mcitedefaultseppunct}\relax
\EndOfBibitem
\bibitem[Riva \latin{et~al.}(2023)Riva, Tamascelli, Dunnett, and
  Chin]{PhysRevB.108.195138}
Riva,~A.; Tamascelli,~D.; Dunnett,~A.~J.; Chin,~A.~W. Thermal cycle and polaron
  formation in structured bosonic environments. \emph{Phys. Rev. B}
  \textbf{2023}, \emph{108}, 195138\relax
\mciteBstWouldAddEndPuncttrue
\mciteSetBstMidEndSepPunct{\mcitedefaultmidpunct}
{\mcitedefaultendpunct}{\mcitedefaultseppunct}\relax
\EndOfBibitem
\bibitem[Warshel and Levitt(1976)Warshel, and Levitt]{Warshel1976}
Warshel,~A.; Levitt,~M. Theoretical studies of enzymic reactions: Dielectric,
  electrostatic and steric stabilization of the carbonium ion in the reaction
  of lysozyme. \emph{J. Mol. Biol.} \textbf{1976}, \emph{103}, 227--249\relax
\mciteBstWouldAddEndPuncttrue
\mciteSetBstMidEndSepPunct{\mcitedefaultmidpunct}
{\mcitedefaultendpunct}{\mcitedefaultseppunct}\relax
\EndOfBibitem
\bibitem[Hirata and Head-Gordon(1999)Hirata, and Head-Gordon]{Hirata1999}
Hirata,~S.; Head-Gordon,~M. {Time-dependent density functional theory within
  the Tamm–Dancoff approximation}. \emph{Chem. Phys. Lett.} \textbf{1999},
  \emph{314}, 291--299\relax
\mciteBstWouldAddEndPuncttrue
\mciteSetBstMidEndSepPunct{\mcitedefaultmidpunct}
{\mcitedefaultendpunct}{\mcitedefaultseppunct}\relax
\EndOfBibitem
\bibitem[Yanai \latin{et~al.}(2004)Yanai, Tew, and Handy]{Yanai2004}
Yanai,~T.; Tew,~D.~P.; Handy,~N.~C. {A new hybrid exchange-correlation
  functional using the Coulomb-attenuating method (CAM-B3LYP)}. \emph{Chem.
  Phys. Lett.} \textbf{2004}, \emph{393}, 51--57\relax
\mciteBstWouldAddEndPuncttrue
\mciteSetBstMidEndSepPunct{\mcitedefaultmidpunct}
{\mcitedefaultendpunct}{\mcitedefaultseppunct}\relax
\EndOfBibitem
\bibitem[Stanton and Bartlett(1993)Stanton, and Bartlett]{Stanton1993equation}
Stanton,~J.~F.; Bartlett,~R.~J. The equation of motion coupled-cluster method.
  A systematic biorthogonal approach to molecular excitation energies,
  transition probabilities, and excited state properties. \emph{J. Chem. Phys.}
  \textbf{1993}, \emph{98}, 7029--7039\relax
\mciteBstWouldAddEndPuncttrue
\mciteSetBstMidEndSepPunct{\mcitedefaultmidpunct}
{\mcitedefaultendpunct}{\mcitedefaultseppunct}\relax
\EndOfBibitem
\bibitem[Comeau and Bartlett(1993)Comeau, and Bartlett]{Comeau1993equation}
Comeau,~D.~C.; Bartlett,~R.~J. The equation-of-motion coupled-cluster method.
  Applications to open-and closed-shell reference states. \emph{Chem. Phys.
  Lett.} \textbf{1993}, \emph{207}, 414--423\relax
\mciteBstWouldAddEndPuncttrue
\mciteSetBstMidEndSepPunct{\mcitedefaultmidpunct}
{\mcitedefaultendpunct}{\mcitedefaultseppunct}\relax
\EndOfBibitem
\bibitem[Krylov(2008)]{Krylov2008equation}
Krylov,~A.~I. Equation-of-motion coupled-cluster methods for open-shell and
  electronically excited species: The hitchhiker's guide to Fock space.
  \emph{Annu. Rev. Phys. Chem.} \textbf{2008}, \emph{59}, 433--462\relax
\mciteBstWouldAddEndPuncttrue
\mciteSetBstMidEndSepPunct{\mcitedefaultmidpunct}
{\mcitedefaultendpunct}{\mcitedefaultseppunct}\relax
\EndOfBibitem
\bibitem[Loos \latin{et~al.}(2018)Loos, Scemama, Blondel, Garniron, Caffarel,
  and Jacquemin]{Loos2018mountaineering}
Loos,~P.-F.; Scemama,~A.; Blondel,~A.; Garniron,~Y.; Caffarel,~M.;
  Jacquemin,~D. A mountaineering strategy to excited states: Highly accurate
  reference energies and benchmarks. \emph{J. Chem. Theory Comput.}
  \textbf{2018}, \emph{14}, 4360--4379\relax
\mciteBstWouldAddEndPuncttrue
\mciteSetBstMidEndSepPunct{\mcitedefaultmidpunct}
{\mcitedefaultendpunct}{\mcitedefaultseppunct}\relax
\EndOfBibitem
\bibitem[Loos \latin{et~al.}(2020)Loos, Lipparini, Boggio-Pasqua, Scemama, and
  Jacquemin]{Loos2020mountaineering}
Loos,~P.-F.; Lipparini,~F.; Boggio-Pasqua,~M.; Scemama,~A.; Jacquemin,~D. A
  mountaineering strategy to excited states: Highly accurate energies and
  benchmarks for medium sized molecules. \emph{J. Chem. Theory Comput.}
  \textbf{2020}, \emph{16}, 1711--1741\relax
\mciteBstWouldAddEndPuncttrue
\mciteSetBstMidEndSepPunct{\mcitedefaultmidpunct}
{\mcitedefaultendpunct}{\mcitedefaultseppunct}\relax
\EndOfBibitem
\bibitem[Humeniuk and Glover(0)Humeniuk, and Glover]{Glover2024}
Humeniuk,~A.; Glover,~W.~J. Multistate, Polarizable QM/MM Embedding Scheme
  Based on the Direct Reaction Field Method: Solvatochromic Shifts, Analytical
  Gradients and Optimizations of Conical Intersections in Solution. \emph{J.
  Chem. Theory Comput.} \textbf{0}, \emph{0}, null\relax
\mciteBstWouldAddEndPuncttrue
\mciteSetBstMidEndSepPunct{\mcitedefaultmidpunct}
{\mcitedefaultendpunct}{\mcitedefaultseppunct}\relax
\EndOfBibitem
\bibitem[Zuehlsdorff \latin{et~al.}(2020)Zuehlsdorff, Hong, Shi, and
  Isborn]{Zuehlsdorff2020}
Zuehlsdorff,~T.~J.; Hong,~H.; Shi,~L.; Isborn,~C.~M. {Influence of Electronic
  Polarization on the Spectral Density}. \emph{J. Phys. Chem. B} \textbf{2020},
  \emph{124}, 531--543\relax
\mciteBstWouldAddEndPuncttrue
\mciteSetBstMidEndSepPunct{\mcitedefaultmidpunct}
{\mcitedefaultendpunct}{\mcitedefaultseppunct}\relax
\EndOfBibitem
\bibitem[Galiana and Lasorne(2023)Galiana, and Lasorne]{galiana23}
Galiana,~J.; Lasorne,~B. {On the unusual Stokes shift in the smallest PPE
  dendrimer building block: Role of the vibronic symmetry on the band origin?}
  \emph{J. Chem. Phys.} \textbf{2023}, \emph{158}, 124113\relax
\mciteBstWouldAddEndPuncttrue
\mciteSetBstMidEndSepPunct{\mcitedefaultmidpunct}
{\mcitedefaultendpunct}{\mcitedefaultseppunct}\relax
\EndOfBibitem
\bibitem[Gallop \latin{et~al.}(2024)Gallop, Maslennikov, Mondal, Goetz, Dai,
  Schankler, Sung, Nihonyanagi, Tahara, Bodnarchuk, \latin{et~al.}
  others]{gallop2024ultrafast}
Gallop,~N.~P.; Maslennikov,~D.~R.; Mondal,~N.; Goetz,~K.~P.; Dai,~Z.;
  Schankler,~A.~M.; Sung,~W.; Nihonyanagi,~S.; Tahara,~T.; Bodnarchuk,~M.~I.
  \latin{et~al.}  Ultrafast vibrational control of organohalide perovskite
  optoelectronic devices using vibrationally promoted electronic resonance.
  \emph{Nat. Mater.} \textbf{2024}, \emph{23}, 88--94\relax
\mciteBstWouldAddEndPuncttrue
\mciteSetBstMidEndSepPunct{\mcitedefaultmidpunct}
{\mcitedefaultendpunct}{\mcitedefaultseppunct}\relax
\EndOfBibitem
\bibitem[Garcia-Vidal \latin{et~al.}(2021)Garcia-Vidal, Ciuti, and
  Ebbesen]{garcia2021manipulating}
Garcia-Vidal,~F.~J.; Ciuti,~C.; Ebbesen,~T.~W. Manipulating matter by strong
  coupling to vacuum fields. \emph{Science} \textbf{2021}, \emph{373},
  eabd0336\relax
\mciteBstWouldAddEndPuncttrue
\mciteSetBstMidEndSepPunct{\mcitedefaultmidpunct}
{\mcitedefaultendpunct}{\mcitedefaultseppunct}\relax
\EndOfBibitem
\bibitem[Sokolovskii \latin{et~al.}(2023)Sokolovskii, Tichauer, Morozov, Feist,
  and Groenhof]{sokolovskii2023multi}
Sokolovskii,~I.; Tichauer,~R.~H.; Morozov,~D.; Feist,~J.; Groenhof,~G.
  Multi-scale molecular dynamics simulations of enhanced energy transfer in
  organic molecules under strong coupling. \emph{Nat. Commun.} \textbf{2023},
  \emph{14}, 6613\relax
\mciteBstWouldAddEndPuncttrue
\mciteSetBstMidEndSepPunct{\mcitedefaultmidpunct}
{\mcitedefaultendpunct}{\mcitedefaultseppunct}\relax
\EndOfBibitem
\bibitem[Dunnett(2021)]{dunnett_angus_2021_5106435}
Dunnett,~A. angusdunnett/MPSDynamics:. 2021;
  \url{https://doi.org/10.5281/zenodo.5106435}\relax
\mciteBstWouldAddEndPuncttrue
\mciteSetBstMidEndSepPunct{\mcitedefaultmidpunct}
{\mcitedefaultendpunct}{\mcitedefaultseppunct}\relax
\EndOfBibitem
\end{mcitethebibliography}

\end{document}